%% file: main.tex
\documentclass[a4paper,11pt]{article}
\usepackage{jheppub} % for details on the use of the package, please see the JINST-author-manual
\usepackage{lineno}

\newcommand{\Bzerodecaymu}{\ensuremath{\Bzb\to\Dstarp\mun\neumb}}
\newcommand{\BzerodecayDdecay}{\ensuremath{\Bzb\to\Dstarp (\to\Dz\pip)\mun\neumb}}
\newcommand{\Bzerodecayswave}{\ensuremath{\Bzb\to\Dz\pip\mun\neumb}}
\newcommand{\Bzerodecayell}{\ensuremath{\Bzb\to\Dstarp\ellm\neulb}}
\newcommand{\Ddecay}{\ensuremath{\Dstarp\to\Dz\pip}}
\newcommand{\Ddoublestar}{{\ensuremath{\D^{**}}}\xspace}
\newcommand{\Dzerostar}{\ensuremath{\D^\ast_0(2300)}}

\newcommand{\Done}{\ensuremath{\D_1(2420)}}
\newcommand{\Dtwo}{\ensuremath{\D^\ast_2(2460)}}
\usepackage{longtable}
\usepackage{calc}
\usepackage{booktabs}
\usepackage{comment}
\usepackage[normalem]{ulem}

\input{symbols-def}

\title{\boldmath The Semileptonic B Frontier: \\LHC, B-factories, Lattice QCD \& the Hunt for Precision}

\author[a]{F. De Fazio,}
\author[b]{P. De Simone,}
\author[c]{M. Fedele,}
\author[b]{G. Finocchiaro,}
\author[b]{F. Mescia,}
\author[b]{M. Rotondo,}
\author[c]{A. Barone,}
\author[d]{B. Capdevila,}
\author[e]{M. De Cian,}
\author[f]{L. Dreyfus,}
\author[g]{M. Fael,}
\author[h]{G. Finauri,}
\author[i]{G. Gagliardi,}
\author[j]{G. Gaudino,}
\author[k,l]{D. Giusti,}
\author[m]{N. Gubernari,}
\author[f]{A-M. Heyn,}
\author[n]{F. Manganella,}
\author[b]{E. Minucci,}
\author[e]{B. Mitreska,}
\author[o]{E. Manoni,}
\author[p]{T. Martinov,}
\author[e]{L. Paolucci,}
\author[q]{Z. Polonsky,}
\author[g]{A. Poluektov,}
\author[b]{M. Palutan,}
\author[i]{F. Sanfilippo,}
\author[i]{S. Simula,}
\author[i]{A. Smecca,}
\author[n]{N. Tantalo,}
\author[q]{A. Tinari,}
\author[r]{A. Vaquero,}
\author[d]{R. Vazquez Gomez,}
\author[s]{M. Valli,}
\author[t]{L. Vittorio,}
\author[f]{V. Vobbilisetti}

\affiliation[a]{INFN, Sezione di Bari, via Orabona 4, I-70126 Bari, Italy}
\affiliation[b]{Laboratori Nazionali di Frascati, Via E. Fermi 54, I-00040 Frascati, Italy}
\affiliation[c]{PRISMA$^{++}$ Cluster of Excellence \& Mainz Institute for Theoretical Physics, Johannes Gutenberg University, D-55099 Mainz, Germany}
\affiliation[d]{ICCUB, Universitat de Barcelona, Mart\'i i Franqu\`es 1, E-08028 Barcelona, Spain}
\affiliation[e]{Department of Physics and Astronomy, University of Manchester, Manchester, United Kingdom}
\affiliation[f]{Aix Marseille Univ, CNRS/IN2P3, CPPM, Marseille, France}
\affiliation[g]{Dipartimento di Fisica e Astronomia Galileo Galilei \& INFN, Universit\`a di Padova, via F. Marzolo 8, I-35131 Padova, Italy}
\affiliation[h]{Dipartimento di Fisica \& INFN, Universit\`a di Torino, Via Pietro Giuria 1, I-10125 Turin, Italy}
\affiliation[i]{INFN, Sezione di Roma Tre, Via della Vasca Navale 84, I-00146 Rome, Italy}
\affiliation[j]{INFN, Sezione di Napoli, Via Cintia, I-80126 Napoli, Italy}
\affiliation[k]{J\"ulich Supercomputing Centre, Forschungszentrum J\"ulich, D-52428 J\"ulich, Germany}
\affiliation[l]{Fakult\"at f\"ur Physik, Universit\"at Regensburg, D-93040, Germany}
\affiliation[m]{Helmholtz-Institut f\"ur Strahlen- und Kernphysik, Universit\"at Bonn, D-53115 Bonn, Germany}
\affiliation[n]{Dipartimento di Fisica \& INFN, Universit\'a di Roma Tor Vergata, Via della Ricerca Scientifica 1, I-00133 Roma, Italy}
\affiliation[o]{INFN, Sezione di Perugia, Via A. Pascoli, I-06123 Perugia, Italy}
\affiliation[p]{INFN, Sezione di Trieste, Via Padriciano 99, I-34012 Trieste, Italy}
\affiliation[q]{Physik-Institut, Universit\"at Z\"urich, CH-8057 Z\"urich, Switzerland}
\affiliation[r]{Departmento de Física Teórica, Universidad de Zaragoza and CAPA, CP50009, Zaragoza, Spain}
\affiliation[s]{INFN, Sezione di Roma, Piazzale A. Moro 2, I-00185 Roma, Italy}
\affiliation[t]{Dipartimento di Fisica \& INFN, Universit\`a di Roma La Sapienza, Piazzale A. Moro 2, I-00185 Roma, Italy}

\emailAdd{fulvia.defazio@ba.infn.it}
\emailAdd{patrizia.desimone@cern.ch}
\emailAdd{mfedele@uni-mainz.de}
\emailAdd{giuseppe.finocchiaro@lnf.infn.it}
\emailAdd{federico.mescia@lnf.infn.it}
\emailAdd{marcello.rotondo@cern.ch}

\newpage

\abstract{
This review document summarizes the main outcomes and key theoretical and experimental developments discussed at the workshop \textit{``The Semileptonic $B$ Frontier: LHC, $B$-Factories, Lattice QCD \& the Hunt for Precision''}, held at Laboratori Nazionali di Frascati on 16--18 February, (\href{https://agenda.infn.it/event/49469/}{workshop agenda}).
\\
Bringing together experimentalists and theorists, the meeting addressed recent advances and outstanding challenges in both charged-current ($b \to c/u\,\ell\nu$) and neutral-current ($b \to s\,\ell^+\ell^-$) semileptonic $b$-hadron decays. We synthesize current experimental measurements from Belle, Belle~II, and LHCb with state-of-the-art theoretical inputs---including Lattice QCD determinations, Heavy Quark Expansion (HQE) frameworks, and QED radiative corrections. In addition to reviewing the status of CKM matrix elements, hadronic form factor extractions, and Lepton Flavor Universality (LFU) tests, this report highlights novel observables and fitting techniques designed to maximize the sensitivity of future analyses and guide upcoming data-taking campaigns.}

\begin{document}
\maketitle
\flushbottom

\clearpage

\section{Table of Abbreviations}

\begin{longtable}[l]{p{0.2\textwidth}p{0.8\textwidth-4\tabcolsep}}
  \textbf{Abbreviation}&\textbf{Meaning}\\
  \midrule\endhead
    BDT   & Boosted Decision Tree \\
    BGL   & Boyd--Grinstein--Lebed \\
    BLNP  & Bosch--Lange-Neubert--Paz \\
    BLPR  & Bernlochner--Ligeti--Papucci--Robinson \\
    BSM   & Beyond the Standard Model \\
    CKM   & Cabibbo--Kobayashi--Maskawa \\
    CL    & Confidence Level \\
    CLN   & Caprini--Lellouch--Neubert \\
    CM    & Center-of-Mass \\
    DM    & Dispersive Matrix \\
    EFT   & Effective Field Theory \\
    FCNC  & Flavour Changing Neutral Current \\
    FEI   & Full Event Interpretation \\
    FF    & Form Factor \\
    GGOU  & Gambino--Giordano--Ossola-Uraltsev \\
    HISQ  & Highly Improved Staggered Quarks \\
    HLT   & Hansen--Lupo--Tantalo \\
    HQE   & Heavy Quark Expansion \\
    HQET  & Heavy Quark Effective Theory \\
    LCSR  & Light-Cone Sum Rules \\
    LFU   & Lepton Flavour Universality \\
    LHC   & Large Hadron Collider \\
    LO    & Leading order \\
    LQCD  & Lattice QCD \\
    NLO   & Next-to-Leading Order \\
    NNLO  & Next-to-Next-to-Leading Order \\
    NP    & New Physics \\
    OPE   & Operator Product Expansion \\
    PID   & Particle Identification \\
    RFA   & Rest Frame Approximation \\
    SM    & Standard Model \\
    SMEFT & Standard Model Effective Field Theory \\
    UV    & Ultraviolet \\
\end{longtable}
\addtocounter{table}{-1}

\clearpage

%%%%%%%%%%%%%%%%%%%%%%%%%%%%%%%%%%%%%%%%%%%%%%%%%%
% Geeral introduction, with overview of the report 
%%%%%%%%%%%%%%%%%%%%%%%%%%%%%%%%%%%%%%%%%%%%%%%%%%
\section{Introduction}
Semileptonic decays of $b$ hadrons offer some of the cleanest experimental
windows onto the CKM matrix elements \Vcb and \Vub, while also serving as
sensitive probes of New Physics at the TeV scale and beyond. Mediated at tree
level by the $W$ boson, these transitions retain a single leptonic current in
the final state and are, at least in the exclusive case,  much less affected by strong-interaction effects than purely hadronic decays. This relative theoretical cleanliness, together with
comparatively large branching fractions, has made semileptonic decays
ideal for determining the magnitudes of the CKM matrix elements
governing $b$-quark transitions, and one of the primary tools with which the
flavour sector of the Standard Model has been tested over the past three
decades.

The importance of measurements using semileptonic decays extends well beyond the determination of
\Vcb and \Vub themselves. Because the CKM matrix is unitary, precise
knowledge of its elements provides a stringent, over-constrained test of the
three-generation Standard Model picture of quark mixing and CP violation: any
inconsistency among independent determinations of the matrix elements, or
between measured and predicted CP-violating observables, would point towards
sources of flavour and CP violation beyond the SM. Moreover, \Vcb and
\Vub are essential normalisation factors in the theoretical predictions
for a wide range of rare and CP-violating processes, from $B$-meson mixing to
radiative and rare semileptonic decays - any bias in their
determination propagates directly into the interpretation of those processes.
Semileptonic decays are therefore not merely a precision-physics exercise in
their own right, but constitute a necessary theoretical input for essentially the entire
programme of indirect New Physics searches in the $b$-quark sector.

The past several years have seen remarkable progress on multiple fronts.
At the experimental level, the Belle~II experiment at SuperKEKB has begun
delivering  high-precision measurements of exclusive and inclusive semileptonic
transitions, exploiting the clean $e^+e^-$ environment and new algorithms to
achieve reconstruction of final states with multiple neutrinos.
The well-defined initial state available at a $B$ factory, together with
almost complete angular coverage, allows the missing four-momentum associated
with one or more neutrinos to be inferred with a precision unmatched at a
hadron collider, and makes possible searches for decay modes -- such as those
with $\tau$ leptons or purely leptonic final states -- that would otherwise
be inaccessible. At the LHC, the \lhcb experiment continues its programme by
studying not only the \Bd and \Bu mesons, but also the \Bs, \Bc, \Lb, and
heavier baryons that are not available at $B$ factories. Thanks to the huge
$b$-hadron production cross-section at the LHC and to the large boost of the
produced hadrons, \lhcb is able to access a broad range of
$b$-hadron species and a large variety of decay channels, and is pushing the frontier of
precision in the hadronic environment despite the challenges posed by missing
neutrinos and large combinatorial backgrounds. The complementarity between
these two experimental approaches -- one offering kinematic closure and
purity, the other statistical power and access to the full spectrum of
$b$-hadron states -- is a recurring theme throughout this Report, and is
central to resolving many of the open questions that are discussed.

On the theoretical side, there has been substantial progress in perturbative
calculations for both inclusive decays and rare processes and in non-perturbative
determinations of hadronic form factors from lattice-QCD for
exclusive $b\to c$ and $b\to u$ transitions. New techniques are also emerging
for direct lattice computation of inclusive decay rates, a programme that
until recently has been considered essentially out of reach. These developments
have been accompanied by increasingly sophisticated frameworks for combining
theoretical and experimental information, such as unitarity- and
analyticity-based form-factor parametrizations, or global fits incorporating
simultaneously Standard Model and New Physics degrees of freedom. They 
have substantially sharpened the interpretation of existing data even in the
absence of qualitatively new measurements.

Nonetheless, a number of longstanding puzzles persist, and much of this
Report is organised around them. The determination of \Vcb from exclusive
decays consistently yields values smaller than inclusive determinations at the
$2$--$3\sigma$ level. This tension has persisted for nearly two decades and
its resolution requires a coherent improvement in both experimental
measurements and lattice form-factor determinations. A closely related,
if historically less severe, tension is also present in the determination of
\Vub. The ratios $R(D)$ and $R(D^*)$, which test LFU in charged-current
$b \to c\tau\nu$ transitions, have in the past shown a combined discrepancy
with the SM prediction at the level of several standard
deviations. As discussed in this Report, the size and even the
existence of this tension now depends sensitively on the theoretical
treatment adopted for the relevant hadronic form factors, and clarifying this
dependence is itself an important open problem. In the neutral-current
sector, measurements of angular observables and differential branching
fractions in $b \to s\ell^+\ell^-$ transitions continue to exhibit systematic
tensions with SM predictions, but the question of whether these reflect some
kind of New Physics, or underestimated long-distance QCD contributions from
charm rescattering and charming-penguin topologies, remains open and is the
subject of intense ongoing theoretical work. In each of these cases, further
progress will require not only larger datasets, but a more complete
theoretical control of the relevant hadronic dynamics -- whether encoded in
exclusive form factors, inclusive structure functions, or non-local
correlators -- and a correspondingly careful assessment of the associated
uncertainties.

This Report aims to provide a comprehensive and timely account of these
developments. It brings together contributions from theorists and
experimentalists working at the frontier of semileptonic $b$-hadron physics,
covering both,  charged-current processes, including exclusive and inclusive $b \to c$ and $b \to u$ transitions,
and neutral-current processes, driven by the $b \to
s\ell^+\ell^-$ flavour-changing neutral current. A particular emphasis is
placed on the interplay between experiment and theory, and, within theory,
between lattice and continuum approaches to the relevant hadronic matrix
elements. Many of the open questions summarised above can only be settled
through the combination of improved measurements, improved non-perturbative
calculations, and a coherent treatment of the associated systematic
uncertainties. The Report also addresses prospective developments, including
the expected impact of future Belle~II and LHCb datasets, and advances in
theoretical frameworks from both lattice and continuum communities, several
of which are only beginning to reach a level of maturity where they can be
confronted quantitatively with data.

The Report is organised as follows: Section~\ref{sec:experiments} describes
the experimental apparatus and the techniques at Belle~II and LHCb, together with
the reconstruction strategies common to the semileptonic analyses discussed
throughout the Report. Section~\ref{sec:overture} then provides a theoretical
overview of the relevant effective field theory framework, introducing the
flavour puzzle and the precision and discovery frontiers that structure the
remainder of the discussion. The experimental and theoretical status of
charged-current semileptonic decays is presented in
Sections~\ref{sec:exp_cc} and~\ref{sec:th_cc}, respectively, followed by the
corresponding experimental and theoretical discussions for neutral-current
decays in Sections~\ref{sec:nc_exp} and~\ref{sec:nc_theo}. Experimental and
theoretical aspects of lepton flavour universality tests and their new
physics implications are reviewed in Sections~\ref{sec:lfu_exp}
and~\ref{sec:lfu_theo}. The present data-taking status and future upgrade
plans are discussed in Section~\ref{sec:future}.
Section~\ref{sec:prospects} collects the overall prospects and conclusions.

\clearpage

%%%%%%%%%%%%%%%%%%%%%%%%%%%%%%%%%%%%%%%%%%%%%%%%%
% LHCb and BFactories: 
% General description of the detectors
% Experimental techniques
% Description of the current datasets
%%%%%%%%%%%%%%%%%%%%%%%%%%%%%%%%%%%%%%%%%%%%%%%%%

\section{Experimental Techniques}
\label{sec:experiments}

{\it Patrizia De Simone}\\

The measurements reviewed in this Report are performed at two complementary experimental facilities: the \belle and \belletwo detectors at the KEKB $e^+e^-$ collider, and the \lhcb detector at the LHC. These experiments operate in different environments and exploit distinct production mechanisms, reconstruction strategies, and systematic uncertainties. Their complementarity is essential: \belle and \belletwo provide a clean, fully constrained initial state with precise control over neutral particles and missing energy, while \lhcb benefits from the enormous $b$-hadron production cross-section at the LHC and has access to the full spectrum of $b$-hadron species. 

The reconstruction of $b$-hadrons undergoing kinematically closed decays, in which the energies and momenta of all daughter particles are measured, as in $B\to K^{(*)}\ell\ell$ decays, provides enough kinematic constraints to achieve a clear separation between signal and background.
By contrast, reconstructing decays involving missing particles, such as neutral current semileptonic $b$-hadron decays, is more challenging and has required the development of specialized techniques to distinguish the signal from the various sources of background.
  
The following subsections describe the key features of each experiment and the principal techniques employed in analyses of semileptonic decays.

\subsection{Semileptonic measurements at \boldmath{\B}~Factories}

At $B$~factories, $B$ mesons are produced through the decay of the \FourS resonance. 
At the SuperKEKB collider the electron and positron beams have asymmetric energies
of 7 \gev and 4 \gev, respectively, so that the \FourS meson moves forward in the laboratory
frame with a Lorentz boost of $\beta\gamma \simeq$ 0.28, causing $B$ mesons to travel 
a measurable distance of about 130$~\mu$m along the beam axis before decaying.
The \FourS meson is the lightest $b\bquarkbar$ resonance with mass above the \BBb pair production threshold and therefore decays almost exclusively in $\Bz\Bzb$ and $B^+B^-$ pairs, where both possibilities occur with about the same rate. 
Due to the small mass difference between the \FourS state and the \BBb pairs, the $B$ mesons are produced in the \FourS centre-of-mass frame with very small momentum  of about $|\vec p_B| \simeq 330\mev$. 
%Specifically, the momentum of the $B$ meson is approximately $|\vec p_B| \simeq 330\mev$. 
This is why the decay products of the two mesons are almost isotropically distributed 
in the \FourS rest frame.

The cross-section of the \FourS production
is approximately 1.06 $\rm {nb}$, resulting in about $1.1\times 10^6/\rm{fb}^{-1}$ \BBb pairs. 
This production represents only about one-fourth of all the hadronic events from the $e^+e^-$ collisions, with the remainder being non-\BBb events, usually called continuum background. In general, they are rejected exploiting the difference between the topology of particles coming from \FourS decays and from $e^+e^-\to q\quarkbar$ processes: tracks from $e^+e^-\to q\quarkbar$ processes are produced in two opposite jets, while they are spherical for $B$ meson decays. 

The $B$ meson pairs from \FourS decays are selected
by reconstructing one of the mesons, labelled as the $B_{tag}$, in exclusive decay modes; the rest of charged tracks and neutrals must come from the other $B$ meson, labelled as the $B_{sig}$.
This technique, known as \B-tagging,  is highly effective in reducing the impact of the continuum background and generally improving event reconstruction.

The Belle~II Collaboration developed a key ingredient of the  \B-tagging technique, the FEI algorithm~\cite{Keck:2018lcd};
a hierarchical reconstruction framework based on multivariate classifiers trained on simulated samples to select the tagged $B_{tag}$ mesons. 
The algorithm reconstructs a large number of hadronic and semileptonic decay channels and provides the four-momentum of the $B_{tag}$ meson together with event-level information.

\noindent
The \B-tagging techniques are classified in two main approaches: 
\begin{enumerate}
\item{The hadronic tagging provides a tight kinematic constraint at the cost of a low efficiency of about $ 0.5\%$} : 
the $B_{tag}$ is fully reconstructed into a combination of many different hadronic decay modes. 
Each decay mode starts from a set of well-reconstructed charm mesons, like $D^0$, $D^+$, $D^{*+}$, $D^{*0}$, $D_s$, $D_s^*$ or $J/\psi$, combined with additional charmless mesons ($\pi^{\pm}$,  $K^{\pm}$, $\pi^0$ and $K_s$) to form possible $B$ candidates.
To optimise efficiency, over a thousand possible decay chains are taken into consideration.
Two kinematic variables are used to test the compatibility of the $B_{tag}$ candidate with a $B$ meson:
	\begin{enumerate}
	\item the beam-energy substituted mass, $m_{ES} = \sqrt{ s/4 - p_B^2}$;	
	\item the energy difference, $\Delta E = E_B - \sqrt{s}/2 $;
	\end{enumerate}
 where $\sqrt{s}$ is the total energy in the \FourS CM system, and $p_B$ and $E_B$ denote the momentum and the energy of the $B_{tag}$ candidate in the CM, respectively.
The reconstruction of the four-momentum of the $B_{tag}$ allows the determination on a event-by.event basis, of the four-momentum of the signal $B$ meson, $B_{sig}$, even in the presence of neutrinos.

\item{The semileptonic tagging reaches about $2\%$ efficiency with looser constraints and larger backgrounds} : 
the $B_{tag}$ is reconstructed using  $B\to D^*\ell\nu_\ell$ and $B\to D\ell\nu_\ell$ decays. These decays have the highest branching fractions of all $B$ decays and together comprise more than $7\%$ of all possible decays of \B mesons. Moreover, the efficiency of reconstructing semileptonic decays is generally higher than the one of their fully hadronic countparts. It can range between 0.5\% and 1.0\% depending on the required purity.
The downside is that the background is higher than for the hadronic tagging and that the presence of the neutrino on the tag side does not allow for tight kinematic constraints.
The key variable to select the tag side is the cosine of the angle between the $B$ candidate and the visible system $Y$, 
\begin{equation}
\cos\theta_{BY}\equiv \frac{2E_{B}E_Y-m_B^2-m_Y^2}{2|p_B||p_Y|},\\
\end{equation}

\noindent
where $Y\equiv D\ell$ or $D^*\ell$. The energies and momenta of the $B$ meson and the $Y$ system are computed in the CM.
For a single massless particle missing, $cos\theta_{BY}$ is limited to the physical range of $-$1 to 1. 
Instead, for an incorrectly identified tag-side, because of $B\to D^{**}\ell\nu$ or other decays with a $D^{*(*)}$ and a lepton in the final state, it can extend beyond these boundaries due to additional missing particles.   
\end{enumerate}

\subsection{Semileptonic measurements at the LHC}

The production mechanisms of \bquark quarks at hadron colliders are quark annihilation (${q \bar q}\to {b \bar b}$) and gluon fusion (${gg}\to {b \bar b}$) processes, with the latter dominating~\cite{Nason:1999ta}. 
Gluon fusion produces ${b\bar b}$ pairs with a small opening angle and a small transverse momentum ($p_T$). 

At the LHC, the LHCb forward spectrometer collects data from the dominant gluon fusion processes, from which 
all \bquark-hadron species originate, such as \Bp, \Bz, \Bs, \Bc, and \Lb.
The LHCb experiment covers the forward acceptance with pseudorapidity $\eta$ in the range of $2<\eta < 5$. 
This corresponds to only 4\% of the full solid angle, but captures about $25\%$ of the total $pp\to b{\bar b}$cross-section. The visible $b$-hadron cross section was measured to be $72~\upmu{\rm b}$ at $7~{\rm TeV}$ and  $144~\upmu{\rm b}$ at 13~{\rm TeV}~\cite{LHCb:2016qpe}. The LHCb detector features a particle identification system, capable of separating charged K and $\pi$ mesons over a large kinematic range and therefore allowing for a precise identification of the final states. Details about the various subdetectors and the overall performances are described in Ref.~\cite{LHCb:2008vvz,LHCb:2014set}.

\noindent
The acceptance of the general purpose experiments CMS and ATLAS is limited to the central region of $|\eta|<2.2$. 
This makes the experiments 40\% efficient for $pp\to b{\bar b}$ processes. However, due to their different triggering strategies, optimized for high-\pt physics, and the lack of K and $\pi$ separation, their reach for semileptonic physics is more limited than for the LHCb experiment.

Both ATLAS and CMS have contributed with various measurements of neutral currect $b\to s\ell\ell$ processes. To date, however, the only semileptonic \bquark\to\cquark measurements performed, are two CMS measurements of $B_c\to J/\psi\tau \nu$ decays ~\cite{CMS:2024seh,CMS:2025jfx}.

\noindent
The \lhcb experiment recorded 3\invfb of data in 2011--2012 at \sqs~=~7--8 TeV (Run 1), 6\invfb in 2015--2018 at \sqs~=~13 TeV (Run 2) and more than 26\invfb in 2024 - 2026 at 13.6 TeV (Run 3).
In 2018, the CMS experiment recorded 59.7 \invfb of \proton\proton data at \sqs~=~13 TeV.
During Run 3, which started in 2022, both \lhcb and \cms were recording data at \sqs~=~13.6 TeV. 
%So far, no LFU measurements have been published using this dataset.

At a proton collider, the reconstruction of the \bquark-hadron, $H_b$, in semileptonic decays cannot rely on kinematic constraints as in $B$ factories. However the physics analyses greatly benefit
from the huge \bquark-hadron production cross section and from the large boost ($\beta\gamma\approx 50$) of the $H_b$, making it travel on the order of 1\cm before decaying.

The \lhcb detector incorporates several features that are essential for measurements involving semileptonic \bquark-hadron decays:
\begin{enumerate}
\item a vertex detector with a very small proper time ($\sim 45$~fs) and impact parameter resolution ($\sim 22~\mu$m for tracks with high $p_T$). This provides excellent separation between primary and (secondary) heavy hadron vertices.
\item tracking detectors with a very good momentum resolution ($\Delta p/p \sim$ 0.5\% at 5\gevc), in order to have good mass resolutions for the reduction of backgrounds and to distinguish individual final states
\item an excellent particle identification system, in order to fight the combinatorial background;
\item a highly efficient and selective trigger, allowing for a large number of semileptonic decays to be written offline in order to reduce the huge background from minimum bias events.
\end{enumerate}

\noindent
During most running periods of the LHC, a significant number of \proton\proton collisions occur simultaneously in the same bunch crossing, so-called pile-up. While CMS and ATLAS ran at pile-ups between 30 and 50 in Run2, and about 60 in Run3. Instead LHCb opted for a smaller pile-up of about 2.0 during Run 1, and about 1.2 during Run 2, in order to ease measurements with \bquark-hadrons, as they need well-reconstructed production and decay vertices. With the start of Run 3, LHCb was able to increase the pile-up to 5.3 thanks to the upgraded detector.
 
The semileptonic signals extraction relies on multidimensional template fits in 
kinematic variables, selected to allow for good discrimination between 
semimuonic and semitauonic channels, as well as the various background sources.
To use these kinematic variables, one needs to proceed to the $H_b$ candidate rest frame. Since the $H_b$ momentum cannot be fully reconstructed in hadron colliders, an approximation must be made. 

Thanks to the excellent performance of the \lhcb~vertex detector, the flight direction of the $H_b$
is measured with good precision from which 
the momentum of the neutrino from the semimuonic $H_b$ channels,
can be deduced up to a twofold ambiguity \cite{Pmu}.
The semitauonic \bquark-hadron channels are selected looking for the fully leptonic or hadronic $\tau$ decays,
the relative procedures to evaluate the missing momentum of the $\nu_{\tau}$ are different \cite{Ptau}:
\begin{itemize}
\item $\tau \rightarrow \mu \nu_{\mu} \nu_{\tau}$: the procedure, named RFA, assumes that the
         proper velocity of the $H_b$ along the beam axis is the same as that of the reconstructed visible final state, the selected pair of a \cquark-hadron, $H_c$, and a muon.
         In the highly boosted regime of \lhcb, the RFA is a good approximation that leads to a resolution on the squared $4$-momentum transferred to
         the lepton pair, \qsq, of about $22\%$.
\item  $\tau \rightarrow \pi \pi \pi \nu (\pi^0)$: in the highly boosted regime of \lhcb~the $H_b$ and the $\tau$ vertices are well separated
         and both their flight directions can be measured with good precision, allowing to evaluate the missing momentum up to a fourfold ambiguity,
         resulting in a overall \qsq resolution of about $19\%$.     
\end{itemize}

\noindent
CMS instead uses the assumption that the four-momentum of the $H_b$ equals the four-momentum of its visible decay products, 
scaled by the nominal mass of $H_b$: $p_{B}= m_B/m_{\rm vis} \cdot p_{\rm vis}$.

\noindent
Finally, as an example, the kinematic variables used to study semimuonic and semitauonic channels when the $\tau$ is selected through its fully leptonic decay, are:  
\begin{itemize}
\item the muon energy in the $H_b$ rest frame, $E_{\mu}^*$, that is much higher for the semimuonic channel than for muons coming from the semitauonic channel, 
         where a significant amount of energy is attributed to the neutrinos in the tau decay; 
\item the missing mass squared, $m_{miss}^2 = (p_{B}-p_{D}-p_{\mu})^2$, that represents the invariant mass of the neutrinos 
         and thus peaks around zero for muon decays, while it has higher values for the semitauonic decays; 
\item the squared four-momentum transfer to the lepton system, $\qsq = (p_{B}-p_{D})^2$, that is higher for the semitauonic channel. 
\end{itemize}

The selection of \bquark-hadron semileptonic decays comes with a large amount of background from partially reconstructed \bquark-hadron decays,
mainly due to inclusive semileptonic decays (feed-down), 
such as $B\to D^{**}\ell\nu$~decays,  
and due to double charm decays where one of the \cquark-hadron decays semileptonically. 
Multivariate analysis techniques,  where an additional charged or neutral particle can be associated to the $H_b$ signal candidate, are heavily used to reject these
backgrounds. Such techniques are also used
to select enriched background samples used to constrain the final fit. 

Histogram templates are used to model the final selected data sample, 
which consists of various background sources, as well as the semitauonic and
the semimuonic channels in varying percentages depending on the specific physics analysis.
The background components due to particle misidentification and due to combinatorial background are often modelled with templates 
based on data control samples.
Most other backgrounds are described with templates based on simulation, which comes with an inherent model dependence, \eg the FF parametrization used at the generation stage. The {\tt HAMMER} software tool \cite{Hammer} is heavily used not only for
reweighting, event-by-event, the simulated samples to different FF models, but also for varying the FF parameters
during the final fit procedure through software tools such as the {\tt RooFit-Hammer} interface\cite{RooHammer} or the python tool \cite{PyHammer}.

\clearpage

%%%%%%%%%%%%%%%%%%%%%%%%%%%%%%%%%%%%%%%%%%%%%%%%%
% Theoretical introduction:
% - status 
%%%%%%%%%%%%%%%%%%%%%%%%%%%%%%%%%%%%%%%%%%%%%%%%%
\section{Theoretical Overview}
\label{sec:overture}

{\it Matteo Fael}\\

This section provides the theoretical framework underlying the rest of the Report. We first review the flavour puzzle of the SM and its interpretation within the SMEFT, before turning to the two complementary avenues through which semileptonic $B$ decays probe this structure: the precision frontier, aimed at accurate determinations of the CKM matrix elements \Vcb and \Vub, and the discovery frontier, focused on processes and observables that are especially sensitive to physics beyond the SM.

\subsection{The flavour puzzle}

The SM of particle physics provides an extremely successful
description of subatomic phenomena. It  describes the electromagnetic, weak, and strong interactions (while omitting gravity) and how the
three families of quarks and leptons interact with each other through the exchange of gauge bosons. The gauge sector based on the symmetry group
$SU(3)_c \times SU(2)_L \times U(1)_Y$ 
is highly constrained and predictive in its structure.

The situation is very different in the Yukawa sector, 
where fermions couple to the Higgs field. 
Here the SM appears much less structured: 
the observed quark and lepton masses display a hierarchical pattern, 
and the origin of the three fermion generations remains unexplained. 
Similarly, we do not yet have a compelling explanation for the structure of the CKM matrix. From this perspective, flavour physics is in a similar situation as chemistry at the end of the nineteenth century: the elements could be 
organized in the periodic table according to their properties, but the underlying organizing principle was still unknown.

At the same time, the SM does not account for several empirical observations,
such as neutrino oscillations, the origin of dark matter, and the baryon
asymmetry of the universe. Together with the theoretical motivations mentioned
above, these facts suggest that the SM is incomplete and 
instead should be regarded as an EFT, 
valid at energies of the order of the electroweak scale,
$v \simeq 246\,\mathrm{GeV}$. From this perspective, the renormalizable SM of
Glashow, Weinberg, and Salam can be interpreted as the leading term of an
effective Lagrangian: the SMEFT~\cite{Buchmuller:1985jz,Grzadkowski:2010es,Brivio:2017vri}. 
This theory contains higher-order operators built out of 
the SM degrees of freedom and respecting the SM gauge symmetry,
\begin{equation}
    \mathcal{L}_{\rm SMEFT}
    =
    \mathcal{L}^{(4)}
    +
    \sum_{d>4}
    \frac{1}{\Lambda_{\rm NP}^{d-4}}
    \mathcal{L}^{(d)} .
\end{equation}

The first term, $\mathcal{L}^{(4)}$, is the usual renormalizable SM
Lagrangian, containing operators of dimension four or less.
There are several accidental features in the leading Lagrangian: 
neutrino masses are absent; charged-lepton flavour violation does not occur; flavour violation in the quark sector follows a very
specific pattern and appears only through charged-current interactions mediated
by the $W$ boson; baryon and lepton numbers are conserved; and CP violation
arises from a single irreducible phase in the CKM matrix, leading to a highly
constrained pattern of CP-violating effects.

The inclusion of higher-dimensional operators relaxes these accidental
properties. At dimension five, the Weinberg operator generates Majorana masses
for neutrinos \cite{Weinberg:1979sa}. At dimension six, new sources of lepton
and quark flavour violation can appear, not necessarily aligned with the CKM
matrix. Baryon and lepton numbers may be violated, and new CP-violating
interactions can arise. In the SMEFT, CP violation is therefore a generic
possibility rather than a feature restricted to the weak charged-current
interactions of the renormalizable SM.

Although the SMEFT allows for a broad range of new interactions, 
many of the corresponding effects are experimentally very strongly constrained. For example, charged-lepton-flavour-violating processes such as $\mu \to e\gamma$~\cite{MEGII:2025gzr} and $\mu \to eee$~\cite{SINDRUM:1987nra,Blondel:2013ia} have not been observed so far. 
Similarly, flavour-changing neutral-current processes in the quark sector, 
such as neutral meson mixing~\cite{Albrecht:2024oyn}, agree remarkably well with SM predictions.

In a generic SMEFT with Wilson coefficients of order unity, dimension-six
operators would typically generate contributions to these observables that are
already excluded unless the scale of new physics is extremely high,
depending on the operator and on its flavour structure. 
This tension is commonly referred to as the ``flavour puzzle.''

From the EFT point of view, the absence of large deviations in flavour
observables suggests that the ultraviolet completion of the SM cannot be
completely generic. Either the scale of new physics lies far above the TeV
scale, or the underlying theory must possess a highly non-trivial flavour
structure that suppresses flavour-violating effects. Understanding the origin of
this structure, together with the mechanism responsible for the observed
hierarchies in fermion masses and mixing angles, remains one of the central open
questions in flavour physics.

\subsection{The precision frontier}

$B$ physics provides a powerful laboratory in which to address the
flavour puzzle. The upper $2\times 2$ block of the CKM matrix is already close
to unitary, and information on the coupling of the third generation to the
lighter quarks is obtained largely through transitions of the form $b\to c$ and
$b\to u$. 
In particular, the CKM elements $V_{cb}$ and $V_{ub}$ are determined
primarily from semileptonic $B$-meson decays. 
The CKM elements involving the top quark, $V_{td}$ and $V_{ts}$, are instead poorly constrained by tree-level top decays at hadron colliders; the strongest information on them is obtained indirectly from kaon and $B$-meson processes.

These CKM elements are essential inputs for the interpretation of
flavour-changing neutral-current observables, such as $B$-$\overline B$ mixing,
$b\to s\gamma$, $b\to s\ell^+\ell^-$, and $b\to s\nu\bar\nu$ decays.

The study of $B$ physics can be broadly divided into two complementary
directions. 
The precision frontier in the study of $B$ physics
aims at accurate determinations of the CKM elements $V_{cb}$ and $V_{ub}$. 
They can, in principle, be extracted
through purely leptonic decays such as 
$B_c \to \tau\bar\nu_\tau$ and $B^+ \to \tau^+\nu_\tau$. 
These modes are theoretically clean, since the relevant hadronic information is 
encoded in decay constants that can be computed with high precision using LQCD~\cite{FlavourLatticeAveragingGroupFLAG:2024oxs}. 
At present, however, their experimental precision is not competitive with that of semileptonic modes~\cite{Belle-II:2025ruy}.

Semileptonic $B$ decays mediated by the charged weak currents $b\to c\ell\bar\nu_\ell$ 
and $b\to u\ell\bar\nu_\ell$ therefore provide the main avenue for determining $V_{cb}$ and $V_{ub}$. 
They have relatively large branching fractions and clean experimental signatures.
The relevant kinematics is commonly described in terms of the invariant mass of the lepton-neutrino pair, $q^2$, and the invariant mass of the hadronic final state, $M_X^2$.

There are two complementary strategies. In exclusive decays, one reconstructs a specific 
hadronic final state, such as $B\to D\ell\bar\nu_\ell$, $B\to D^*\ell\bar\nu_\ell$, or $B\to \pi\ell\bar\nu_\ell$. 
The non-perturbative QCD dynamics is encoded in FFs, which are functions of $q^2$. These FFs can be constrained by a combination of LQCD, symmetry arguments, analyticity, unitarity, and experimental data.

A central issue in exclusive determinations is therefore how to parametrize the FFs
without introducing unnecessary theoretical bias. 
Historically, the CLN parametrization has played a very important role in the analysis of $B\to D^*\ell\bar\nu_\ell$ decays \cite{Caprini:1997mu}. 
It exploits heavy-quark symmetry and provides a compact description in terms of a small number of parameters. 
More recently, however, the increasing precision of both experimental data and 
LQCD calculations has motivated the use of more flexible parametrizations, such as BGL \cite{Boyd:1995sq,Boyd:1997kz}. 
These approaches rely more directly on general principles such as analyticity and unitarity, and allow for a more systematic treatment of truncation uncertainties~\cite{Grinstein:2017nlq,Bordone:2019vic,Bernlochner:2017jka,Gambino:2019sif,Bernlochner:2019ldg}.

The difference between these parametrizations is not merely technical. 
Since the value of $|V_{cb}|$ is extracted from the decay rate at non-zero recoil,
assumptions on the shape of the FFs can propagate directly 
into the CKM determination. 
For this reason, modern analyses aim to combine experimental differential rate~\cite{BaBar:2019vpl,Belle:2023bwv,Belle-II:2023okj,Belle-II:2025rna}
with LQCD determinations of FFs at non-zero recoil~\cite{FermilabLattice:2021cdg,Aoki:2023qpa,Harrison:2023dzh}
using the BGL parametrization, in order to keep parametrization uncertainties under control. 
This is one of the main output of the recent discussion on exclusive $|V_{cb}|$.

Inclusive semileptonic decays provide a different and complementary method. 
In processes such as $B\to X_c\ell\bar\nu_\ell$ and $B\to X_u\ell\bar\nu_\ell$, 
one does not select a specific hadronic final state, but sums over all states compatible 
with the underlying quark-level transition. 
The theoretical description is based on the heavy-quark expansion~\cite{Bigi:1992su,Bigi:1993fe,Blok:1993va,Bigi:1994wa,Manohar:1993qn}, 
which expresses sufficiently inclusive observables as an expansion in powers 
of $1/m_b$ and $\alpha_s$. 
At leading order, the decay is described by the decay of a free bottom quark, 
while non-perturbative effects enter through a finite set of heavy-quark parameters. 
Perturbative QCD and power corrections have been computed to 
higher orders along the years~\cite{Manohar:1993qn,Blok:1993va,Gremm:1996df,Bigi:1996si,Czarnecki:1997sz,Aquila:2005hq,Dassinger:2006md,Pak:2008qt,Pak:2008cp,Dowling:2008mc,Biswas:2009rb,Mannel:2010wj,Alberti:2012dn,Alberti:2013kxa,Fael:2018vsp,Fael:2020iea,Colangelo:2020vhu,Fael:2020njb,Fael:2020tow,Fael:2022frj,Mannel:2023yqf,Egner:2023kxw,Mannel:2021zzr,Fael:2024gyw,Finauri:2025ost}, and recently implemented in the 
python package \texttt{Kolya}~\cite{Fael:2024fkt}.

Experimentally, inclusive analyses use not only the total semileptonic rate, 
but also moments of kinematic distributions, such as the charged-lepton energy spectrum, the hadronic and leptonic invariant-mass distribution~\cite{CLEO:2004bqt,BaBar:2009zpz,Belle:2006jtu,Belle:2006kgy,DELPHI:2005mot,CDF:2005xlh,Belle:2021idw,Belle-II:2022evt}. 
A global fit to these moments allows one to determine simultaneously the heavy-quark parameters and the CKM element $|V_{cb}|$~\cite{Bernlochner:2022ucr,Finauri:2023kte}. 
The main theoretical challenge is the assessment of uncertainties from missing higher-order
perturbative and power corrections, as well as the theoretical  correlations among different observables.
Since many moments originate from the same underlying differential rate, 
their theoretical uncertainties are not independent, and their treatment can have  an important impact on the final error budget. 

In recent years, LQCD has also started to address inclusive semileptonic decays directly~\cite{Gambino:2020crt}. 
First studies of inclusive rates and moments in charm~\cite{DeSantis:2025yfm,DeSantis:2025qbb} and beauty systems~\cite{Gambino:2022dvu,Barone:2023tbl,Kellermann:2025pzt} have appeared, opening the possibility of a future non-perturbative validation of the heavy-quark expansion in semileptonic processes. 
Although these calculations are still at an early stage, they represent a promising development, especially in view of the precision expected from Belle II.

The extraction of $|V_{ub}|$ from inclusive decays is more challenging than that of $|V_{cb}|$. 
The reason is that the dominant $b\to c\ell\bar\nu_\ell$ background is much larger than the charmless $b\to u\ell\bar\nu_\ell$ signal. 
As a result, experimental analyses often impose cuts that restrict the accessible phase space. In these regions, the standard local heavy-quark expansion has to be supplemented by 
additional non-perturbative information, usually encoded in shape functions~\cite{Neubert:1993ch,Neubert:1993um,Bigi:1994wa}.
Inclusive determinations of $|V_{ub}|$ from $B\to X_u\ell\bar\nu_\ell$ in the endpoint regions of phase space are commonly performed using the BLNP framework, based on QCD factorization and shape functions in the endpoint region \cite{Bosch:2004th,Bosch:2004cb}, or the GGOU framework~\cite{Gambino:2007rp,Gambino:2016fdy}, based on a Wilsonian treatment of the heavy-quark expansion in the kinetic scheme~\cite{Gambino:2007rp}.

A long-standing feature of the semileptonic precision program is the tension between 
inclusive and exclusive determinations of both $|V_{cb}|$ and $|V_{ub}|$~\cite{HeavyFlavorAveragingGroupHFLAV:2024ctg}. 
For $|V_{cb}|$, inclusive determinations from $B\to X_c\ell\bar\nu_\ell$ tend to give values higher than those obtained from exclusive $B\to D^{(*)}\ell\bar\nu_\ell$ decays. 
For $|V_{ub}|$, a similar pattern is observed between inclusive $B\to X_u\ell\bar\nu_\ell$ analyses and exclusive $B\to \pi\ell\bar\nu_\ell$ determinations. 
The numerical significance of these discrepancies depends on the precise inputs, 
FF treatment, and averaging procedure, but their persistence has made them  one of the central issues in flavour physics.

The importance of this tension goes beyond the determination of CKM parameters. 
The values of $|V_{cb}|$ and $|V_{ub}|$ enter global CKM fits and provide essential normalization factors for many rare $B$-meson decays~\cite{Bobeth:2013uxa,Brod:2010hi,Buras:2022wpw}. 
A shift in these inputs can affect the interpretation of flavour-changing neutral-current observables, CP-violating quantities, and possible deviations from the SM. 
Clarifying the origin of the inclusive-exclusive discrepancy is therefore an important 
task for exploiting the discovery potential of flavour physics fully.

\subsection{The discovery frontier}
The discovery frontier of semileptonic $B$ decays focuses on processes that are suppressed in the SM by the GIM mechanism, or on observables that test
accidental SM properties like lepton flavour universality. 
Such processes are especially sensitive probes of new physics and of possible new sources of flavour and CP violation.

In the SM, the charged-current interaction is universal in the lepton flavour, 
apart from the effects induced by the different masses of the leptons. 
Ratios of semileptonic decay rates are therefore particularly clean probes of new physics, 
because many hadronic and CKM uncertainties cancel~. 
The most prominent examples are~\cite{Bordone:2019vic,Gambino:2019sif,Bernlochner:2022ywh,Martinelli:2021myh,FermilabLattice:2021cdg,Harrison:2023dzh,Aoki:2023qpa,Harrison:2021tol}
\begin{equation}
R(D^{(*)})
=
\frac{
\mathrm{Br}(B\to D^{(*)}\tau\bar\nu_\tau)
}{
\mathrm{Br}(B\to D^{(*)}\ell\bar\nu_\ell)
},
\qquad \ell=e,\mu .
\end{equation}

The modes with $\tau$ leptons are experimentally more challenging, due to the presence of
additional neutrinos in the final state~\cite{BaBar:2012obs,BaBar:2013mob,Belle:2015qfa,Belle:2016dyj,Belle:2019rba,Belle-II:2024ami,LHCb:2023cjr,LHCb:2023zxo,LHCb:2024jll,Belle-II:2025yjp}, but they are especially interesting from the 
theoretical point of view. 
They probe at low energies the charged-current decay $b\to c\tau\bar\nu_\tau$ that appears in the SM, while being sensitive to possible interactions beyond the SM. 
For this reason, observables such as $R(D)$ and $R(D^*)$ provide a natural bridge between the
precision frontier and the discovery frontier: they require precise control of semileptonic 
FFs and CKM inputs, but at the same time they directly test one of the accidental 
properties of the SM, namely lepton flavour universality.

A second class of observables at the discovery frontier is provided by rare
decays mediated by flavour-changing neutral currents. In the SM, transitions
such as $b\to s\gamma$, $b\to s\ell^+\ell^-$ and $b\to s\nu\bar\nu$ are absent
at tree level and arise only through loops. They are therefore suppressed both
by the CKM elements $|V_{qb} V_{qs}^\star|$ and by the GIM mechanism. 
This makes them particularly sensitive to new virtual particles or interactions.

At energies below the electroweak scale, these processes are described by the
weak effective theory, obtained after integrating out the heavy degrees of
freedom of the SM. For $b\to s$ transitions one can write schematically~\cite{Buchalla:1995vs}
\begin{equation}
    \mathcal{L}_{\rm eff}
    =
    -\frac{4G_F}{\sqrt{2}}
    \left[
    \lambda_c \left(C_1 \mathcal{O}_1^c + C_2 \mathcal{O}_2^c\right)
    +
    \lambda_u \left(C_1 \mathcal{O}_1^u + C_2 \mathcal{O}_2^u\right)
    -
    \lambda_t \sum_{i=3}^{10} C_i \mathcal{O}_i
    \right],
    \label{eq:L_eff_fcnc}
\end{equation}
where $\lambda_q=V_{qb}V^*_{qs}$. The operator basis contains
current-current operators, QCD penguin operators, electromagnetic and
chromomagnetic dipoles, and semileptonic operators:
\begin{equation}
\begin{aligned}
\mathcal{O}_1^q =& (\bar{s}^\alpha_{L}\gamma_\mu q^\beta_L)(\bar{q}^\beta_L\gamma^\mu b^\alpha_L)\,, &
\mathcal{O}_2^q =& (\bar{s}_{L}\gamma_\mu q_L)(\bar{q}_L\gamma^\mu b_L)\,, \\
\mathcal{O}_3 =& (\bar{s}_{L}\gamma_{\mu}b_{L})\sum_q(\bar{q}\gamma^{\mu}q)\,, &
\mathcal{O}_4 =& (\bar{s}_{L}\gamma_{\mu}T^ab_{L})\sum_q(\bar{q}\gamma^{\mu}T^aq)\,, \\
\mathcal{O}_5 =& (\bar{s}_{L}\gamma_{\mu_1}\gamma_{\mu_2}\gamma_{\mu_3}b_{L})
\sum_q(\bar{q}\gamma^{\mu_1}\gamma^{\mu_2}\gamma^{\mu_3}q)\,, &
\mathcal{O}_6 =& (\bar{s}_{L}\gamma_{\mu_1}\gamma_{\mu_2}\gamma_{\mu_3}T^ab_{L})
\sum_q(\bar{q}\gamma^{\mu_1}\gamma^{\mu_2}\gamma^{\mu_3}T^aq)\,, \\
\mathcal{O}_7 =& \frac{e}{16\pi^2}m_b(\bar{s}_{L}\sigma^{\mu\nu} b_R)F_{\mu\nu}\,, &
\mathcal{O}_8 =& \frac{g_s}{16\pi^2}m_b(\bar{s}_{L}\sigma^{\mu\nu}T^a b_R)G_{\mu\nu}^a\,, \\
\mathcal{O}_9 =& \frac{e^2}{16\pi^2}(\bar{s}_{L}\gamma_\mu b_L)(\bar{\ell}\gamma^\mu \ell)\,, &
\mathcal{O}_{10} =& \frac{e^2}{16\pi^2}(\bar{s}_{L}\gamma_\mu b_L)(\bar{\ell}\gamma^\mu\gamma_5 \ell)\,.
\end{aligned}
\end{equation}

In particular, the dipole
operator $\mathcal{O}_7$ controls the short-distance contribution to radiative decays,
while the semileptonic operators $\mathcal{O}_9$ and $\mathcal{O}_{10}$ enter in the
$b\to s\ell^+\ell^-$ transitions. 
The semileptonic decay with two final state neutrinos is described
by a single operator usually denoted by $\mathcal{O}_\nu$. 
The matching at the electroweak scale
and the renormalization-group evolution down to the scale $m_b$ are known to
high perturbative accuracy, making these modes precision probes of the SM
flavour structure \cite{Buchalla:1995vs,Buras:1989xd,Chetyrkin:1996vx,Misiak:2004ew,Gorbahn:2004my,Czakon:2006ss,Brod:2026yxw}.

The radiative transition $b\to s\gamma$ is dominated by the dipole
operator $O_7$, but receives important contributions from the current-current
operators $C_1$ and $C_2$. 
Both inclusive and exclusive modes are
studied experimentally~\cite{CLEO:2001gsa,BaBar:2007yhb,Belle:2009nth,BaBar:2012fqh,BaBar:2012eja,Belle:2014nmp,BaBar:2008txc,Belle:2008imr,Belle-II:2024kbc,CLEO:1999psv}. 
The inclusive decay $B\to X_s\gamma$ is theoretically
particularly clean because the heavy-quark expansion can be used.
Recent improvements in the SM prediction focus on the non-trivial charm-mass
effects in the loops~\cite{Misiak:2015xwa,Czakon:2015exa,Misiak:2020vlo,Greub:2023msv,Fael:2023gau,Czaja:2023ren,Greub:2024mwp} and the photon-energy cut dependence~\cite{Melnikov:2005bx,Asatrian:2006sm,Asatrian:2010rq,Ligeti:2008ac,Bernlochner:2020jlt,Dehnadi:2022prz,Fael:2026fxp}. 
Exclusive modes such as $B\to K^*\gamma$ and related channels instead 
require hadronic FFs at $q^2=0$~\cite{Becirevic:2006nm,Bouchard:2013eph,Bailey:2015dka,Parrott:2022rgu},
but provide additional information on helicity structure and possible
right-handed currents.

The semileptonic rare decays $b\to s\ell^+\ell^-$ are richer, because they
probe several short-distance couplings at the same time. 
The dilepton invariant mass $q^2$ allows one to separate different kinematic regions and to construct angular observables~\cite{Matias:2012xw}. 
In exclusive modes such as $B\to K\ell^+\ell^-$ and
$B\to K^*\ell^+\ell^-$, the theoretical description involves local FFs
multiplying the short-distance coefficients, together with non-local matrix
elements generated by the insertion of current-current and penguin operators.
These non-local effects are especially important because the intermediate
$c\bar c$ states produce the narrow charmonium resonances, which strongly affect
the physical spectrum. 
A typical object entering the theoretical description is
the time-ordered product of the electromagnetic current with the four-quark
operators and the electromagnetic current (with $\mathcal{O}_1, \mathcal{O}_2$ giving the largest contribution):
\begin{equation}
\mathcal{H}_\mu^{B \to M} \propto \int d^4x\, e^{iqx} 
\langle M | T \{ j_\mu^{\text{em}}(x), \sum_{i=1}^6 C_i \mathcal{O}_i(0) \} | B \rangle .
\label{eq:nonlocal}
\end{equation}
The control of these long-distance contributions is one of the central
theoretical challenges in rare exclusive semileptonic decays~\cite{Khodjamirian:2010vf,Khodjamirian:2012rm,Hurth:2020ehu,Gubernari:2020eft,Gubernari:2022hxn,Ciuchini:2022wbq,Isidori:2025dkp,Isidori:2024lng,Ciuchini:2026hxy}.

Inclusive modes such as $B\to X_s\ell^+\ell^-$ provide a complementary
strategy. Since one sums over the hadronic final state, the heavy-quark
expansion can again be used, and the sensitivity to exclusive FFs is
reduced. At the same time, inclusive rare decays retain a non-trivial dependence
on the charm mass and on the treatment of perturbative and non-perturbative
corrections~\cite{Huber:2020vup,Huber:2024rbw}. They therefore play a role analogous to inclusive semileptonic
decays in the precision frontier: they can be systematically improved, 
but require a careful treatment of cuts, perturbative corrections and power corrections.

Rare semileptonic decays are also important probes of lepton flavour
universality. Ratios comparing electron and muon final states test whether the
short-distance coefficients are independent of the charged-lepton flavour, as
predicted in the SM up to very small corrections. In this sense, observables in
$b\to s\ell^+\ell^-$ transitions provide a neutral-current counterpart to
$R(D)$ and $R(D^*)$ in charged-current decays. 
Angular observables in
$B\to K^*\mu^+\mu^-$ and related channels further allow one to disentangle
different short-distance structures, for example vector, axial-vector and
dipole interactions. Their interpretation, however, always requires a careful
assessment of hadronic uncertainties, in particular those associated with
non-local charm contributions.

The modes with neutrinos in the final state, such as $B\to K\nu\bar\nu$,
$B\to K^*\nu\bar\nu$ and $B\to X_s\nu\bar\nu$, are theoretically very clean~\cite{Parrott:2022rgu,Becirevic:2023aov,Fael:2025xmi}.
Contrary to $b\to s\ell^+\ell^-$, they are not affected by long-distance
photon exchange or narrow charmonium resonances. 
In the SM, they are governed by a single short-distance coefficient~\cite{Altmannshofer:2009ma,Buchalla:1993bv,Buchalla:1998ba,Misiak:1999yg,Brod:2010hi}, 
which makes them especially transparent probes of new physics. 
Experimentally, however, they are challenging because of
the missing energy carried by the neutrino pair~\cite{Belle-II:2023esi,Belle-II:2025lfq,Belle-II:2025bho}. 
These modes are also sensitive to possible new invisible particles, and therefore provide a bridge between flavour physics and more general searches for hidden sectors.

Rare decays thus complete the logic of the semileptonic frontier. The
determination of $|V_{cb}|$ and $|V_{ub}|$ fixes important CKM inputs for loop
processes, while rare flavour-changing neutral-current decays test whether the
short-distance flavour structure follows the highly constrained SM pattern.
Together with LFU ratios in charged-current modes, they form a coherent
programme: precision semileptonic decays determine the parameters of the SM
related to the third generation of quarks,
and rare semileptonic decays use this information to search for
new sources of flavour, CP violation, and possibly new invisible degrees of
freedom.

\clearpage

%%%%%%%%%%%%%%%%%%%%%%%%%%%%%%%%%%%%%%%%%%%%%%%%%
% b->c l nu
%  Esperimental status at b-factories and LHCb
% form factors, Vcb and Vub
%%%%%%%%%%%%%%%%%%%%%%%%%%%%%%%%%%%%%%%%%%%%%%%%%

\section{Experimental status of charged-current semileptonic b-hadron decays}
\label{sec:exp_cc}

Charged-current semileptonic decays of $b$ hadrons, mediated at tree level by the
$b \to c\ell\nu$ and $b \to u\ell\nu$ transitions, provide the principal experimental
approaches for determining the CKM matrix elements \Vcb and \Vub. 
These determinations are performed using exclusive channels, in which a specific hadronic final state is reconstructed, and inclusive measurements, where all possible hadronic final states are considered. 
These two approaches are theoretically complementary but have historically yielded values of $|V_{cb}|$ and $|V_{ub}|$ in tension at the $2-3\sigma$ level~\cite{HeavyFlavorAveragingGroupHFLAV:2024ctg}, one of the most persistent puzzles in flavour physics. This section reviews the current experimental status at B-Factories and \lhcb, including recent measurements of branching fractions, FF parameters, as well as the key challenges and future prospects.

\subsection{Charged-current semileptonic and leptonic $b$-hadron decays at Belle~II}

{\it Giovanni Gaudino}\\

Belle~II operates at the SuperKEKB asymmetric-energy $e^+e^-$ collider at the \FourS resonance, providing a particularly clean environment for the study of semileptonic and leptonic $B$ decays. Compared to hadron-collider experiments, the well-defined initial state and the absence of additional particles from the interaction greatly simplify the reconstruction of final states containing neutrinos. This feature is especially important for measurements of purely leptonic decays and semileptonic transitions used to determine the CKM matrix elements $|V_{ub}|$ and $|V_{cb}|$.  In the analyses discussed in this section, hadronic tagging is employed. Although it has a lower efficiency than semileptonic tagging, it provides significantly higher purity and strong suppression of continuum and combinatorial backgrounds, which is crucial in channels with multiple neutrinos in the final state.

\paragraph{Measurement of $\mathcal{B}(B^+\to\tau^+\nu_\tau)$}

The decay $B^+\to\tau^+\nu_\tau$ is the most experimentally accessible purely leptonic $B$-meson decay because the large $\tau$ mass partially compensates for the helicity suppression present in leptonic decays mediated by charged weak currents. In the SM, the decay proceeds through the annihilation of the constituent quarks into a virtual $W^+$ boson, and its branching fraction is given by
\[
\mathcal{B}(B^+\to\tau^+\nu_\tau)=
\frac{G_F^2 m_B m_\tau^2}{8\pi}
\left(1-\frac{m_\tau^2}{m_B^2}\right)^2
f_B^2 |V_{ub}|^2 \tau_{B^+}.
\]\label{eq:b2tau}
As a consequence, the measurement provides direct sensitivity to the product $f_B|V_{ub}|$ and constitutes an important test of the SM. In addition, this decay is sensitive to possible contributions from charged Higgs bosons or to other new-physics scenarios affecting charged-current interactions. Belle~II measured this decay using the full Run~1 dataset corresponding to an integrated luminosity of $365~\mathrm{fb}^{-1}$ collected at the \FourS resonance~\cite{taunu}. The analysis reconstructs one $B$ meson with the FEI hadronic-tagging algorithm, while the signal side is identified from the remaining particles in the event. The $\tau$ lepton is reconstructed in the channels $\tau^+\to e^+\nu_e\bar{\nu}_\tau$, $\mu^+\nu_\mu\bar{\nu}_\tau$, $\pi^+\bar{\nu}_\tau$, and $\rho^+\bar{\nu}_\tau$, which together provide a good balance between efficiency and purity. The dominant backgrounds arise from continuum $e^+e^-\to q\bar q$ production and combinatorial from $B\bar B$ decays. These contributions are suppressed using event-shape observables, requirements on the residual neutral energy in the calorimeter, and dedicated boosted decision tree classifiers. Signal yields are extracted through a simultaneous fit to the extra calorimetric energy and the missing-mass-squared distributions in all reconstructed $\tau$ channels. Belle~II measures
\[
\mathcal{B}(B^+\to\tau^+\nu_\tau)
=
[1.24 \pm 0.41\,(\mathrm{stat})
\pm 0.19\,(\mathrm{syst})]\times10^{-4},
\]
with a significance of $3.0\sigma$. Assuming the SM prediction for the decay constant $f_B$, the corresponding value of the CKM matrix element is
\[
|V_{ub}| =
[4.41^{+0.74}_{-0.89}] \times 10^{-3}.
\]
The result is consistent with previous measurements and with SM expectations, demonstrating the capability of Belle~II to perform precision measurements in channels with multiple neutrinos in the final state.

\paragraph{Search for $B^+\to\mu^+\nu_\mu$}
The leptonic decay $B^+ \to \mu^+ \nu_\mu$ is highly suppressed within the SM. Its branching fraction, which is given by Eq. \ref{eq:b2tau} changing $m_\tau$ with $m_\mu$, is proportional to the square of the CKM matrix element $|V_{ub}|$ and is further constrained by helicity suppression, resulting in a very low predicted rate. The previous best experimental result was a 90\% CL upper limit of $\mathcal{B}(B^+ \to \mu^+ \nu_\mu) < 8.7 \times 10^{-7}$, established by the Belle experiment. An improved measurement of the $B^+ \to \mu^+ \nu_\mu$ branching fraction has been recently performed by analysing the combined Belle and Belle~II $B\bar{B}$ datasets \cite{Belle:2026lfg}. Following a strategy similar to the original Belle analysis, the tag-side $B$ meson is reconstructed inclusively. In this approach, the tag-side candidate is inferred from all tracks and calorimeter clusters remaining after the selection of the signal muon, thereby maximizing the signal efficiency for this rare process. The four-momentum of the tag-side $B$ meson is used to estimate the signal $B$ meson rest frame, where the signal is characterized by a monochromatic muon momentum distribution ($p^*_\mu$). To enhance the signal purity, several advanced techniques were implemented:
\begin{itemize}
    \item \textbf{MC-driven corrections:} These account for unreconstructed particles, such as $K^0_L$ mesons or neutrinos, which affect the resolution of the reconstructed tag-side $B$ momentum.
    \item \textbf{Multivariate Classifiers (MVA):} Specialized MVA-based classifiers were trained to mitigate the large backgrounds from continuum processes ($e^+e^- \to q\bar{q}$). 
    \item \textbf{Off-resonance matching:} An additional continuum-based MVA was employed to better align the simulated background distributions with off-resonance data collected below the $\Upsilon(4S)$ resonance.
\end{itemize}

A significant improvement over the original Belle study include updated modelling of semileptonic $b \to u$ transitions. These transitions constitute a major background in the $p^*_\mu$ signal region, and their modelling represents a leading source of systematic uncertainty in the final measurement. The signal yield is extracted via a binned maximum likelihood fit to the boosted muon momentum distribution. The analysis yields a branching fraction of $(4.4 \pm 1.9_{\text{stat}} \pm 1.0_{\text{syst}}) \times 10^{-7}$, corresponding to a 90\% C.L. upper limit of \[
\mathcal{B}(B^+ \to \mu^+ \nu_\mu) < 6.7 \times 10^{-7}.\] 
This result represents the most stringent limit to date on this decay channel. 

\paragraph{$|V_{ub}|$ from inclusive $B\to X_u\ell\nu$ decays}

Belle~II has also performed measurements of the inclusive semileptonic decays $B\to X_u\ell\nu_\ell$, with $\ell=(e,\mu)$, which provide an alternative determination of $|V_{ub}|$~\cite{vub}. Inclusive measurements are theoretically robust because they rely on heavy-quark expansions and are less sensitive to hadronic FFs than exclusive decays. Experimentally, however, they are challenging because of the overwhelming background from Cabibbo-favoured $B\to X_c\ell\nu_\ell$ decays. In this analysis, the tag-side $B$ meson is reconstructed with hadronic FEI tagging, while the remaining particles in the event are used to inclusively reconstruct the hadronic system $X_u$. Dedicated multivariate classifiers are employed to suppress continuum and charm backgrounds. The analysis exploits the distributions of the lepton energy, hadronic invariant mass, and momentum transfer squared $q^2$ to isolate the signal contribution. The signal yield is extracted from simultaneous fits to the kinematic distributions in signal-enriched and control regions. In the phase-space region defined by $E_\ell^{(B)}>1~\mathrm{GeV}$, Belle~II measures
\[
\Delta \mathcal{B}(B\to X_u\ell\nu_\ell)
=
(1.54 \pm 0.08\,(\mathrm{stat})
\pm 0.12\,(\mathrm{syst}))\times10^{-3}.
\]
Using the GGOU theoretical framework~\cite{Gambino:2007rp}, the extracted value of the CKM matrix element is
\[
|V_{ub}| =
(4.01 \pm 0.11\,(\mathrm{stat})
\pm 0.16\,(\mathrm{syst})
^{+0.09}_{-0.07}\,(\mathrm{th}))
\times10^{-3}.
\]
This result is consistent with previous inclusive determinations and remains larger than the values obtained from exclusive measurements, reflecting the long-standing tension between inclusive and exclusive extractions of $|V_{ub}|$~\cite{HeavyFlavorAveragingGroupHFLAV:2024ctg}.

\paragraph{$|V_{cb}|$ from $B\to D\ell\nu$ decays}

Belle~II determined $|V_{cb}|$ from the exclusive semileptonic decays $B\to D\ell\nu_\ell$ using the Run~1 dataset corresponding to $365~\mathrm{fb}^{-1}$~\cite{vcb}. The channels $B^0\to D^-\ell^+\nu_\ell$ and $B^+\to\bar D^0\ell^+\nu_\ell$ are reconstructed, while the neutrino momentum is inferred from the missing momentum in the event. The differential decay rate is measured as a function of the recoil variable $w=v_B\cdot v_D$, which describes the kinematics of the decay. Signal yields are extracted through fits to $\cos\theta_{BY}$ distributions in bins of $w$, where $\theta_{BY}$ is the angle between the reconstructed $D\ell$ system and the parent $B$ meson. The measured branching fractions are
\begin{align*}
\mathcal{B}(B^0\to D^-\ell^+\nu_\ell)
&=
(2.06 \pm 0.05\,(\mathrm{stat})
\pm 0.10\,(\mathrm{syst}))\%,\\
\mathcal{B}(B^+\to\bar D^0\ell^+\nu_\ell)
&=
(2.31 \pm 0.04\,(\mathrm{stat})
\pm 0.09\,(\mathrm{syst}))\%.
\end{align*}\\
The extraction of $|V_{cb}|$ is performed using the BCL parameterization of the hadronic FF~\cite{bcl}, yielding
\[
|V_{cb}| =
(39.2 \pm 0.4\,(\mathrm{stat})
\pm 0.6\,(\mathrm{syst})
\pm 0.5\,(\mathrm{th}))
\times10^{-3}.
\]
The analysis also probes LFU through the ratio
\[
\frac{\mathcal{B}(B\to De\nu_e)}
{\mathcal{B}(B\to D\mu\nu_\mu)}
=
1.020 \pm 0.020\,(\mathrm{stat})
\pm 0.022\,(\mathrm{syst}),
\]
which is in excellent agreement with the Standard Model expectation.

\subsection{Charged-current semileptonic and leptonic $b$-hadron decays at \lhcb}

{\it Biljana Mitreska}\\

The LHCb experiment contributed in various measurements using semileptonic \bquark-hadron decays. In the following a brief overview of differential and CKM measurements from LHCb is given, excluding LFU measurements which are reviewed in section~\ref{LFU-LHCb}.
Differential observables in $B^0 \to D^{-} \mu^{+} \nu_{\mu}$ decays, such as the decay angles and the squared momentum transfer to the lepton pair ($q^2$), provide a powerful probe of the full decay dynamics. These observables are sensitive to both hadronic effects and possible contributions from NP. They enable a wide range of measurements, including determinations of the $D^{*}$ longitudinal polarization, FFs and CKM matrix elements, as well as Wilson coefficients and angular observables. 

\paragraph{Measurement of the $D^{*}$ longitudinal polarization}

Semileptonic decays of $B$ mesons into final states with a $\tau$ lepton are a useful test of the SM and its extensions. Studies of
their angular observables such as the longitudinal $D^{\ast}$ polarization
fraction, $F_{L}^{D^\ast} = \frac{d\Gamma_L/dq^2}
{d\Gamma/dq^2}$, provide additional sensitivity to possible NP scenarios. The
value of $F_{L}^{D^\ast}$ could be enhanced by new scalar operators or
suppressed by new tensor operators. $F_{L}^{D^\ast}$ was measured by LHCb using a dataset corresponding to an integrated luminosity of $5~\mathrm{fb}^{-1}$~\cite{LHCb:2023ssl}. The analysis was performed in
two $q^2$ regions, one below and one above $7~\mathrm{GeV}^2/c^4$. The signal sample is composed of $B^0 \to D^{\ast -}\tau^+\nu_\tau$ decays, where the $D^{\ast -}$ meson
decays into the $D^0\pi^-$ final state, and the $\tau^+$ lepton decays to
three charged pions and a neutrino. The value of the longitudinal
polarization is calculated from the signal yield obtained from a
multi-dimensional binned fit including $\cos\theta_d$, $q^2$, and the $\tau$ decay time of the lepton. The averaged value of the measurement over the two $q^2$ bins is
$F_{L}^{D^\ast} = 0.41 \pm 0.06 \pm 0.03$,
which is compatible with the SM predictions and previous measurements from
Belle~II~\cite{Belle:2018ezy}. 

\paragraph{Measurement of FF parameters with a $B^0 \to D^{*-} \mu^{+} \nu_{\mu}$ angular analysis}

In this analysis the kinematic information of $B^0 \to D^{*-} \mu^{+} \nu_{\mu}$ decays is used to measure the hadronic FFs via a five-dimensional fit to the decay angles, $q^2$, and the squared invariant mass missing from the visible system ($m_{miss}^2$), using data with with an integrated luminosity of 3.0 fb$^{-1}$. This is the first measurement of hadronic FF parameters with $B^0 \to D^{*-} \mu^{+} \nu_{\mu}$ decays at LHCb using the CLN~\cite{Caprini_1998}, BGL~\cite{Boyd:1997kz}, and the BLPR~\cite{Bernlochner:2017jka} formalism.

\begin{figure}[!tbp] 
  \begin{center}
    \includegraphics[width=0.32\textwidth]{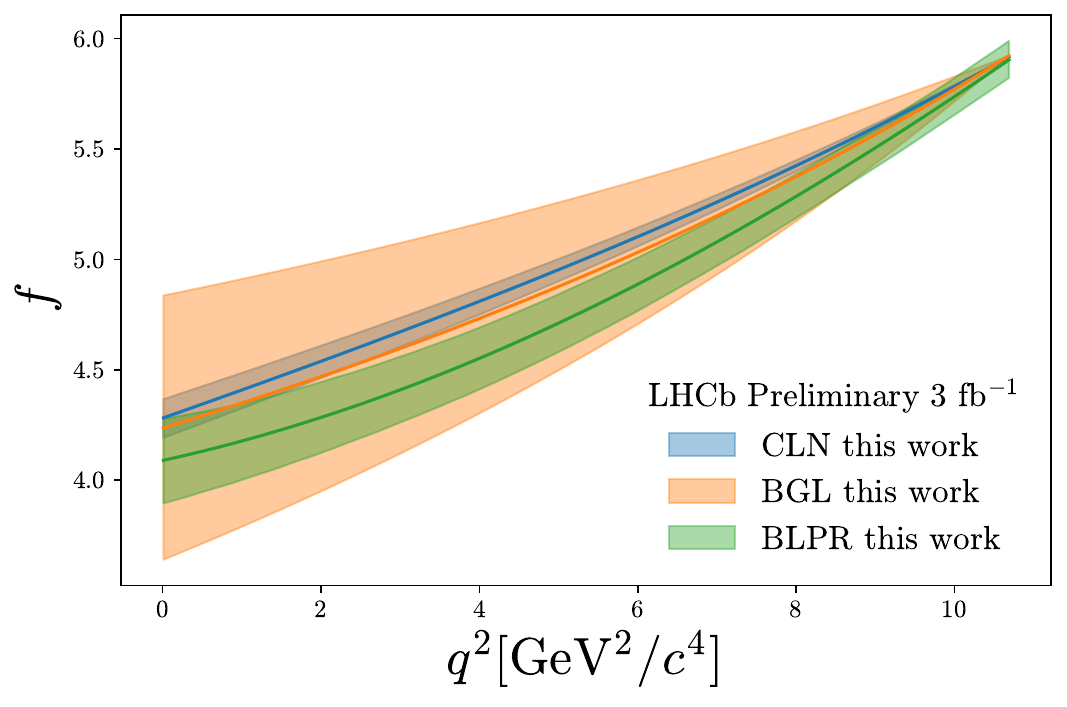}
    \includegraphics[width=0.32\textwidth]{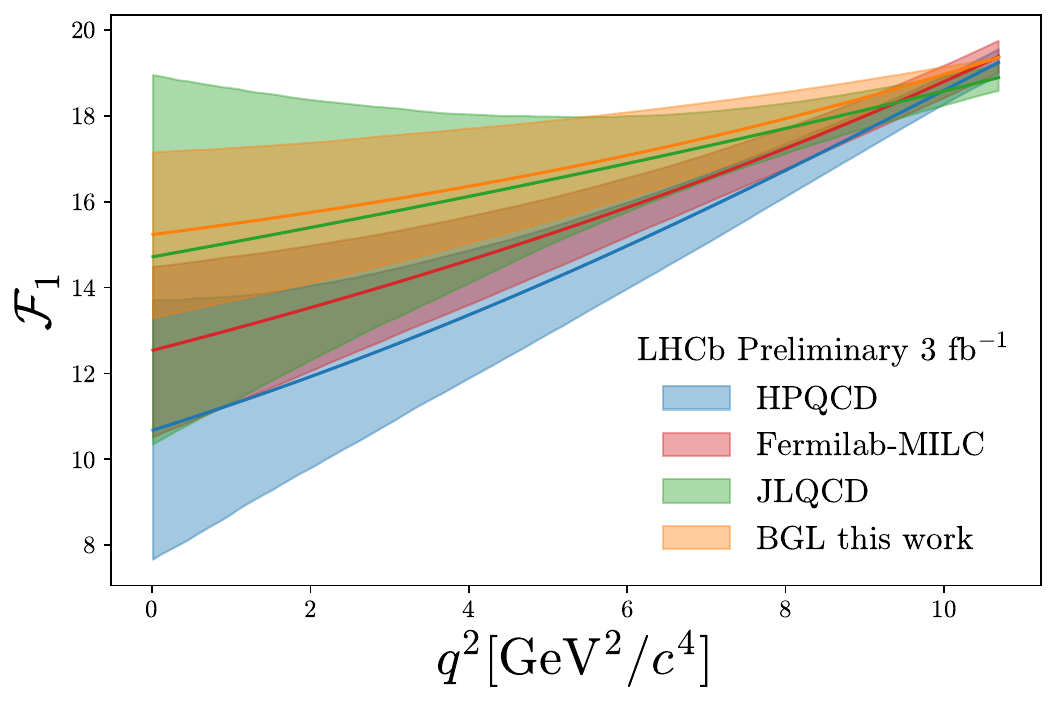}
    \includegraphics[width=0.32\textwidth]{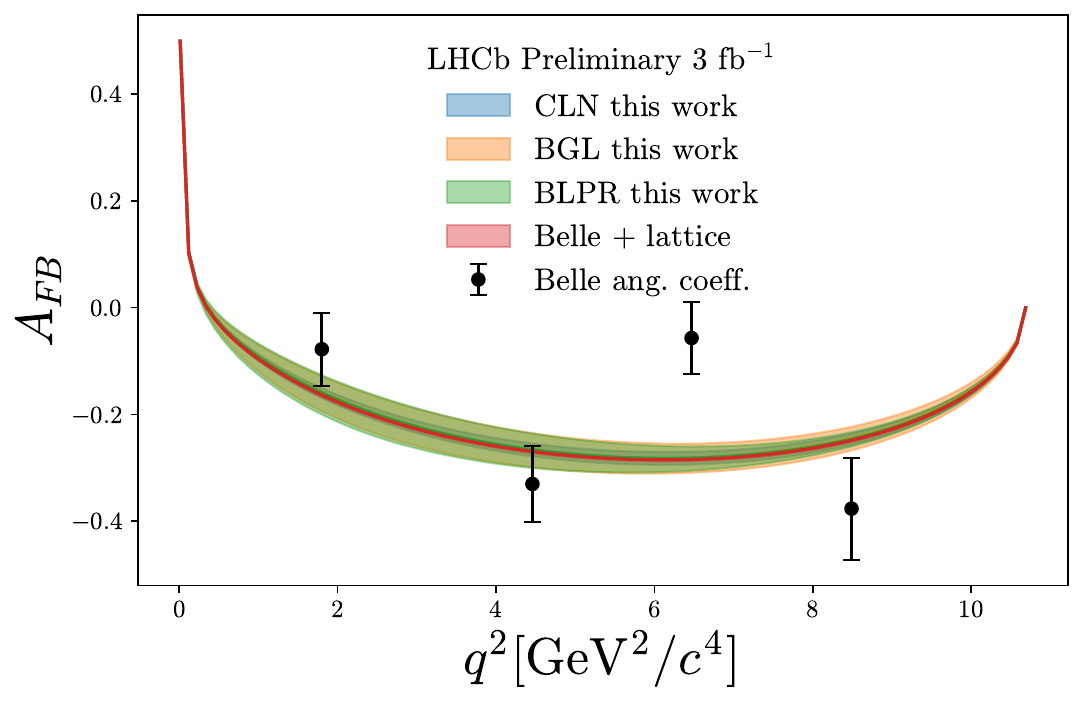}
  \end{center} 
  \caption{Left: Comparison of the $f$ results obtained in this FF measurement using CLN (blue), BGL (orange), and BLPR (green) parametrisations in the BGL basis. The uncertainty bands assume uncorrelated systematic uncertainties. Middle: Comparison of the results obtained for the $\mathcal{F}_{1}$ FF using the BGL to LQCD calculations from HPQCD (blue), Fermilab-MILC (red) and JLQCD (green). The uncertainty bands for this work are derived assuming that systematic uncertainties are uncorrelated. Right: Distribution of forward-backward asymmetry of the muon ($A_{\rm FB}$) compared with Belle results from Ref.~\protect\cite{Bernlochner:2017jka} (red) and Ref.~\protect\cite{Belle:2023xgj} (black points).}
  \label{fig:formfactor_parametrisations}
\end{figure}

The measured BGL parameters measured with this parametrisation are: $a_0$, $a_1$, $b_1$, $c_1$, $c_2$, and they satisfy the unitarity constraint. The measured values for the hadronic FF parameters are given in Ref~\cite{LHCb:2026amx} together with their uncertainties and statistical correlations. Fig.~\ref{fig:formfactor_parametrisations} (left) shows a comparison of the results obtained in this measurement in the BGL basis. A tension is seen between the CLN and BLPR $f(q^2)$ FF that can be expected given the restrictive CLN assumptions. Fig.~\ref{fig:formfactor_parametrisations} (middle) compares $\mathcal{F}_1$ results obtained using the BGL parametrisation to the LQCD calculations by the HPQCD~\cite{Harrison:2023dzh}, the Fermilab-MILC ~\cite{FermilabLattice:2021cdg} and the JLQCD collaboration~\cite{Aoki:2023qpa}. The FFs $f(q^2)$ and $g(q^2)$ are compatible while in $\mathcal{F}_1(q^2)$ the results agree better with Fermilab-MILC and JLQCD than HPQCD. A similar situation can be seen for CLN and BLPR from Ref~\cite{LHCb:2026amx}. In Fig.~\ref{fig:formfactor_parametrisations} (right) the muon forward-backward asymmetry, $A_{\rm FB}$, is calculated using the hadronic FF parameter results. It shows excellent agreement with the results from the BLPR fit~\cite{Bernlochner:2017jka} and as well with the Belle measurement of the angular coefficients~\cite{Belle:2023xgj}. The results of the FF parameters from BGL, BLPR, and CLN give mutually consistent results and are compatible with lattice calculations and theoretical predictions, while achieving improved precision.

\paragraph{Measurement of the shape of $B_s^{0} \to D_s^{*-} \mu^{+} \nu_{\mu}$}

A measurement of the shape of the $B_s^{0} \to D_s^{*-} \mu^{+} \nu_{\mu}$ differential decay rate at LHCb, obtained as a function of the hadron recoil parameter $w$, was performed in Ref.~\cite{LHCb:2020hpv}. The fitted signal yields in the seven $w$ bins are unfolded using a response matrix that accounts for model efficiency and detector resolution. The unfolded results support consistent CLN and BGL results.

\paragraph{Measurement of $V_{cb}$ from $B_s^0 \to D_s^{(*)-} \mu^{+} \nu_{\mu}$} 

The $V_{cb}$ CKM element was measured at LHCb using $B_s \to D_s \mu \nu$ and $B_s \to D_s^* \mu \nu$ decays~\cite{LHCb:2020cyw}. The strategy involves a 2D binned fit model using the corrected mass variable $m_{corr}$ and $p_{\perp}$, which is the component of the momentum transverse to the $B_s$ flight direction. For each iteration of the minimisation the values of FF parameters (CLN and BGL) and $V_{cb}$ are probed. The main backgrounds feature the semileptonic $B^0$ feed down modes, modelled with simulation, and the combinatorial background, modelled with random combinations from data. The $V_{cb}$ value is  measured as \[
|V_{cb}|_{\mathrm{CLN}} = \left(41.4 \pm 0.6\,(\mathrm{stat}) \pm 0.9\,(\mathrm{syst}) \pm 1.2\,(\mathrm{ext})\right)\times 10^{-3},
\]
\[
|V_{cb}|_{\mathrm{BGL}} = \left(42.3 \pm 0.8\,(\mathrm{stat}) \pm 0.9\,(\mathrm{syst}) \pm 1.2\,(\mathrm{ext})\right)\times 10^{-3},
\] where compatible values using BGL and CLN are obtained. Furthermore, an independent phenomenological analysis using the efficiency and resolution of the signal mode from this measurement was carried out in Ref.~\cite{Earnshaw:2025dhd}, where information on the BCL FF parameters was inferred using only the perpendicular momentum distribution. 

\paragraph{$V_{ub}/V_{cb}$ from $B_s \to K \mu \nu$}

The ratio $V_{ub}/V_{cb}$ was probed by LHCb accompanied by the discovery of the $B_s \to K \mu \nu$ decay~\cite{LHCb:2020ist}. The decay $B_s^0 \to D_s^- \mu^+ \nu_\mu$ was used as a normalization channel. The measurement of the branching fraction is performed in two regions of the invariant mass squared of the muon and the neutrino,
$q^2$, as well as integrated over the full $q^2$ range. The ratio
$|V_{ub}|/|V_{cb}|$ is derived in the two $q^2$ regions using calculations of the FFs of the $B_s^0 \to K^-$ and $B_s^0 \to D_s^-$ transitions
based on both light cone sum rules~\cite{Khodjamirian:2017fxg} and LQCD methods~\cite{FermilabLattice:2019ikx}. The measured ratio is 

\[
\left|\frac{V_{ub}}{V_{cb}}\right|_{\mathrm{low\ q^2}}
=
0.0607
\pm 0.0015\,(\mathrm{stat})
\pm 0.0013\,(\mathrm{syst})
\pm 0.0008\,(D_s)
\pm 0.0030\,(\mathrm{FF}),
\]

\[
\left|\frac{V_{ub}}{V_{cb}}\right|_{\mathrm{high\ q^2}}
=
0.0946
\pm 0.0030\,(\mathrm{stat})
{}^{+0.0024}_{-0.0025}\,(\mathrm{syst})
\pm 0.0013\,(D_s)
\pm 0.0068\,(\mathrm{FF}).
\]

\paragraph{Prospects for differential measurements, $V_{ub}$ and $V_{cb}$}

This includes measurements of $|V_{ub}|$ using $B \to \rho \mu \nu$ decays, as well as $B \to K \mu \nu$ decays that provide the differential decay spectrum in approximately ten bins of $q^2$. Using a similar strategy baryon decays such as $\Lambda_b \to \Lambda_c \mu \nu$ can be used to extract $V_{cb}$ at LHCb. For the angular analyses, the measurements of the Wilson and angular coefficients in $B_{(s)} \to D_{(s)}^{*} \ell \nu$ decays can further provide insight into the LFU anomalies. 
Within the next planned upgrades of LHCb higher instantaneous luminosities will allow to collect large datasets improving the statistical precision of measurements. The expected precision on semileptonic measurements over the years is estimated at around is estimated at around 6 $\%$ on $V_{ub}$ and 1.7 $\%$ on $V_{cb}$~\cite{ATLAS:2025lrr} if in the years to come all the different challenges in computation, background modelling and estimation are met. Overall, the LHCb programme of $b \to c \ell \nu$ measurements plays a central role in flavour physics, offering complementary information to $B$-factory experiments and continuing to probe possible deviations from SM expectations.

\subsection{Future developments and challenges}
This subsection summarize new approaches to analyse $b\to c$ semileptonic decays and new observables.\\

\subsubsection{Full angular analysis of $\Bs \to \Dss \mun \nu_{\mu}$ decays at \lhcb}
\label{sec:federico}

{\it Federico Manganella}\\
% \subsubsection{The four-dimensional analysis}

This analysis upgrades the previous LHCb measurement, based on a one-dimensional fit in $q^2$ with $1.7 \invfb$ of Run 2 data~\cite{q2AnalysisBs2Dsst}. This new analysis uses the entire Run 2 dataset, corresponding to approximately $6 \invfb$ of integrated luminosity, and aims to perform a full four-dimensional angular analysis in the phase-space variables \{$q^2$, $\cos\theta_{\ell}$, $\cos\theta_d$, $\chi$\}. The increased statistics enables, for the first time at LHCb, a simultaneous determination of the angular structure of the decay amplitude in multiple bins of all relevant kinematic variables. Here, $q^2 = (p_{\Bs} - p_{\Dss})^2$ is the squared momentum transfer to the lepton pair, $\theta_\ell$ is the helicity angle between the direction of the lepton and the direction of the $\Dss$ in the virtual $W$ lepton rest frame, $\theta_d$ is the helicity angle between the direction of the $\Ds$ meson and the direction of the $\Dss$ in the $\Dss$ rest frame, and $\chi$ is the angle formed by the $\Dss$ meson decay and the $W$ boson decay, defined in the $\Bs$ meson rest frame. The full phase-space is divided into six bins in $q^2$, three in $\cos\theta_{\ell}$, three in $\cos\theta_d$, and five in $\chi$, yielding a total of $270$ bins in which signal yields are extracted.
\\

The full decay chain is $\Bs \to \Dss \mun \nu_{\mu}$, with $\Dss \to \Ds \gamma$, and the $\Ds$ subsequently reconstructed in the three-body final state $\Ds \to \Kp \Km \pip$. The $\Ds$ hadron is selected requiring a series of PID cuts and mass hypotheses vetoes, additionally selecting the resonances $\Dspm \to \phi\pipm$ and $\Dspm \to \Kstarz \Kpm$ on Dalitz plots. The doubly-charmed background decays, $H_b \to \Dss H_c$ with the $\Dss$ decaying semileptonically, are suppressed by a dedicated topological BDT, which is then used to define a control (signal) region at lower (higher) BDT values, to constrain the background normalisation. The combinatorial background related to the $\Dss$ photon emission is finally statistically evaluated using the $\sPlot$ technique~\cite{sPlot}, fitting the $\Ds\gamma$ invariant mass.
\\

Starting from the effective Hamiltonian~\cite{Colangelo:2018cnj}, the theoretical differential decay rate can be written as
\begin{equation}
    \frac{d^4\Gamma}{dq^2 d\cos\theta_{\ell} d\cos\theta_d d\chi} \propto \vert V_{cb} \vert^2 \sum_{i=1}^{12} I_i(q^2) \Xi_i(\theta_{\ell}, \theta_d, \chi)
\label{eq:diffRate}
\end{equation}
where the $\Xi_i$ are known analytic angular functions and the $I_i(q^2)$ are linear combinations of hadronic FFs, which are parametrised using the CLN \cite{Caprini:1997mu} or BGL \cite{Boyd:1997kz} models. NP contributions may be incorporated either through reweighting, using for instance the HAMMER package~\cite{Bernlochner:2020tfi}, or by introducing explicit NP coupling constants $\epsilon_{NP}$ in the FFs description. A notable feature of the SM prediction is that three of the angular coefficient functions - $I_7$, $I_8$ and $I_9$ - vanish. Thus, any significant non-zero measurement of these observables would constitute direct model-independent evidence for BSM physics.
\\

The signal yields in all $270$ phase-space bins are extracted through a two-step binned template fit to the corrected mass $m_{corr}$, defined in terms of visible observables as
\begin{equation}
    m_{corr} = \sqrt{m_{\Dss\mun}^2 + \vert p_{\Dss\mun}^\perp \vert^2} + \vert p_{\Dss\mun}^\perp \vert
\end{equation}
In the first step, a simultaneous fit in the six $q^2$ bins, integrating over the angular variables, is performed jointly on the signal and control regions; this step allows primarily  to determine  the normalization of the doubly-charmed and combinatorial background templates. In the second step,  these background normalizations are used as Gaussian constraints to the full fit in the signal region, which extracts the 270 per-bin signal yields as independent free parameters.

From the extracted yields, three different approaches are used to achieve three complementary physical results. The first one is a direct comparison of the unfolded angular distributions with theoretical predictions, based on CLN parameters from the HFLAV world average~\cite{HeavyFlavorAveragingGroupHFLAV:2024ctg} and BGL FFs from HPQCD lattice calculations~\cite{davies24}. In agreement with recent Belle II results, some tension with the lattice-based prediction is anticipated. The second approach is a direct fit of the CLN and BGL hadronic parameters to the (un)folded distributions, where various truncation orders of the BGL expansion have been studied to assess the sensitivity to the parametric choice. The third and most novel approach is a fully model-independent extraction of the twelve independent angular coefficients functions $I_i(q^2)$ in bins of $q^2$, following the methodology employed by Belle II in the $\Bd \to \Dstar \ell \nu$ analysis~\cite{B2DstBelle24}. This is possible since in a given bin of the phase-space ($k$-th bin in $q^2$, $p$-th bin in $\cos\theta_{\ell}$, $q$-th bin in $\cos\theta_d$ and $r$-th bin in $\chi$), we expect from \ref{eq:diffRate} a number of signal events equal to 
\begin{align*}
    N^{k, p, q, r} &= \int_{(\Delta q^2)_k}\int_{(\Delta\cos\theta_{\ell})_p}\int_{(\Delta\cos\theta_{d})_r}\int_{(\Delta\chi)_r} \frac{d^4\Gamma}{dq^2 d\cos\theta_{\ell} d\cos\theta_d d\chi} dq^2 d\cos\theta_{\ell} d\cos\theta_d d\chi  \\
        &\propto \sum_i \int_{(\Delta q^2)_k} \left( 1 - \frac{m_{\mu}^2}{q^2} \right)^2 \vert \Vec{p}_{\Dss}(q^2) \vert I_i(q^2) dq^2 \cdot \int_{(\Delta\Omega)_{p,q,r}} \Xi_i(\theta_{\ell}, \theta_d, \chi) d\Omega \\
        &\propto \sum_i J_i^k(q^2) \cdot \zeta_i^{p,q,r}(\theta_{\ell}, \theta_d, \chi)
\end{align*}
where the $\zeta_i^{p,q,r}(\theta_{\ell}, \theta_d, \chi)$ integrals are analytically computable, and the $q^2$-dependent integrals $J_i^k$ are treated as free parameters, thus providing access to the FFs behaviour in bins of $q^2$. This last approach makes no assumptions about the FF parametrisation and provides, in principle, the most direct and unbiased access to the hadronic physics.
\\
 
% \subsubsection{Status and outlook}
The analysis is currently in an advanced stage of development. The full fit framework has been assembled and is finalised with respect to numerical stability and convergence reliability. Discrepancies in the post-fit distributions are currently under investigation, and the evaluation of systematic uncertainties will be carried out once all the data-simulation comparisons are understood. In summary, the full four-dimensional angular analysis of $\Bs \to \Dss \mun \nu_{\mu}$ at LHCb represents a qualitative step forward in the experimental programme of semileptonic $b$ physics. By exploiting the full angular structure of the decay, the analysis provides simultaneous sensitivity to the hadronic FFs and — through the SM-forbidden angular coefficients — to potential New Physics in charged-current $b \to c$ transitions, all within a single, self-consistent measurement framework.
\\

\subsubsection{Fitting methods for semileptonic $\B\to \Dstar\ell\neulb$ angular analyses at \lhcb}

{\it Lea Dreyfus} \\

As part of the upcoming LHCb measurements focused on angular analyses of the semileptonic decays $\B\to \Dstar\ell\neulb$,
a comprehensive study of different fitting approaches is being performed to find the best method in terms of statistical precision and modelindependendence. 
In addition to the folding method discussed above, variations of the weighting method proposed in Ref.~\cite{Hill:2019zja} are also investigated. 

The basic idea of the weighting method is to build weighted binned templates for each of the 12 angular terms of the full decay rate, and then to fit the data distribution as a linear combination of these templates. The weights are assigned based on the true values of the angular variables $\cos\theta_D, \cos\theta_\ell$, and $\chi$ in the simulated events, while the templates are built using the reconstructed angles. This way, the method automatically accounts for detector effects and resolution. However, the instrumentation effects in $q^2$ are not taken into account, and thus the angular coefficients obtained from the fit in each $q^2$ bin can be biased. 

To compare the different methods, a simulation of the $\Bzb\to\Dstarp\ellm\neulb$ decay is performed using the \texttt{RapidSim} package~\cite{Cowan:2016tnm}, with acceptance and resolution effects realistic for LHCb Run 3. To study the statistical precision, events are generated according to the SM predictions using CLN FFs~\cite{Caprini:1997mu} implemented in EvtGen~\cite{Lange:2001uf}. For the model dependence, events are reweighted to the BLPR FFs~\cite{Bernlochner:2017jka} with various New Physics contributions using the \texttt{Hammer} package~\cite{Bernlochner:2020tfi}. The range of NP variations is chosen conservatively such that the magnitude of each NP amplitude, pseudoscalar (P), right-handed vector (VR) and tensor (T), is equal to the SM left-handed vector amplitude. In addition to the pure SM scenario, 8 NP scenarios are tested: single P, VR, and T amplitudes being purely real or purely imaginary, as well as P and T interference being purely real or purely imaginary.

Using the same close-to-minimal binning for both methods (3 bins in $\cos\theta_{\ell}$ and $\cos\theta_D$, 6 bins in $\chi$, 4 bins in $q^2$ and 5 bins in the missing mass squared $m_{miss}^2 \equiv(p_B - p_{D^*} - p_\ell)^2$), the folding method is found to have a 1.5--2 times larger statistical uncertainty on the angular coefficients compared to the weighting method. Both methods are somewhat biased even in the SM scenario. However, the folding method has a bias few times larger than the weighting method in some extreme NP scenarios. 

In an attempt to further reduce the model dependence, a modification of the weighting method is introduced, where the $q^2$
dependence is also included in the construction of the templates by introducing a polynomial parametrisation. Since a polynomial of order at least four is needed for an accurate description of the $q^2$ dependence of each angular term, the number of templates increases from 12 to about 60. This technique also requires finer binning in $q^2$, which, combined with the rectangular binning in the angles, would lead to a very large total number of bins. Therefore, a non-rectangular binning in angular space is proposed, together with the technique to optimise it for the best statistical precision. As a result, the polynomial weighting method can reach a similar level of statistical precision compared to the $q^2$-integrated weighting method, while providing a direct model-independent $q^2$ dependence of each angular coefficient. 
\\

\subsubsection{CPV in \Bzb $\to$ \Dstarp  \mun \neumb: Angular analysis using triple products}

{\it Anna-Maria Heyn}\\

A specific observable related to the studies of the fully differential decay density in $b\to c$ semileptonic transitions is \CP asymmetry. While \CP-violating terms are measured, in particular, in the full angular analysis of \Bzerodecayell\ decays, a dedicated analysis is beneficial. 
Generally, \CP violation arises from a weak phase difference. In semileptonic $\B$ decays  of the kind $\Bzb\to D^{(*,**)} \ellm \neulb$, CPV can only occur in the presence of NP, as there is only one SM amplitude and an NP admixture is needed to generate the weak phase difference. CPV thus model-independently shows the presence of NP. In the case of NP, CP violation can show up in triple products or in the decay rate if multiple hadronic states interfere to generate an additional strong phase difference. Both effects can of course appear in tandem. 

In decays with a \Dstar meson, there is only one strong amplitude and no interference, hence the CPV will only show in the triple products
in different combinations of NP couplings, see \cite{Bhattacharya:2019olg}. In the decay \Bzerodecayell, where the leptons are light (\ie electrons or muons), the triple products are sensitive to $Im(g_R)$ and $Im(g_Pg_T^*)$ at leading order. Looking at $\ell = \tau$, the terms proportional to $ m_\ell / \sqrt{q^2}$ are not suppressed. This introduces a sensitivity to $Im(g_T)$ and $Im(g_P)$ individually. The behaviour of decays with \Ddoublestar mesons depends on the quantum numbers of the excited $\D$ states: decays with $D^*_2$ mesons are sensitive to the same combinations of couplings as \Bzerodecayell decays, whereas the decays with $D_1$ mesons are sensitive to $g_S$ instead of $g_P$, see Ref.~\cite{Bernlochner:2017jxt}.

For CPV to be apparent in the \emph{decay} directly, a strong phase is also needed. This is possible if there are multiple states that can interfere (see Ref.~\cite{Aloni:2018ipm}). This is not possible in the decay \Bzerodecayell, but in the presence of higher \Dstar excitations. For example, the decays with interfering \Dtwo\ and $\Done\to \Dstar\pi$ decays are sensitive to the same combinations of couplings as above, but they are experimentally different and could thus be a complementarily analysed. For the interference of the \Dzerostar\ and \Ddecay decays, the expected effect is small due to the small width of the \Dstarp meson. 

Looking at $B_s \to D_s^{**}\ell\nu$ decays instead can also be fruitful. The $D^{**}_s$ states are more separated compared to the $D^{**}$ ones and thus do not interfere. Without the interference a CPV measurement in the decay is not possible. But with cleaner separation between the peaks, the experimental procedure is easier for triple products. Thus, the triple products can also be measured for example in the $B_s \to D_{s1}(2536)\ell\nu$, $D_{s1}(2536)\to D^*K$ decay, 
%, where the $D_{s1}$ is much narrower than the $D_1$ that overlaps with the $D^*_2$
allowing an additional avenue to $Im(g_S)$\cite{Bhattacharya:2019olg}. This could be possible for example with LHCb after Run 3. 

For the \BzerodecayDdecay\ decay, \lhcb is aiming to measure the \CP-violating parameters in a dedicated analysis using Run 2 data samples. More details of this decay and an analysis strategy are detailed here. 

At leading order, the terms generating CPV are shown in Table~\ref{tab:ang_funcs} (from Ref.~\cite{Bhattacharya:2019olg}). 
The three angles $\theta_D$, $\theta_\ell$ and $\chi$ are defined analogously to the first part of  Sec.\ref{sec:federico}. 
In general, the differential decay rate of this decay is the sum of $P$-even and $P$-odd components: 
\begin{equation}
        P(q^2, \theta_\ell, \theta_D, \chi) = P_\text{even}(q^2, \theta_\ell, \theta_D, \chi) + P_\text{odd}(q^2, \theta_\ell, \theta_D, \chi)\label{eq:P_evenodd}, 
\end{equation}
where the $P$-odd part is the sum of two terms
\begin{equation}
        P_\text{odd}(q^2, \theta_\ell, \theta_D, \chi) = P_{\text{odd}}^{(1)}(q^{2}, \theta_{\ell}, \theta_{D}) \sin\chi
        + P_{\text{odd}}^{(2)}(q^{2}, \theta_{\ell}, \theta_{D}) \sin 2\chi. 
\end{equation}
The $P$-odd component can be extracted from the total decay rate as 
\begin{equation}
    P_{\text{odd}}^{(\{1,2\})}(q^{2}, \theta_{\ell}, \theta_{D})
        = \frac{1}{\pi} \int_{-\pi}^{\pi}
        P(q^{2}, \theta_{\ell}, \theta_{D}, \chi) \, 
        \{\sin \chi, \sin 2\chi \} \, \text{d}\chi, \label{eq:P_cancel}
\end{equation}
such that the $P$-even contribution is fully cancelled out. 
This is beneficial because in this way the $P$-even systematic effects that make a large part of full angular analyses, such as those from FFs and background mismodelling, can be avoided.  Thus, tighter constraints for the determination of the NP couplings is possible. The statistical precision for this method is expected to be on the level of 1\% for $Im(g_R)$ to 0.1\% for $Im(g_Pg_T^*)$, see Ref.~\cite{Dedu:2023yxx}. 

\begin{table}
\centering
\resizebox{\linewidth}{!}{
\begin{tabular}{c|c|c}
\textbf{Coefficient} & \textbf{Coupling} & \textbf{Angular function} \\
\hline
Im$(A_{\perp} A_0^*)$ & Im$[(1 + g_L + g_R)(1 + g_L - g_R)^*]$ & $- \sqrt2\sin 2 \theta_\ell \sin 2 \theta_D \sin \chi$ \\
% \hline
Im$(A_{\parallel} A_{\perp}^*)$ & Im$[(1 + g_L - g_R)(1 + g_L + g_R)^*]$ & $2 \sin^2 \theta_\ell \sin^2 \theta_D \sin 2\chi$ \\
% \hline
Im$(A_{SP} A_{\perp,T}^*)$ & Im$(g_P g_T^*)$ & $ - 8 \sqrt2\sin \theta_\ell \sin 2\theta_D \sin \chi$ 
\end{tabular}}
\caption{Triple products, NP couplings and their angular functions, from Ref.~\cite{Bhattacharya:2019olg}.}
\label{tab:ang_funcs}
\end{table}

Effects that are $P$-odd can be either \CP-odd or \CP-even, depending on their behaviour under charge conjugation. The \CP-odd contribution is the main goal of the analysis, but looking at the \CP-even counterparts of the NP parameters from the same data can still be instructive. 
%The latter are not expected in this decay, and this feature can be used to obtain a control sample. 
There is no \CP-even $P$-odd contribution expected because those would be coming from a strong phase difference, which is impossible in this decay mode. Therefore, this contribution is zero in the signal, even in the presence of NP. This component can thus serve as a control sample. Other effects, such as certain backgrounds or instrumentation effects, can produce nonzero contributions. 
The most significant backgrounds are detailed in Ref.~\cite{Dedu:2023yxx}; none of them are expected to significantly impact the uncertainties. 

Additionally, $P$-odd terms can be generated by certain instrumentation effects. Systematic biases due to non-uniform efficiencies and misalignment of the tracker are currently under investigation.

Another $P$-odd effect that can influence the control channel is the \emph{$S$-wave interference}. 
In data, not only \BzerodecayDdecay\ is present, 
but also the $S$--wave: \Bzerodecayswave. 
The amplitudes of those two decays add directly. 
Simulation is performed using the \BzerodecayDdecay\ amplitude parametrisations from Ref.~\cite{Ligeti:2016npd}
and the \Bzerodecayswave\ amplitude from Ref.~\cite{Bernlochner:2017jxt}. 
The relative magnitude between the amplitudes is determined by using approximations of their respective branching fractions give in Ref.\cite{Gustafson:2023lrz}.
The branching fraction for the decay $\Bzb \to \D_0^*(2300) \mun \neumb$ is taken from the estimate in Ref.~\cite{Du:2025beb}, while that for the decay \Bzerodecaymu\ is taken from the PDG~\cite{ParticleDataGroup:2024cfk}. 
Simulation shows that the fraction of $P$-odd contribution from this interference effect is $<0.1\%$, which is smaller than the expected statistical precision.

\clearpage

%%%%%%%%%%%%%%%%%%%%%%%%%%%%%%%%%%%%%%%%%%%%%%%%%
% b->c l nu
% Theoretical status at b-factories and LHCb
% form factors, Vcb and Vub
% lattice and continnum approaces
%%%%%%%%%%%%%%%%%%%%%%%%%%%%%%%%%%%%%%%%%%%%%%%%%

\section{Theoretical status of charged-current semileptonic b hadrons decays}
\label{sec:th_cc}

A precise experimental programme in charged-current semileptonic decays fully relies on theoretical predictions for the relevant hadronic matrix elements. For exclusive decays, the main theoretical ingredients are the FFs describing the $B_{(s)} \to D_{(s)}^{(*)}$ and $B \to \pi$ transitions, which must be computed non-perturbatively, primarily through LQCD and LCSR. For inclusive decays, HQE allows to write the decay rate as a systematic series in $\Lambda_\text{QCD}/m_b$, with perturbative coefficients that are now known to NNLO for the leading-power contributions. The expansion also involves non-perturbative inputs, namely the matrix elements of HQET operators, which are universal hadronic parameters and can be determined from experimental data. This section reviews the state of the art in both approaches. We highlight the open questions and the theoretical improvements needed to match the precision anticipated from future experimental datasets.

\subsection{Extraction of the $B \to D^*$ FFs from experimental data}

{\it Silvano Simula}\\

The semileptonic $B \to D^* \ell \nu_\ell$ decay represents a golden channel for testing the SM as well as for phenomenological studies of the effects of NP. 
As is well known, within the SM the hadronic ingredients are four FFs, namely the vector FF $g(w)$, the axial FFs $f(w)$ and $F_1(w)$, and the pseudoscalar FF $F_2(w)$, while the NP contributions require the knowledge of three tensor FFs.

The angular distributions of the experimental data sets available from Refs.~\cite{Belle:2018ezy, Belle:2023bwv, Belle-II:2023okj} for light leptons in the final state have been analysed for the first time in Ref.\,\cite{Martinelli:2024vde}. Following a straightforward approach, first proposed in Ref.\cite{Bobeth:2021lya}, the problem is reduced to the determination of five basic parameters, which encode in the most general way the contributions to the single-differential decay rates coming from operators present in the effective Hamiltonian either in the SM or from NP. 
The analysis is model independent and never requires knowledge of $\vert V_{cb}\vert$. 

The findings of Ref.\,\cite{Martinelli:2024vde} allow a direct comparison among the results obtained from different experiments as well as with the theoretical predictions based on the hadronic FFs obtained in Ref.\,\cite{Martinelli:2023fwm} by applying the unitary DM approach\,\cite{DiCarlo:2021dzg} to all available results from the LQCD simulations of Refs.~\cite{FermilabLattice:2021cdg, Harrison:2023dzh, Aoki:2023qpa}. 
It turns out that for the forward-backward asymmetry $A_{FB}$, the longitudinal $D^*$-polarization fraction $F_L$ and the transverse asymmetry $A_{1c}$ there are visible differences between different experimental data sets within about two standard deviations. Similar differences exist between the SM predictions, based on all available LQCD FFs, and some sets of data. In the rest of this contribution, the results corresponding to the FFs obtained in Ref.\,\cite{Martinelli:2023fwm} using simultaneously all the lattice points from Refs.~\cite{FermilabLattice:2021cdg, Harrison:2023dzh, Aoki:2023qpa} will be referred to as the LQCD predictions. 

Remarkably good agreement is observed between the experimental data of Ref.\,\cite{Belle:2023xgj}, given in terms of the twelve angular coefficients $J_i(w)$ of the four-fold differential decay rate (analogous to those defined in Eq.~\eqref{eq:diffRate}), and the LQCD predictions. Such a consistency, once confirmed by further experiments, may leave little room for NP effects in the light lepton sector. In this respect, the forthcoming results from \lhcb\,\cite{Manganella:2025jfg} concerning the coefficients $J_i(w)$ in the case of $B_s^0 \to D_s^* \mu \nu_\mu$ decays will be very valuable.

A further analysis of the semileptonic  $B \to D^* \ell \nu_\ell$ decays has been performed in Ref.\,\cite{Martinelli:2024bov} by extending the work of Ref.\,\cite{Martinelli:2024vde}, limited to light leptons in the final state, through the inclusion of semitauonic decays.

In addition to the experimental data on light leptons from Refs.~\cite{Belle:2018ezy, Belle:2023bwv, Belle-II:2023okj} the measurements of the ratio of branching fractions $R(D^*)$, the longitudinal $D^*$-polarization fraction $F_{L, \tau}$ and the $\tau$-polarization $P_\tau(D^*)$ have been considered. For all lepton channels the analysed quantities do not require the knowledge of $|V_{cb}|$. The hadronic FFs relevant in the SM are extracted directly from the experimental data using a fitting procedure based on the BGL $z$-expansions of the FFs\,\cite{Boyd:1997kz} with built-in constraints from unitarity. The only theoretical inputs are the susceptibilities (derivatives of the vacuum polarization functions), computed on the lattice in Ref.\,\cite{Martinelli:2021frl}, required to impose the constraints of unitarity on the extracted FFs. 

The main results of Ref.\,\cite{Martinelli:2024bov} are as follows:
\begin{itemize}

\item The value of $F_2(w) / f(1)$ necessary to reproduce the experimental values of $F_{L, \tau}$, determined recently in two $w$-bins by the \lhcb Collaboration\,\cite{LHCb:2023ssl}, is in good agreement with LQCD calculations. In contrast, to fit the experimental determination of $R(D^*)$ a much larger value of $F_2(w) / f(1)$ is necessary in the whole relevant kinematical range.  
This difference may be due to some fluctuation/systematic effect in the experimental measurements or to the presence of NP that affects in a different way $F_{L, \tau}$ and $R(D^*)$ (see also Refs.~\cite{Blanke:2018yud, Blanke:2019qrx}). In any case the difference is within 2$\sigma$ deviations.

\item The large value of $F_2(w) / f(1)$, required by the addition of the experimental value of $R(D^*)$, yields values of other observables in semitauonic decays, namely $F_{L, \tau}$, $P_\tau(D^*)$, $A_{FB, \tau}$ and $A_{1c, \tau}$, in tension with the corresponding LQCD predictions up to the level of $\approx 4\sigma$. In particular, the SM implies a strong anti-correlation between $R(D^*)$ and $A_{FB, \tau}$, an observable not yet measured, see also Ref.~\cite{Fedele:2023ewe}.

\item There is an apparent tension between the helicity amplitude $H_0$, obtained from the analysis of the light-lepton data of Refs.~\cite{Belle:2018ezy, Belle:2023bwv, Belle-II:2023okj}, and the LQCD FF $F_1$, used to compute $H_0$, obtained in Ref.\,\cite{Martinelli:2023fwm}. However, there is good consistency between $H_0$, obtained using only the data from Ref.\,\cite{Belle:2023bwv}, and $F_1$ (see also Ref.\,\cite{Martinelli:2024vde}). Finally, there is also good consistency between $H_0$, obtained from the analysis of the light-lepton data of Refs.~\cite{Belle:2018ezy, Belle:2023bwv, Belle-II:2023okj}, and $F_1$ from JLQCD only.
The differences among the FFs determined by each of the lattice Collaborations (in particular, the so-called {\it issue of the slope of $F_1$}\footnote{In Ref.\,\cite{Martinelli:2024bov} it is shown that an improvement in the lattice determination of the so-called HQET ratio $R_2(1)$ at zero recoil will be crucial to address the issue of the slope of $F_1$.}) are extensively discussed in Ref.\,\cite{Martinelli:2023fwm}, where their impact on the determination of both $\vert V_{cb} \vert$ and $R(D^*)$ is properly analysed.

\item The simultaneous addition of all the tauonic quantities is dominated by the experimental value of $R(D^*)$ due to its current, higher precision.

\end{itemize}

Moreover, using the reduced FFs extracted\,\cite{Martinelli:2024bov} from the light-lepton data of Refs.~\cite{Belle:2018ezy, Belle:2023bwv, Belle-II:2023okj} and the total decay rate for the semileptonic $B \to D^* \ell \nu_\ell$ with $\ell = e, \mu$ from the latest PDG review\,\cite{ParticleDataGroup:2024cfk}, one obtains $|V_{cb} f(1)| = 0.2318 (46)$\,GeV. Assuming for $f(1)$ the LQCD prediction $f(1) = 5.845(50)$\,GeV, one has $|V_{cb}| = 39.7 (8) \cdot 10^{-3}$. This value agrees well with the result $|V_{cb}| = 39.92 (64) \cdot 10^{-3}$, obtained by the bin-per-bin analysis of Ref.\,\cite{Martinelli:2023fwm} carried out using the same experimental data sets and lattice points. It also agrees with the result $|V_{cb}| = 39.46 (53) \cdot 10^{-3}$ from the FLAG review\,\cite{FlavourLatticeAveragingGroupFLAG:2024oxs}, while it deviates by $\approx 2.1 \sigma$ from the latest determination $|V_{cb}| = 41.64 (47) \cdot 10^{-3}$\,\cite{Carvunis:2025vab} from inclusive semileptonic decays and  by $\approx 2.7 \sigma$ from the updated prediction $|V_{cb}| = 42.22 (51) \cdot 10^{-3}$ obtained within the SM by UTfit\,\cite{UTfit:2022hsi}.

The results of the analysis of Ref.\,\cite{Martinelli:2024bov} indicate a clear difficulty in simultaneously reproducing the experimental values of $R(D^*)$ and of other $\tau$-observables within the SM. In particular, an important correlation has been found between $R(D^*)$ and $F_{L,\tau}$ and a relevant anti-correlation between $R(D^*)$ and $A_{FB, \tau}$, a quantity not yet measured.

In the next future it will be important to improve the current precision of the measurements of the $\tau$-observables as well as to perform a first-time determination of the forward-backward $A_{FB, \tau}$ and transverse $A_{1c, \tau}$ asymmetries. That will be crucial in learning more about the suggested inconsistency found within the SM between the experimental values of $R(D^*)$ and of other $\tau$-observables.
On the theoretical side it will be important to include in the BGL $z$-expansion the effects of sub-threshold branch-cuts (i.e., below the pair-production one), like in the cases of the semileptonic $B \to D^* \ell \nu_\ell$ or $B_c \to J / \Psi \ell \nu_\ell$ decays. This can be done following the recent  approach of multiple dispersive bounds described in Refs.\,\cite{Simula:2025lpc,  Simula:2025fft}.\\

\subsection{Extraction of the $B \to D^*$ FFs from LQCD}

{\it Alejandro Vaquero}\\

\begin{figure}[h!]
\centering
\includegraphics[width=0.95\textwidth,angle=0]{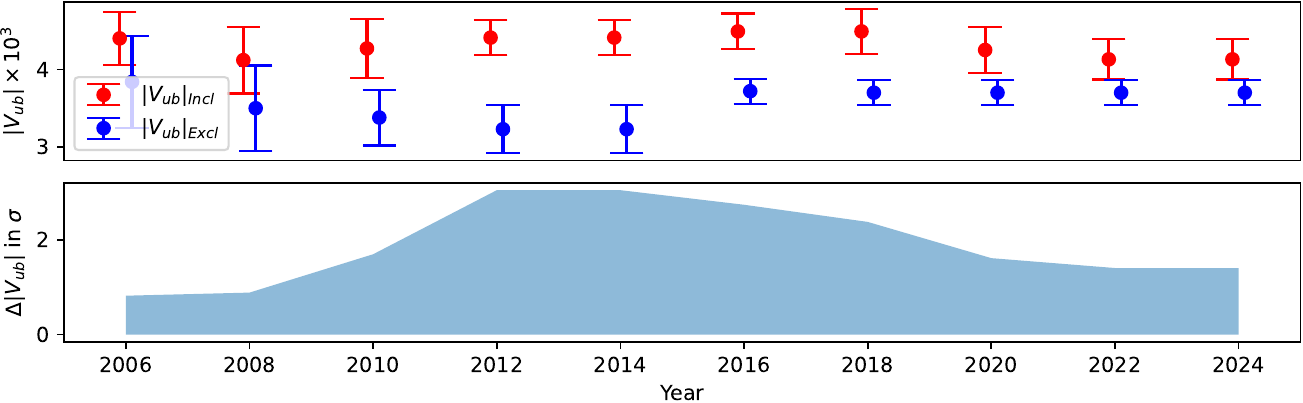}

\includegraphics[width=0.95\textwidth,angle=0]{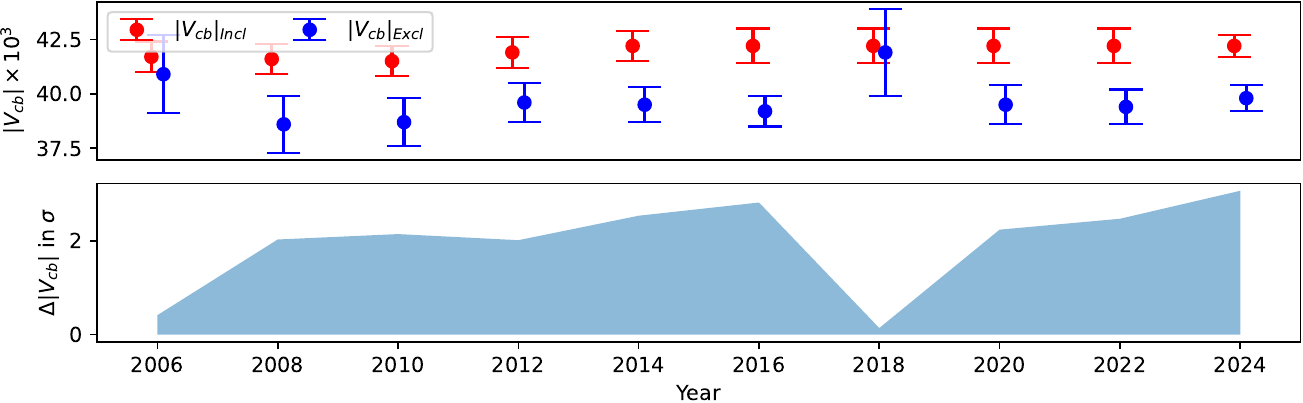}
\caption{Current tensions between the inclusive and exclusive determinations of $V_{xb}$. Data extracted from the PDG review of each year as marked.}
\label{fig:CKMtensions}
\end{figure}

There are two powerful reasons to explore semileptonic $B$ decays.
The first one is the current discrepancy between inclusive and exclusive determinations of the $V_{ub}$ and $V_{cb}$ CKM matrix elements (see Fig.~\ref{fig:CKMtensions} for an historical overview).
Decomposing the discrepancy into its two components, we observe that the tension in $V_{ub}$ is not really significant, standing below $2\sigma$. In contrast, $V_{cb}$ displays significant differences since 2008 at the $2\sigma-3\sigma$ level. The tension is large enough to be uncomfortable, but small enough so it cannot be taken as an unmistakeable sign of NP, and we have not managed to improve this situation for almost 20 years. It is certainly an annoying tension.

\begin{figure}[h!]
\centering
\includegraphics[width=0.6\textwidth,angle=0]{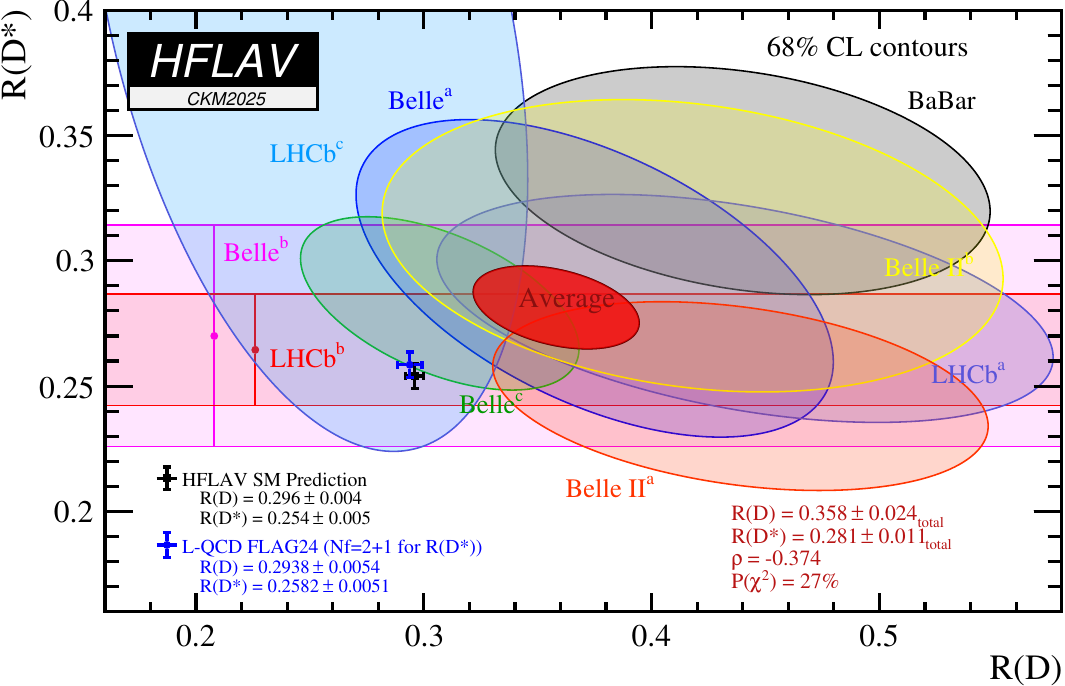}
\caption{Current tensions in the $R(D)$ vs $R(D^\ast)$ plane according to HFLAV~\cite{HeavyFlavorAveragingGroupHFLAV:2024ctg}, as presented in the CKM25 conference.}
\label{fig:LFUratios}
\end{figure}

The second part of the motivation comes from the LFU ratios $R(D)$ and $R(D^*)$, see Fig.~\ref{fig:LFUratios}.
Although there has been great progress during the last few years, the current tensions between theory and experiment in the $R(D)-R(D^\ast)$ plane are around $3\sigma-4\sigma$, mainly coming from $R(D)$. Recently, a new theoretical result coming exclusively from lattice calculations has been published~\cite{FlavourLatticeAveragingGroupFLAG:2024oxs}, that is in very good agreement with previous theoretical estimates. However, these new results have been unable to solve the question at hand.
LQCD calculations play a key role in the solution of both issues.

The CKM matrix elements are SM inputs.
This means that the SM, as a theory, has little to say about their values, except for certain global properties, like the unitarity of the matrix, or their complex nature.
Thus, in order to determine a CKM matrix element, experimental information is necessary.
The SM provides useful expressions, that in the case of a $B\to X\ell\nu$ semileptonic decay, can be summarized in eq.~\eqref{eq:genSLdecay}:
\begin{equation}
\frac{d\Gamma}{dw} = K(w,m_\ell)\left|\mathcal{F}(w)\right|^2\left|V_{xb}\right|^2.
\label{eq:genSLdecay}
\end{equation}
In this formal expression, $\tfrac{d\Gamma}{dw}$ is the differential decay rate, measured in experiment; $\ell$ is the lepton involved in the decay, which is usually assumed to be massless for the cases $\ell=e,\mu$; $K(w,m_\ell)$ contains known kinematic factors provided by the theory; $V_{xb}$ is the CKM matrix element we desire to determine, and $\mathcal{F}(w)$ is a decay amplitude of hadronic nature, that must be computed non-perturbatively.
LQCD can provide the necessary FFs that contribute to the decay amplitude, thus completing the rhs of the equation, whereas experimental measurements provide the lhs.

The LFU ratios, as opposed to the CKM matrix elements, are fully determined  by the theory\footnote{Obviously they depend on other parameters of the theory, i.e., the quark masses, the couplings, etc.}.
This is obvious once we use the general expression in Eq.~\eqref{eq:genSLdecay} to reconstruct the branching fractions: since the CKM matrix elements are independent of the recoil parameter, they cancel in the ratio, and the result does not require any extra experimental input. The only knowledge needed is the FFs over the whole recoil range, which can be provided by LQCD.
Pure LQCD determinations of LFU ratios allow for clear precision tests of the SM, where theory calculations and experimental data are not intermingled.

\begin{figure}[h!]
\centering
\includegraphics[width=0.67\textwidth,angle=0]{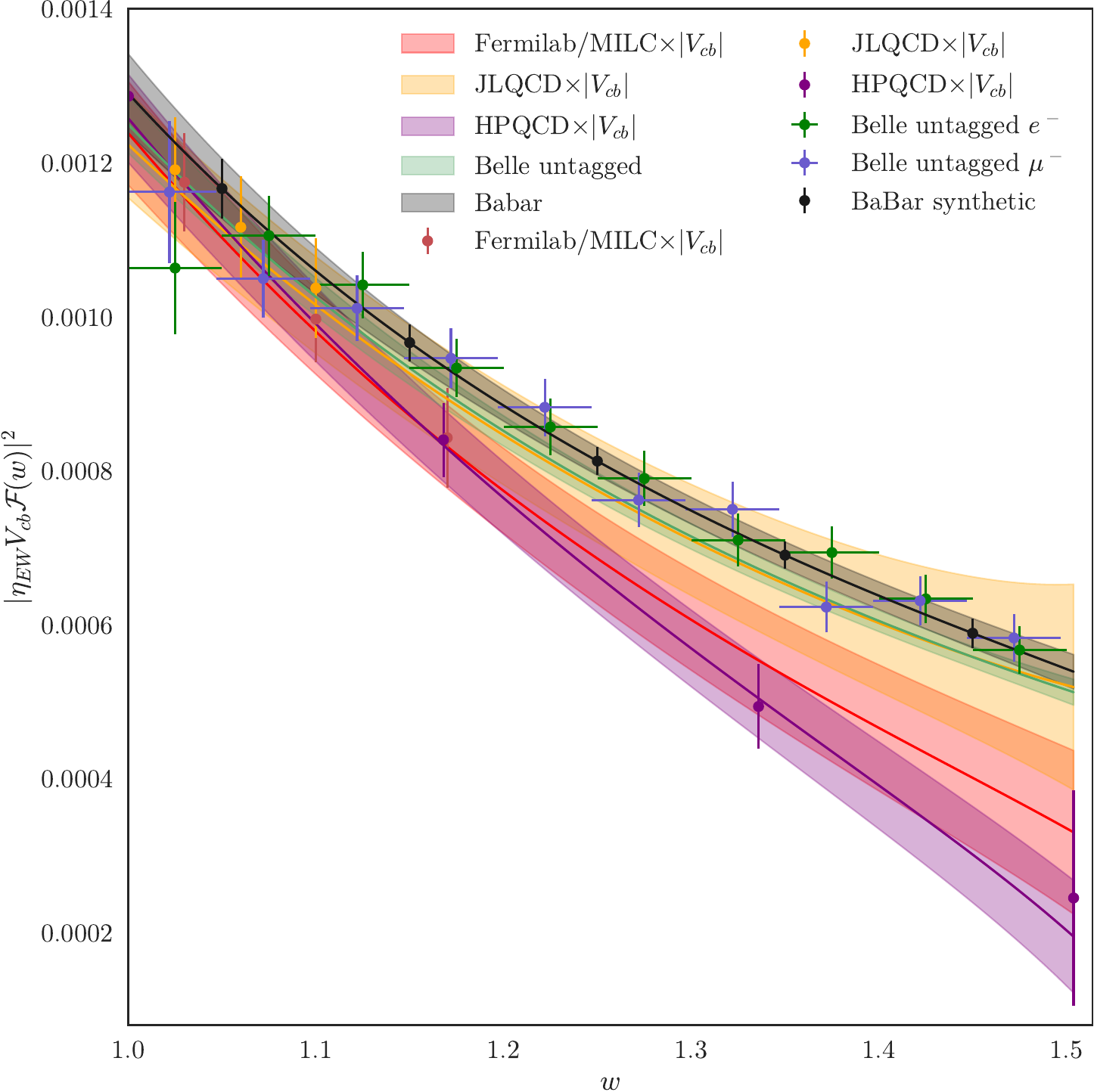}
\caption{Comparison of scaled decay amplitudes $\eta^2_{\textrm{EW}}\left|V_{cb}\right|^2\left|\mathcal{F}(w)\right|^2$ for several experiments and LQCD calculations. Each LQCD calculation is scaled by its own $\left|V_{cb}\right|$ value.}
\label{fig:decayAmplitude}
\end{figure}

Among all the relevant $B_{(s)}$ semileptonic decay channels, $B\to D^\ast\ell\nu$ is arguably the most interesting one.
Experimentally, the pseudoscalar to vector channel is much more accessible than the pseudoscalar to pseudoscalar one, especially at small values of the recoil parameter, where LQCD calculations are more precise.
On the LQCD side, there has been extensive progress in the FF calculations for this particular channel, with three completely independent calculations from three different collaborations recently made available~\cite{FermilabLattice:2021cdg,Harrison:2023dzh,Aoki:2023qpa}.
This surge of LQCD data, however, has not translated into clear solutions for either the tensions between inclusive and exclusive determinations of $V_{cb}$, or the tensions in the $R(D)-R(D^\ast)$ plane.
The main reason is the lack of precision: the FFs are determined within a few percent for small values of the recoil parameter, where the experimental data features larger errors due to the extinction of the phase space. To reach large recoil values, a fit to a suitable parametrization, like BGL~\cite{Boyd:1997kz}, must be done.
Comparing the scaled decay amplitude $\eta^2_{\textrm{EW}}\left|V_{cb}\right|^2\left|\mathcal{F}(w)\right|^2$ obtained from such fits for the three LQCD calculations with the experimental curves, one can see a wide dispersion of results, that suggest the existence of discrepancies among all the data. Fig.~\ref{fig:decayAmplitude} summarizes the situation, where data from the three LQCD calculations, along with data from the 2018 untagged Belle analysis~\cite{Belle:2018ezy} and the 2019 BaBar analysis~\cite{BaBar:2019vpl} for this channel are included.
A careful analysis reveals that the LQCD data shows a good agreement among collaborations, and combined fits are possible mixing the FFs of all the different calculations without incurring in any serious tensions, see also Ref.~\cite{Bordone:2025jur}. This is not surprising: the LQCD FFs are generally compatible within $2\sigma$ across collaborations.
In contrast, a simple combined BGL fit of the Belle and BaBar datasets results in a fit with relatively low quality\footnote{My own results are $p\sim 0.04$ for a 223 quadratic fit for $g$ and $f$, and a cubic fit for $\mathcal{F}_1$.}.
This fact does not imply that there is anything wrong with experimental data, but that we need to improve our datasets.
The same must be said in the case of LQCD data. Even though the data from different calculations is consistent, the errors are still too large to solve any of the issues that motivated these works.
Clearly additional efforts are needed.

The case of $B_{(s)}$ decays to light mesons much more serious, in the sense that different LQCD calculations disagree among each other beyond the $2\sigma$ level.
The last FLAG report~\cite{FlavourLatticeAveragingGroupFLAG:2024oxs} assigns large reduced $\chi^2$ values to their combined fits to the FFs of the calculations that pass all the quality requirements, reaching $\chi^2 > 3$ for $B\to\pi\ell\nu$ and $B_s\to K\ell\nu$, and $\chi^2\approx 2$ for $B\to K\ell\ell$.
The FLAG report compensates for this disagreement by appropriately increasing the resulting errors, but the situation remains at least uncomfortable.
Hence, improved LQCD calculations are also needed in the $B_{(s)}$ to light decays to clarify the situation.

Although the LQCD community is very sensitive to these problems, there are just a few collaborations that devote a significant amount of resources to explore the relevant heavy-to-heavy and heavy-to-light channels.
Among them, the Fermilab Lattice and MILC collaborations are outstanding because of the extension of the program, covering four heavy-to-heavy $B_{(s)}\to D^\ast_{(s)}\ell\nu$, and three heavy-to-light $B\to\pi\ell\nu$, $B_s\to K\ell\nu$ and $B\to K\ell\ell$ channels, and because of the quality of the data.
The collaboration proposes two different analyses, mainly differing in how to treat the heavy quarks.
\begin{figure}[h!]
\centering
\includegraphics[width=0.49\textwidth,angle=0]{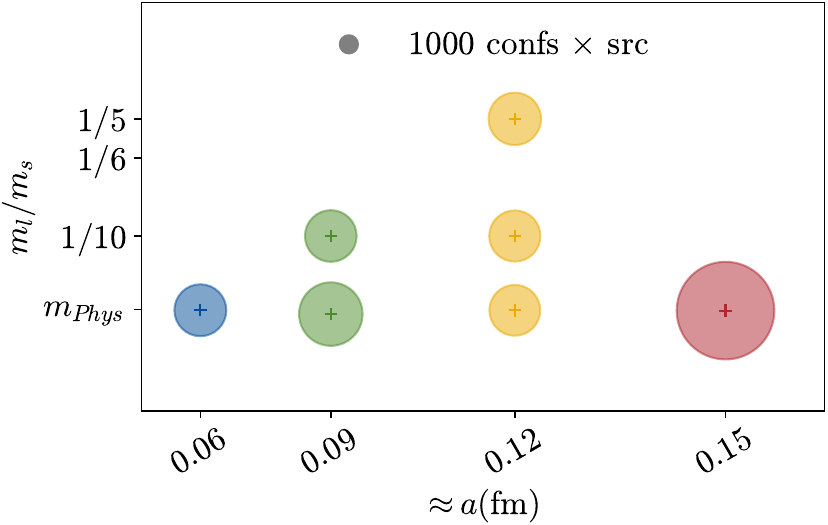}
\includegraphics[width=0.49\textwidth,angle=0]{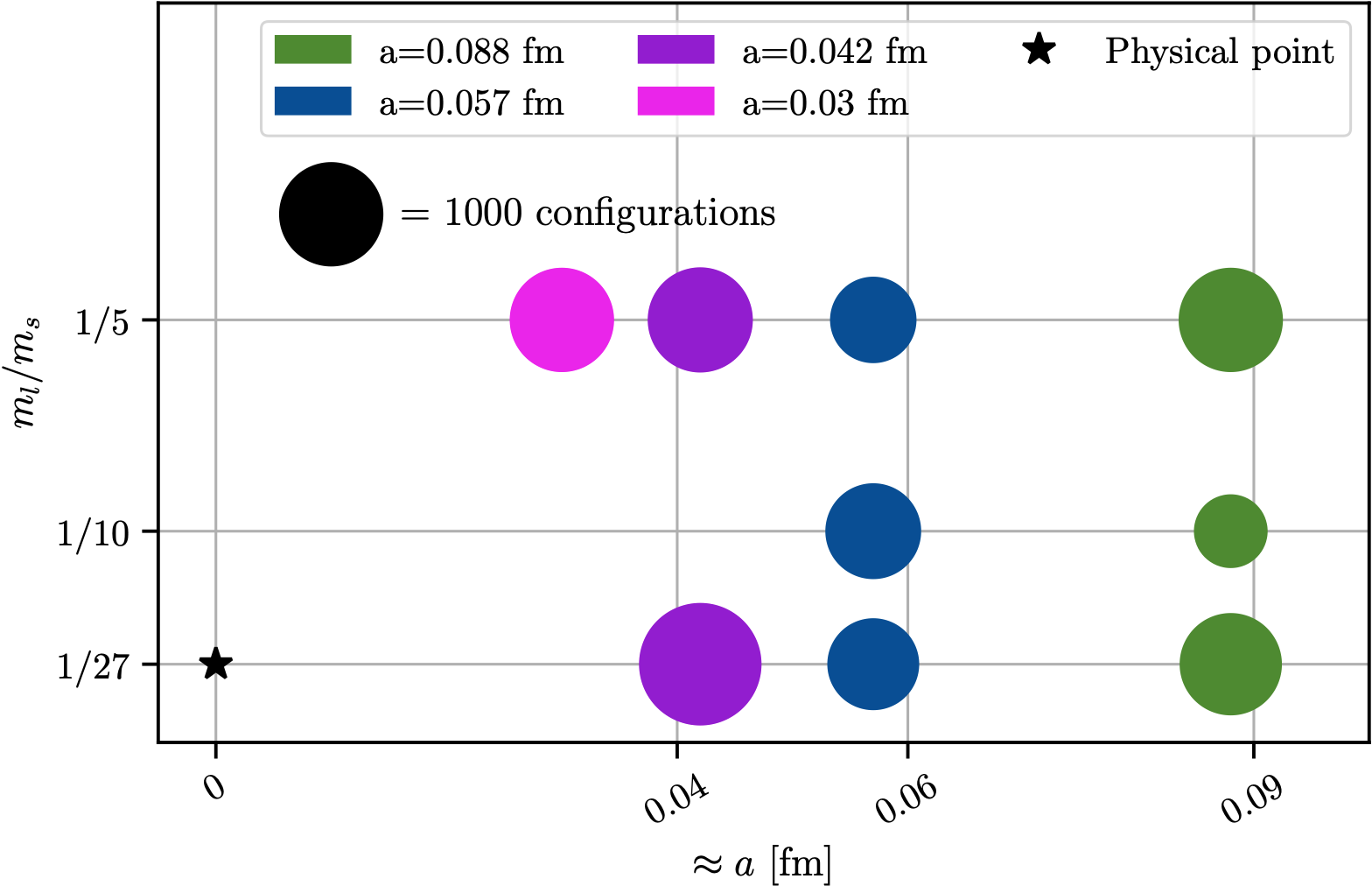}
\caption{Left: Ensembles contributing to the analysis with Fermilab heavy fermions. Right: Ensembles contributing to the all-HISQ analysis.}
\label{fig:ensembles}
\end{figure}
The first analysis uses 7 ensembles with $N_f=2+1+1$ of HISQ fermions~\cite{Follana:2006rc}, whereas the valence $b$ and $c$ quarks are described using an EFT, the so-called Fermilab fermions~\cite{El-Khadra:1996wdx}.
This setup allows to reach physical heavy quark masses, at the cost of requiring a perturbative matching between the EFT and LQCD, resulting in potentially large systematic errors in the renormalization of the axial and vector currents.
On the other hand, more than half the ensembles feature physical pion masses, which should noticeably reduce the errors associated to the chiral extrapolation, as shown in the left panel of Fig.~\ref{fig:ensembles}.
On the whole we expect a moderate decrease in the systematic errors with respect to the previous calculation of the collaboration~\cite{FermilabLattice:2021cdg}.
Moreover, the analyses of all the heavy-to-heavy and the heavy-to-light channels are being carried out simultaneously, in order to keep all the correlations between the FFs.
As a result, one of the products of the analysis is planned to be a correlated result in the $R(D)-R(D^\ast)$ plane, and in the $\left|V_{ub}\right|-\left|V_{cb}\right|$ plane.
This analysis is in a relatively advanced state, and we expect to finalize it by the end of this year.

The second analysis uses the HISQ action for both, heavy and light quarks, as shown in the right panel of Fig.~\ref{fig:ensembles}. 
Although this approach allows us to employ a non-perturbative renormalization scheme, almost eliminating this particular systematic error, it requires the usage of non-physical $b-$quark masses:
the lattice formalism introduces an ultraviolet regulator, the lattice spacing $a$. Particles whose masses are of the same order as the regulator suffer from large discretization effects.
The calculation can still be performed by reducing the mass of the $b$ quark employed in the simulation. A heavy-quark extrapolation can recover physical results, in the same spirit as the chiral extrapolation can recover physical results from simulations at unphysical pion masses, at the cost of introducing a new systematic error in the calculation.
Since heavy-quark discretization errors are usually the largest systematic errors in these calculations, the problem is particularly serious.
This calculation promises to reduce these errors by combining the particularities of the HISQ action with low values of the UV regulator, that allow us to reach physical bottom quark masses.
The HISQ action has been engineered to perform particularly well for heavy quarks, and shows reduced discretization errors $O(v\alpha_s a^2m^2)$, where $v$ is the velocity of the heavy quark, with respect to other actions, that feature an error that scales as $O(a^2m^2)$.
This means that for the same value of the lattice spacing, the heavy quark mass can be larger, before the discretization errors become large enough to make the calculation unfeasible.
If we combine this feature with very fine ensembles, we obtain a calculation with a very smooth extrapolation —or even interpolation— to the physical heavy quark mass.
As a result, the largest source of systematic errors in the error budget is expected to be greatly reduced.
But this calculation, that would again cover the four heavy-to-heavy and the three heavy-to-light channels is a few years away, since we are still at the stage of data generation.

Summing up, as experimentalist improve their measurements of differential decay rates and LFU ratios in the $B$ factories and the LHCb, the LQCD community must raise to the challenge to maximize the utility of experimental results.
The status of the LQCD calculations in the heavy-to-heavy decays has steadily improved, but better calculations are required to sort our the $\left|V_{ub}\right|$ vs $\left|V_{cb}\right|$ puzzle.
The picture in the heavy-to-light decays is objectively worse, with clear tensions between different LQCD calculations that must be urgently explained.
The Fermilab Lattice and MILC collaborations is committed to contribute positively to solve these issues, and to this end it is currently working on two different analyses, to be finalized in the following years.\\

\subsection{Extraction of the $B_{(s)} \to D_{(s)}^*$ FFs from HQE}

{\it Nico Gubernari}\\

Semileptonic $b \to c \ell \bar\nu$ transitions play a central role in precision flavour physics. They provide one of the main avenues for the extraction of the CKM matrix element $|V_{cb}|$ and, more broadly, for testing the SM at the precision frontier. As direct searches have not revealed clear signs of NP so far, indirect probes based on precise comparisons between theory and experiment have become increasingly important. In this context, reducing theoretical uncertainties in hadronic matrix elements is essential in order to fully exploit the present and future experimental programs at LHCb and Belle II.

The main theoretical challenge in exclusive semileptonic decays is the description of the hadronic matrix elements, which encode non-perturbative QCD effects. These matrix elements are conventionally decomposed into FFs, functions of the momentum transfer $q^2$, whose precise determination is required for phenomenological applications. For the processes $\bar B_{(s)} \to D_{(s)}^{(*)}$, increasingly accurate LQCD results are now available~\cite{Bailey:2012rr,Atoui:2013zza,Na:2015kha,MILC:2015uhg,McLean:2019qcx,FermilabLattice:2021cdg,Harrison:2023dzh,Aoki:2023qpa}, while light-cone sum rules provide complementary information, especially at low $q^2$~\cite{Gubernari:2018wyi,Bordone:2019guc}. Since these inputs are typically obtained at discrete kinematic points, a parametrization of the FFs is needed to interpolate and extrapolate them consistently over the full physical region. 
Such a parametrization should not merely be convenient, but should also incorporate as much theoretical information as possible. Two key ingredients we incorporate are unitarity and heavy-quark symmetry~\cite{Boyd:1997kz,Caprini:1997mu}. On the one hand, unitarity implies that the coefficients of the considered are constrained. On the other hand, HQE provides non-trivial relations among different FFs by organizing them in powers of $\Lambda_{\mathrm{QCD}}/m_c$ and $\Lambda_{\mathrm{QCD}}/m_b$. Combining these two ingredients leads to a more predictive and more strongly constrained description than a purely kinematic expansion. 

In the HQE framework, all $\bar B_{(s)}^{*} \to D_{(s)}^{*}$ FFs can be expressed in terms of a limited set of Isgur--Wise functions: one leading function, three subleading functions, and six subsubleading functions. An important message is that corrections of order $\Lambda_{\mathrm{QCD}}^2/m_c^2$ are numerically relevant and must be retained; simpler truncations, such as the traditional CLN setup, are not sufficient for present-day precision studies~\cite{Bordone:2019vic,Bordone:2019guc}. This framework then correlates a large set of vector, axial, and tensor FFs across the various $\bar B_{(s)}^{(*)} \to D_{(s)}^{(*)}$ channels. 

The new ingredient discussed in this work is the systematic inclusion of tensor FFs and their unitarity bounds~\cite{Bordone:2025jur}. In particular, we define for the first time the tensor FFs for $\bar B_{(s)}^{*} \to D_{(s)}^{(*)}$ transitions and derive their contribution to the unitarity bounds. This allows all $\bar B_{(s)}^{(*)} \to D_{(s)}^{(*)}$ FFs to be treated on the same footing within a combined HQE analysis. The inclusion of tensor channels is valuable even for SM phenomenology, because heavy-quark symmetry relates tensor FFs to the vector and axial ones, thereby improving the overall theoretical control. 

Our strategy is therefore to combine all available theoretical inputs in a global fit of the Isgur--Wise functions, while imposing the unitarity bounds directly on the HQE parametrization. In practice, we considered different truncations of the expansion, denoted by models such as $2/1/0$ and $3/2/1$, and fitted them to LQCD inputs, optionally supplemented by light-cone sum-rule information. A particularly useful implementation is the $3/2/1$ model with a mild $SU(3)$-flavour assumption for the subsubleading functions, which serves as our nominal fit.

The main result is that the currently available LQCD calculations are mutually compatible within the HQE description. In all fit models considered, the global $p$-values are excellent, and the fitted HQE parameters remain of natural size, indicating no pathological behaviour up to order $\Lambda_{\mathrm{QCD}}^2/m_c^2$. In this sense, the present lattice data support a coherent heavy-quark-expansion picture for $\bar B_{(s)} \to D_{(s)}^{(*)}$ FFs. Moreover, the fitted results are compatible with earlier analyses that relied more heavily on QCD sum-rule information. 

Concerning unitarity, we find that some channels---notably the $0^-$ and $0^+$ sectors---are already close to saturation. This shows that dispersive information is phenomenologically relevant and that the combination of HQE relations with strong unitarity bounds provides a non-trivial consistency test of the theory inputs. At the same time, in the present analysis the newly derived tensor bounds do not yet significantly constrain the fit parameter space beyond the existing information, although they establish the formal framework for future, more precise applications.
A related issue concerns subthreshold branch cuts, which are usually neglected in standard form-factor parametrizations such as BGL, CLN, or HQE-based implementations. However, for percent-level precision this effect should be taken into account, since it modifies the analytic structure underlying the dispersive expansion; a systematic treatment of these cuts has recently been developed and provides a way to retain rigorous control over truncation uncertainties~\cite{Gopal:2024mgb,Gubernari:2026sqc}.

Overall, this analysis provides a comprehensive framework for studying $\bar B_{(s)} \to D_{(s)}^{(*)}$ FFs at the precision level required by ongoing and future flavour experiments. It brings together LQCD results, heavy-quark symmetry, and unitarity constraints in a unified treatment, and it extends this program for the first time to the complete set of tensor FFs. Besides clarifying the structure of semileptonic $b \to c$ transitions, these results supply improved theory input for phenomenological predictions, including angular observables and lepton-flavour-universality ratios $R(D_{(s)})$ and $R(D_{(s)}^*)$, and contribute to the broader effort to sharpen exclusive determinations of $|V_{cb}|$.

\subsection{Inclusive semileptonic $B\to X_u\ell\nu$ decays}

{\it Bernat Capdevila}\\

Inclusive semileptonic $B$ decays remain one of the main tools for determining $|V_{ub}|$, but they also pose a delicate theoretical problem. While inclusivity with the hadronic final state allows one to organise the calculation in terms of quark-level quantities through the optical theorem and the HQE, the experimental separation of the $B\to X_u\ell\nu$ signal from the much larger $B\to X_c\ell\nu$ background requires phase-space cuts that enhance the sensitivity to non-perturbative effects. Several frameworks have been developed to deal with this regime, most notably BLNP, GGOU, and DGE, and they continue to be used in phenomenological extractions of $|V_{ub}|$ \cite{Lange:2005yw,Gambino:2007rp,Andersen:2005mj}. In this contribution, I will focus on the GGOU framework and on two recent developments that are particularly relevant for future inclusive determinations of $|V_{ub}|$: the calculation of NNLO perturbative corrections to the fully differential partonic decay~\cite{Broggio:2026edk}, and the perturbative corrections to the $1/m_b^2$ terms~\cite{Capdevila:2021vkf}.

The starting point is the triple differential decay rate,
\begin{equation}
\frac{d^3\Gamma}{d\hat E_\ell\, d\hat q_0\, d\hat q^2}
\propto L_{\mu\nu}W^{\mu\nu},
\end{equation}
where the hatted variables are normalised to the $b$-quark mass, $\hat x=x/m_b$. The leptonic tensor $L_{\mu\nu}$ is fixed by the kinematics of the leptons in the final state. The hadronic tensor $W^{\mu\nu}$ is decomposed in terms of structure functions $W_i$, and for massless leptons only $W_{1,2,3}$ contribute to the physical decay distribution. Within the HQE one writes
\begin{equation}
W_i=W_i^{(0)}+W_i^{(\pi)}\frac{\mu_\pi^2}{m_b^2}
+W_i^{(G)}\frac{\mu_G^2}{m_b^2}
+W_i^{(D)}\frac{\rho_D^3}{m_b^3}
+W_i^{(LS)}\frac{\rho_{LS}^3}{m_b^3}+\ldots,
\end{equation}
where $\mu_\pi^2$, $\mu_G^2$, $\rho_D^3$, and $\rho_{LS}^3$ are the non-perturbative HQE matrix elements associated with the kinetic, chromomagnetic, Darwin, and spin-orbit operators, respectively, and each coefficient $W_i^{(j)}$ admits a perturbative expansion in $\alpha_s$. In the original GGOU setup, the leading-power contribution $W_i^{(0)}$ was known up to $O(\beta_0\,\alpha_s^2)$, the so-called BLM NNLO effects, while power-suppressed terms were included up to $O(1/m_b^3)$~\cite{Aquila:2005hq,Gambino:2006wk,Gambino:2007rp}. This situation has now improved.

A major recent result is a semi-analytic determination of the NNLO corrections to the fully differential partonic decay $b\to u\ell\bar{\nu}$~\cite{Broggio:2026edk}. The singular terms entering the leading-power NNLO structure functions are known exactly from EFT factorisation~\cite{Bosch:2004th,Becher:2006qw,Asatrian:2008uk}. To determine the remaining regular terms, we adapted to the semileptonic $b\to u\ell\bar{\nu}$ decay a numerical parton-level code originally developed for the NNLO calculation of the top-quark decay $t\to bW$ with an on-shell $W$ boson~\cite{Gao:2012ja}, and generalised it to the case of a variable leptonic invariant mass $q^2$. The calculation uses the dimensionless hadronic invariant mass
\begin{equation}
\tau=\hat m_X^2=\frac{(p_u+p_X)^2}{m_b^2}
\end{equation}
as resolution variable. Below a cut $\tau^{\rm cut}$ one uses the EFT-controlled singular contribution, while above the cut one integrates numerically the contributions from $b\to u\ell\bar{\nu}\, +\, \mathrm{jet}$ at NLO and $b\to u\ell\bar{\nu}\, + 2\, \mathrm{jets}$ at LO. Infrared singularities in the real-emission contributions are handled by Catani--Seymour dipole subtraction \cite{Catani:1996vz}. In this way one obtains precise numerical results for the double-differential distributions in $(\hat m_X^2,\hat q^2)$ and for the first two lepton-energy moments.

The NNLO structure functions are then written as
\begin{equation}
W_i^{(2)}(\hat q_0,\hat q^2)=\widetilde W_i^{(2)}(\hat q_0,\hat q^2)+R_i^{(2)}(\hat q_0,\hat q^2),
\end{equation}
where $\widetilde W_i^{(2)}$ contains the analytically known singular terms, together with the known BLM contribution, and $R_i^{(2)}$ denotes the remaining regular part. The latter is expanded in a basis of integrable functions chosen so as to reproduce the expected analytic structure, and its coefficients are fitted to the numerical input from the double-differential $(\hat m_X^2,\hat q^2)$ distribution and the corresponding first two lepton-energy moments. The fit is further constrained by analytic results for the total rate, the first two total lepton-energy moments, and the $q^2$ spectrum \cite{vanRitbergen:1999gs,Pak:2008qt,Chen:2022wit}. This procedure yields semi-analytic expressions for the NNLO structure functions, suitable for phenomenological applications.

\begin{figure}[t]
    \centering
    \includegraphics[width=0.82\linewidth]{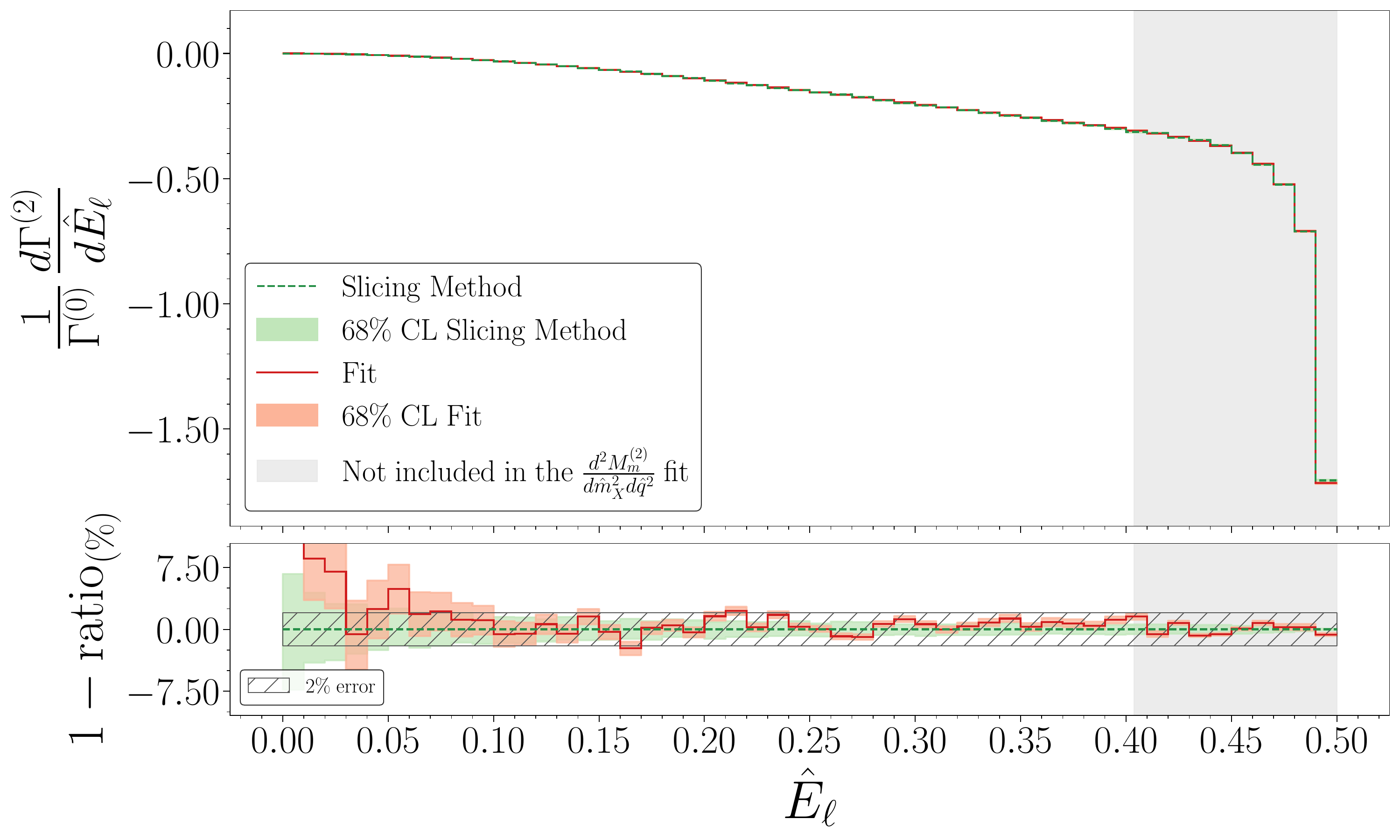}
    \caption{The figure compares the NNLO correction obtained from the fitted structure functions with the direct numerical result obtained with the $\tau$-slicing method described in the text. The agreement is excellent across the full kinematic range.}
    \label{fig:nnlo_validation}
\end{figure}

An important check is provided by observables that are not used as direct fit input. A representative example is the charged-lepton energy spectrum, shown in Fig.~\ref{fig:nnlo_validation}. The agreement between the result obtained from the fitted structure functions and the direct numerical calculation is very good across the full kinematic range. Similar agreement is found for the $\hat m_X^2$ spectrum and for the $q^2$ distribution, where an independent analytic result is also available \cite{Chen:2022wit}. The same conclusion emerges from integrated quantities: Table~\ref{tab:nnlo_moments} shows the NNLO corrections to the total rate and to the first leptonic and hadronic energy moments. The entries for $H_1^{(2)}$ and $H_2^{(2)}$ were not included in the fit and therefore provide a genuine validation. The agreement is at the per-mille level \cite{vanRitbergen:1999gs,Pak:2008qt}.

\begin{table}[t]
    \centering
    \begin{tabular}{c|cc}
         & This work & Comparison \\ \hline
         $\Gamma^{(2)} / \Gamma^{(0)}$  & $-21.2979 \pm 0.0037$ & $-21.2955$ \\
         $L_{1}^{(2)}$   & $-8.1822  \pm 0.0013$ & $-8.1819$ \\
         $L_{2}^{(2)}$   & $-3.3812  \pm 0.0006$ & $-3.3806$ \\
         $H_{1}^{(2)}$   & $-6.1133  \pm 0.0018$ & $-6.1150$ \\
         $H_{2}^{(2)}$   & $-1.7893  \pm 0.0009$ & $-1.7910$ \\
    \end{tabular}
    \caption{NNLO corrections to the total rate and to the first leptonic and hadronic energy moments, compared to the analytic results of Refs.~\cite{vanRitbergen:1999gs,Pak:2008qt}. The hadronic moments $H_{1,2}^{(2)}$ were not used as fit input.}
    \label{tab:nnlo_moments}
\end{table}

Progress has also been made on the power-suppressed terms in the HQE. In particular, the $O(\alpha_s\Lambda_{\rm QCD}^2/m_b^2)$ corrections to the kinetic and chromomagnetic operators were obtained starting from the massive-charm calculation for $B\to X_c\ell\nu$ and then taking the $m_c\to 0$ limit \cite{Alberti:2012dn,Alberti:2013kxa,Capdevila:2021vkf}. This limit is non-trivial because collinear singularities become manifest under phase-space integration and must be extracted in terms of plus distributions and derivatives of delta functions. Writing $\rho=m_c^2/m_b^2$, one encounters for instance
\begin{align}
\dfrac{1}{(\hat{u}+\rho)^3}&\to \left[\dfrac{1}{\hat{u}^3}\right]_{+}+\dfrac{1}{2}\left(\dfrac{1}{\rho^2}-1\right)\delta(\hat{u})-\left(\dfrac{1}{2\rho}-1\right)\delta^\prime(\hat{u}) \nonumber\\
&\quad -\dfrac{1}{4}(3+2\ln\rho)\,\delta^{\prime\prime}(\hat{u})\, ,
\end{align}
which makes explicit the singular terms that ultimately cancel in physical observables \cite{Capdevila:2021vkf}. 

Together with the new NNLO result at leading power, this sharpens the present theory status considerably: the leading-power structure functions are now available at NNLO in semi-analytic form, while the $1/m_b^2$ contributions proportional to $\mu_\pi^2$ and $\mu_G^2$ are available at NLO.

These perturbative developments are directly relevant to phenomenology. Because the kinematic cuts used experimentally push the decay into a region where the local OPE breaks down, the GGOU framework describes the structure functions in terms of perturbative kernels convoluted with $q^2$-dependent shape functions $F_i$ defined in the kinetic scheme \cite{Gambino:2007rp},
\begin{equation}
W_i(q_0,q^2)=m_b^{n_i}(\mu)\int dk_+\,F_i(k_+,q^2;\mu)\, W_i^{\rm pert}\!\left[q_0-\frac{k_+}{2}\left(1-\frac{q^2}{m_bM_B}\right),q^2;\mu\right]\,. 
\end{equation}
The moments of these shape functions are constrained by the OPE, and the $O(\alpha_s/m_b^2)$ corrections just discussed improve precisely those constraints \cite{Capdevila:2021vkf}. Different parametric ans\"atze have traditionally been used to estimate the uncertainty associated with the functional form of the shape functions.

The longer-term goal is to move beyond this setup towards a more flexible and less model-dependent treatment. This is the motivation for the NN$V_{ub}$ programme~\cite{Gambino:2016fdy}, where the shape functions are parametrised by neural networks, in analogy with the philosophy of modern PDF fits. In practice, one writes the shape functions as an underlying function enforcing support and endpoint behaviour, multiplied by a dense neural network depending on $(k_+,\,q^2)$. This makes it possible to incorporate theoretical constraints from the OPE, to include one-dimensional experimental spectra directly in the training, and ultimately to exploit the full NNLO perturbative kernels discussed above.

A further practical ingredient is the construction of fast surrogates for the neural network representation of the shape functions, based on spline or Chebyshev interpolation, so that the many high-dimensional integrals required during training can be evaluated through efficient linear-algebra operations. This is particularly attractive in a \texttt{JAX}-based implementation, where discrete cosine transforms provide a natural route to fast Chebyshev surrogates.

To summarise, inclusive $B\to X_u\ell\nu$ phenomenology has recently advanced on two fronts. First, the fully differential partonic decay is now under NNLO control through a semi-analytic determination of the structure functions, combining exact singular terms, known BLM corrections, and a numerical determination of the remaining regular contributions~\cite{Broggio:2026edk}. Second, the perturbative corrections to the $1/m_b^2$ terms that further constrain the shape functions are also available \cite{Capdevila:2021vkf}. Taken together, these results provide a significantly stronger theoretical basis for future inclusive determinations of $|V_{ub}|$, and they open the way to a next-generation analysis combining full NNLO kernels with neural-network descriptions of the non-perturbative shape functions.

\subsection{Inclusive semileptonic $B\to X_c\ell\nu$ decays}

{\it Gael Finauri}\\

The inclusive $b \to c \ell \bar{\nu}$ decays are treated within an OPE, made possible by the hierarchy $m_b \gg \lqcd$.
The decay rate is therefore written in the HQE as
\begin{align}
\label{eq:GammaInc}
    \Gamma(\bar{B} \to X_c \ell^- \bar{\nu}) =\;& |\Vcb|^2 G_F^2 \frac{m_b^5}{192\pi^3}\biggl( f^{\rm LP}(\rho) + f^{\text{NLP},\pi}(\rho) \frac{\mu^2_\pi}{m_b^2} + f^{\text{NLP},G}(\rho) \frac{\mu^2_G}{m_b^2}\nonumber\\
    &+ f^{\text{NNLP},D}(\rho) \frac{\rho^3_D}{m_b^3} +f^{\text{NNLP},LS}(\rho) \frac{\rho^3_{LS}}{m_b^3} + \mathcal{O}(\lqcd^4/m_b^4)  \biggr)\,,
\end{align}
where the parameters $\mu^2_\pi,\mu^2_G,\rho^3_D,\rho^3_{LS}$ are non-perturbative forward matrix elements between $\bar{B}$ meson states of respectively dimension-5 and dimension-6 operators.
On the other hand, the coefficient functions $f^{\text{P},i}(\rho)$ can be computed in perturbation theory as expansions in $\as$ and depend on the heavy quark masses ratio $\rho = m_c^2/m_b^2$.
The inclusive extraction of $|\Vcb|$ relies on the determination of the HQE non-perturbative parameters. 
The common strategy is to extract them through a global fit to normalized moments of the differential spectrum of $\bar{B} \to X_c \ell^- \bar{\nu}$ decays
\begin{equation}
    \hat{M}_{ijk} = \frac{1}{\Gamma} \int dE_\ell dq^2 dm_X^2 \, E_\ell^i (q^2)^j (m_X^2)^k\frac{d^3 \Gamma}{d E_\ell dq^2 dm_X^2}\,, 
\end{equation}
which depend on the HQE parameters but do not depend on the prefactor of the~\eqref{eq:GammaInc} with $|\Vcb|^2$.
Using the experimental information on lepton energy, $q^2$ and hadronic invariant mass moments, together with the sustained efforts in advancing the theoretical calculations of the OPE, the determination of $|\Vcb|$ reached the $1.1\%$ level of precision.

In this context, it seems natural to wonder how strongly the present data is able to constrain the presence of NP, in a model-independent way.
We do this by extending the calculation of the OPE to the full set of dimension-6 weak effective operators, which through their Wilson coefficients encode the unknown physics of scales $\Lambda_{\rm NP} \gg m_W$.
The weak effective Hamiltonian reads
\begin{align}
    \mathcal{H}_{\rm eff} =&\; \frac{4G_F}{\sqrt{2}}\Vcb \biggl( C_{V_L} [\bar{c} \gamma^\mu P_L b\; \bar{\ell} \gamma_\mu P_L \nu_\ell] + C_{V_R} [\bar{c} \gamma^\mu P_R b\; \bar{\ell} \gamma_\mu P_L \nu_\ell]\nonumber\\
   & C_{S} [\bar{c} b\; \bar{\ell}  P_L \nu_\ell] + C_{P} [\bar{c} \gamma^5 b\; \bar{\ell} P_L \nu_\ell] + C_T [\bar{c} \sigma^{\mu\nu} P_L b\; \bar{\ell} \sigma_{\mu\nu} P_L \nu_\ell]\biggr)\,,
\end{align}
where in the SM $C_{V_L} = 1 + \mathcal{O}(\alpha_{\rm em})$ and the rest of the Wilson coefficients are zero.
We have computed analytically the contribution of the dimension-6 operators to the kinematic moments up to order $\mathcal{O}(\lqcd^3/m_b^3)$ and $\mathcal{O}(\as)$.
The normalized moments are expanded for small values of the absolute values of the NP Wilson coefficients $a_i = |C_i/C_{V_L}|$ up to quadratic order $\mathcal{O}(a_i^2)$, as well as in $\as$ and inverse powers of $m_b$ up to $\mathcal{O}(\as)$ and $\mathcal{O}(\lqcd^3/m_b^3)$ respectively. 
We performed a global fit simultaneously to the HQE parameters (including the quark masses and the branching ratio) and the relevant 7 NP parameters (4 absolute values and 3 relative phases).
The result of the global fit is shown in Table~\ref{tab:fitNPas} where we refer to Ref.~\cite{Carvunis:2026fbx} for details.
From the fit parameters we extract $|\tilde{V}_{cb}| = |C_{V_L} \Vcb| =  35.5^{+3.5}_{-2.9}\cdot 10^{-3}$, where a strong anti-correlation has been observed between $|\tilde{V}_{cb}|$ and the Wilson coefficient of the tensor operator $a_T$.
We expect that the inclusion of new observables, like the ones from exclusive decays (or $q^2$ moments measured with a lower cut on $E_\ell$), could be crucial in eliminating this blind direction and further constrain the parameter space.
\begin{table}
    \centering
\begin{tabular}{ccccccc}
 \hline
&&&&&&\\[-4.3mm]
$m_b\,[\text{GeV}]~$ & $m_c\,[\text{GeV}]~$ & $\mu_\pi^2\,[\text{GeV}^2]~$ & $\mu_G^2\,[\text{GeV}^2]~$ & $\rho^3_D\,[\text{GeV}^3]~$ & $\rho^3_{LS}\,[\text{GeV}^3]~$ & $\rm{BR}_{c\ell\bar{\nu}}\,[\%]$ \\[1.2mm]
\hline
&&&&&&\\[-4.5mm]
4.567 & 1.093 & 0.449 & 0.342 & 0.196 & $-0.136$ & 10.72 \\[1.1mm]
$+0.016\atop{-0.015}$ & $+0.011 \atop -0.011$ & $+0.049 \atop -0.048$ & $+0.067 \atop -0.064$ & $+0.024 \atop -0.027$ & $+0.092 \atop -0.099$ & $+0.16\atop -0.16$\\[1.5mm]
\hline
&&&&&&\\[-4.8mm]
$a_{V_R}$ & $a_S$ & $a_P$ & $a_T$ & $\cos(\delta_{V_R})$ & $\cos(\delta_{ST})$ & $\cos (\delta_{PT})$ \\[1.mm]
\hline
&&&&&&\\[-4.5mm]
0.25 & 0.48 & 0.15 & 0.14 & 1.00 & $-0.99$ & 0.93 \\[1.1mm]
$+0.18\atop{-0.15}$ & $+0.09 \atop -0.12$ & $+0.36 \atop -0.15$ & $+0.06 \atop -0.09$ & $+0.00 \atop -0.63$ & $+0.84 \atop -0.01$ & $+0.07\atop -1.93$\\[1.5mm]
\hline
\end{tabular}
    \caption{Results for the global fit with full set of NP operators with complete $\mathcal{O}(\as)$ corrections ($\chi^2/\text{d.o.f.} = 35.47/67$). Central values on the first line, $68.3\%$ confidence level intervals on the second line.}
    \label{tab:fitNPas}
\end{table}

\subsection{Inclusive semileptonic decays on Lattice: status and prospects}

{\it Antonio Smecca and  Alessandro Barone}\\

LQCD provides accurate and systematically improvable predictions for a wide range of quantities, including decay constants and FFs relevant for heavy-meson decays. These quantities can then be combined with experimental measurements to determine the relevant CKM matrix elements.
Modern LQCD estimates have reached an impressive level of precision which has helped to constrain the uncertainty on the exclusive determination of $|V_{cb}|$~\cite{FlavourLatticeAveragingGroupFLAG:2024oxs}.

Until recently, LQCD calculations were restricted to exclusive decays. This limitation arises because standard lattice methodologies make it exceedingly difficult to handle the multi-hadron final states characteristic of inclusive processes.
A key development was presented in Refs.~\cite{Hansen:2017mnd,Hashimoto:2017wqo,Gambino:2020crt}, where it was shown how to compute inclusive decay rates from smeared spectral densities reconstructed from lattice correlation functions\footnote{Smeared spectral densities can also be used to calculate scattering amplitudes starting from Euclidean correlation functions~\cite{Barata1991,Bulava:2019kbi,Patella:2024cto}}.

For inclusive semileptonic decays, the relevant spectral density is the hadronic tensor\footnote{We use the dimensionless notation introduced in Ref.~\cite{DeSantis:2025yfm}}
\begin{align}
W_{\mu\nu}(p,\omega) = \frac{(2\pi)^4}{2m_H}\langle H(p)|J^\dagger_{\mu}(0)\delta^4(\mathbb{P}-\omega)J_{\nu}|H(p)\rangle
\end{align}
where $m_H$ is the mass of a generic heavy meson, $\mathbb{P}=(\hat{H},\hat{\boldsymbol{P}})$ is the QCD four-momentum operator and $J_{\mu}(x)=\overline{\psi}\gamma_{\mu}(1-\gamma_5)\psi$ is the V-A hadronic weak current. The heavy meson can be either the $D_{(s)}$ meson for $D_{(s)}\rightarrow Xl\nu$ decays or the $B_{(s)}$ meson for the more phenomenologically interesting $B_{(s)}\rightarrow X l\nu$ decays.

Preliminary LQCD studies~\cite{Gambino:2022dvu,Barone:2023tbl} demonstrated that the inclusive $B$-meson decay rate can be computed directly from Euclidean correlation functions measured on the lattice. While these works do not perform the continuum limit and are limited to a single lattice volume, they show good agreement with the OPE and, for heavy quark masses close to the physical bottom mass, a clear separation from the exclusive decay rate.

Simulating the bottom quark relativistically on the lattice remains particularly challenging. To avoid large discretisation artefacts, the heavy fermion mass must remain well below the ultraviolet cut-off, $i.e.$ the inverse lattice spacing. While computational advances have made fully relativistic charm quark simulations feasible, further reductions of the lattice spacing face serious obstacles such as
critical slowing down~\cite{DelDebbio:2004xh,Schaefer:2010hu} and topological freezing~\cite{Alles:1996vn}. 
Furthermore, heavy mesons suffer from a worse signal-to-noise degradation compared to light mesons~\cite{Parisi:1983ae}.
Several strategies are therefore employed: simulating the $b$-quark using an EFT~\cite{Christ:2006us,Lepage:1992tx}, use the same relativistic action as the light quarks with lighter masses of the heavy quark and then extrapolate to the physical point, or construct suitable ratios of hadronic observables and interpolate to the bottom quark sector guided by HQET to better control the systematics~\cite{ETM:2009sed}.

\subsubsection{Inclusive $D_s$ decays\label{subsec:inclusiveDs}}

Given these challenges, a more natural first step is the study of inclusive $D_{s}$-meson decays which require no additional extrapolations or the use of effective actions and are phenomenologically relevant through comparison with existing experimental results~\cite{CLEO:2009uah,BESIII:2021duu}.

Working in the rest frame of the $D_s$-meson, following an analytic integration over the lepton energy, the differential inclusive decay rate reads~\cite{Gambino:2020crt,DeSantis:2025yfm}
\begin{align}
    \frac{1}{\bar{\Gamma}}\frac{d\Gamma}{d\boldsymbol{\omega}^2} = \sum_{p=0}^2 \int_{\omega_0^{\min}}^{\infty}d\omega_0~\Theta^{(p)}(\omega_0^{\max}-\omega_0)~Z^{(p)}(\omega_0,\boldsymbol{\omega}^2),
    \label{eq:diffRate_latt}
\end{align}
where $\bar{\Gamma}$ collects the decay-specific factors such as the Fermi constant $G_F$ and the appropriate CKM matrix element, $Z^{(p)}$ are linear combinations of the hadronic tensor structure functions, and $\Theta^{(p)}(x)=x^p\theta(x)$ is an integration kernel defined in terms of the Heaviside step-function which enforces the phase space integration limits.

On the lattice, the hadronic tensor is not directly accessible. Instead, one computes the amputated correlator for discrete Euclidean times 
\begin{align}
\label{eq:Gmunu}
    G_{\mu\nu}(t) = \int d^3x~e^{i\boldsymbol{\omega}\cdot \boldsymbol{x}}\langle D_s(p)|T\{J^\dagger_{\mu}(t,\boldsymbol{x})J_{\nu}(0,\boldsymbol{0})\}|D_s(p)\rangle~e^{-tm_{D_s}\omega_0},
\end{align}
which, after taking the required linear combinations~\cite{Gambino:2020crt,DeSantis:2025yfm} becomes
\begin{align}
    \hat{Z}^{(p)}(t,\boldsymbol{\omega}^2) = \int_{0}^{\infty}d\omega_0~e^{-tm_{D_s}\omega_0}~Z^{(p)}(\omega_0,\boldsymbol{\omega}^2).
    \label{eq:latt_correlator}
\end{align}
The structure functions of Eq.~\eqref{eq:diffRate_latt} and the lattice correlators are thus related via a Laplace transform.

Performing the inverse Laplace transform on lattice correlators is an ill-posed inverse problem~\cite{Hadamard}. Moreover, since lattice simulations are confined to a finite volume, the spectral density probed by the correlators is not a continuous function, as it would be in infinite volume, but rather a \textit{distribution} of $\delta$-functions.
To handle this distribution in a meaningful way, one needs to convolute the finite-volume spectral density with a smearing kernel. Indeed, smeared spectral densities can be extracted from lattice correlators using spectral reconstruction methods~\cite{BackusGilbert,Hansen:2019idp,Bailas:2020qmv,JARRELL1996133,Burnier:2013nla,Bergamaschi:2023xzx,Kades:2019wtd,Karpie:2019eiq,Buzzicotti:2023qdv,Medrano:2025cmg}, and their infinite volume limit is well behaved for reasonable values of the smearing width. For inclusive semileptonic decays, the smearing kernel is the smeared version of the kernel $\Theta(x) \mapsto \Theta_{\sigma}(x)$.
The physical quantity is then recovered taking the infinite volume limit followed by the limit in which the smearing width is sent to zero\footnote{Phenomenologically meaningful comparisons between lattice results and experimental data can be done at non-zero smearing width~\cite{ExtendedTwistedMassCollaborationETMC:2022sta}}~\cite{Hansen:2017mnd}.

Our approach to the spectral reconstruction is based on Stone-Weierstrass theorem to represent the smearing kernel according to
\begin{align}
    \Theta_{\sigma}^{(p)}(\omega_0^{\max}-\omega_0)=\lim_{N\rightarrow\infty} \sum_{n=1}^{N} g_n^{(p)}(N)e^{-an\omega_0m_{D_s}},
\end{align}
where $t=an$, with basis-function given by the exponential and with coordinates given by the vector of the coefficients $\boldsymbol{g}^{(p)}$.
The coefficients can be found in an efficient way using the HLT method~\cite{Hansen:2019idp}.

The differential decay rate is then obtained as~\cite{Gambino:2020crt,Gambino:2022dvu}
\begin{align}
    \frac{1}{\bar{\Gamma}}\frac{d\Gamma}{d\boldsymbol{\omega}^2} = \lim_{\sigma\rightarrow0}\lim_{a\rightarrow0}\lim_{N\rightarrow \infty}g_n^{(p)}(N)\hat{Z}^{(p)}(an,\boldsymbol{\omega}^2),
    \label{eq:latt_decay_rate}
\end{align}
where $\sigma$ is the smearing width, $a$ is the lattice spacing, $N$ is the temporal extent of the lattice in lattice units and $\hat{Z}^{(p)}(an,\boldsymbol{\omega}^2)$ are the correlators computed in Eq.~\eqref{eq:latt_correlator}.

\begin{figure}
    \centering
    \includegraphics[width=0.7\linewidth]{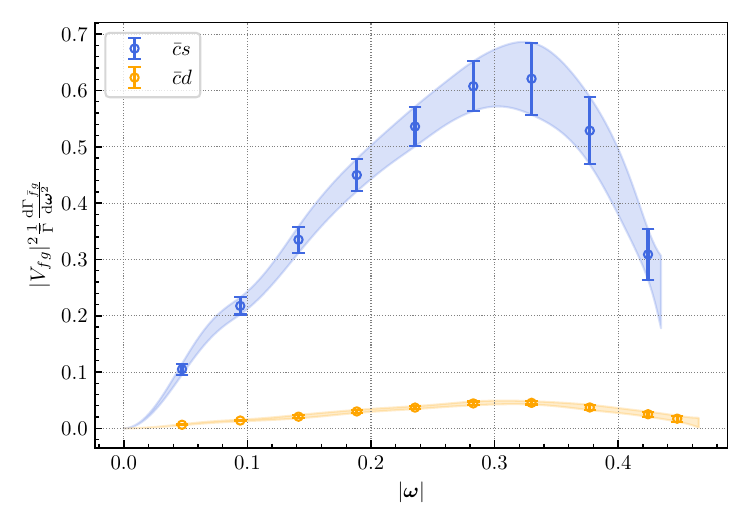}
    \includegraphics[width=0.7\linewidth]{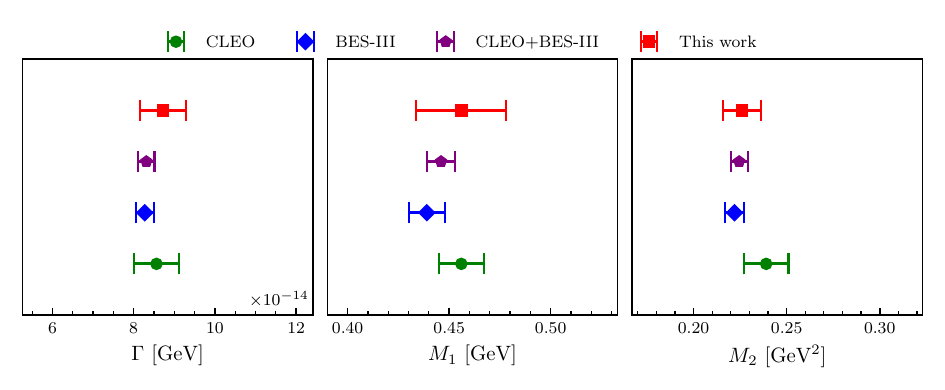}
    \caption{Top: Contributions to the differential decay rate of the dominant $\bar{c}s$ and $\bar{c}d$ channels (multiplied by the squared moduli of the corresponding CKM elements), shown together with a cubic-spline interpolation to the simulated momenta. Bottom: Comparison between the experimental results from~\cite{CLEO:2009uah,BESIII:2021duu,ParticleDataGroup:2024cfk} and our theoretical prediction (red points), for the decay rate (left-panel) and for the first (middle-panel) and second (right-panel) lepton moment}
    \label{fig:latt_Ds_decays}
\end{figure}

\begin{figure}
    \centering
    \includegraphics[width=0.7\linewidth]{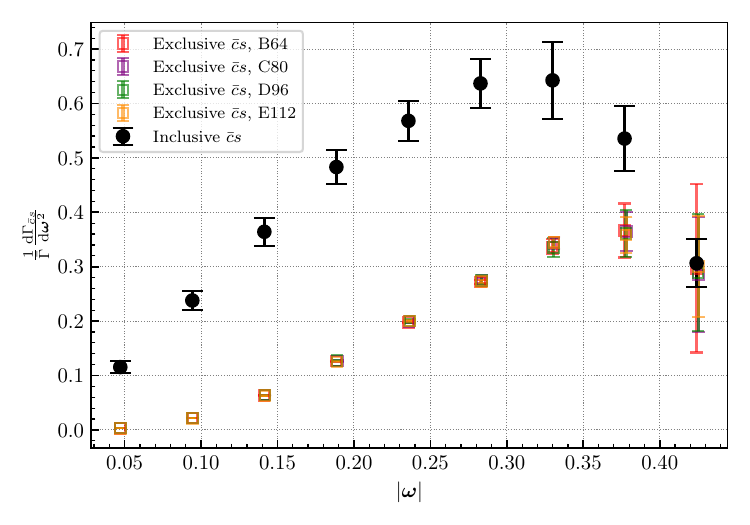}
    \caption{Comparison between the inclusive (black) and the exclusive differential decay rates for the different ensembles in the dominant $\bar{c}s$ flavour channel.}
    \label{fig:latt_excl_inclusive}
\end{figure}

Our results for the inclusive semileptonic $D_s$-meson decays are shown in Fig.~\ref{fig:latt_Ds_decays} together with the total inclusive decay rate compared with the available experimental results~\cite{CLEO:2009uah,BESIII:2021duu}, and are taken from our latest publications~\cite{DeSantis:2025qbb,DeSantis:2025yfm}.

Fig.~\ref{fig:latt_excl_inclusive} shows a clear difference between the inclusive and exclusive semileptonic decays of the $D_s$-meson across the bulk of the phase space, with good agreement within the uncertainties at the kinematical endpoint.

Using our inclusive decay rate alongside experimental results, we determine $|V_{cs}|=0.951(35)$~\cite{DeSantis:2025yfm}. While the uncertainty is currently larger than that of the current PDG value obtained from exclusive decays~\cite{ParticleDataGroup:2024cfk}, we remark that the uncertainty in our results is dominated by the statistical error and will improve in future calculations.

Our work then demonstrates that inclusive semileptonic decays of heavy mesons can now be studied on the lattice from first-principles at a phenomenologically relevant level of accuracy.
Similar conclusions have also been recently reported by a different collaboration~\cite{Kellermann:2026sgp}.
The lattice calculation of inclusive $B_{(s)}$-meson decays requires a more careful treatment of all sources of systematic errors, including those dependent on the strategy adopted to simulate the bottom quark. We expect that lattice calculations will provide reliable results in the near future, hopefully shedding light on the $|V_{cb}|$-puzzle.

\subsubsection{Inclusive $B_s$ decays}

Inclusive $B_s$-meson decays pose additional challenges for LQCD, but numerical calculations are now becoming feasible.
We present ongoing work towards a full calculation of the inclusive $B_s$ decay rate~\cite{Barone:2026awe}, using an effective-theory treatment of the bottom quark.
In particular, the calculation is performed on RBC/UKQCD gauge ensembles~\cite{RBC-UKQCD:2008mhs} with domain-wall fermions for the light, strange and charm quarks, while the bottom quark is simulated using the relativistic heavy-quark action~\cite{El-Khadra:1996wdx} in the Columbia formulation~\cite{Lin:2006ur,Christ:2006us}.
The quark masses are tuned such that mesons containing bottom, charm and
strange valence quarks have masses close to their physical values.

Here we focus on a single ensemble of size $24^3\times 64$, lattice spacing $a\approx 0.11\,\fm$ and a pion mass $M_\pi\approx 330\,\mev$ to demonstrate and study the systematics associated with the
analysis strategy. The calculation is based on Euclidean four-point functions
with two insertions of the weak currents, where one current is kept fixed,
and the other varied in Euclidean time.
After isolating the initial $B_s$ state using $B_s$ two-point functions (see Eq.~\eqref{eq:Gmunu}), 
these correlators are used to reconstruct the (differential) inclusive decay rate. We test the stability of the analysis by varying
the source--sink separation, the smearing of the $B_s$ interpolating operators,
and the position $t_{\rm ins}$ of the fixed current $J_\nu$.

In contrast to the HLT method discussed in Sec.~\ref{subsec:inclusiveDs}, here we employ
the Chebyshev-polynomial reconstruction~\cite{Barata1991,Bailas:2020qmv}. The two approaches have been shown to
give compatible results in a pilot study of inclusive $B_s\to X_c\ell\nu_\ell$
decays~\cite{Barone:2023tbl}, but they differ in the way the reconstruction is regularised. A useful
way to formulate the problem is to start from the polynomial approximation of
the smeared kernel $\Theta_\sigma^{(p)}$ defined in Sec.~\ref{subsec:inclusiveDs}.
For a given smearing width $\sigma$, we denote the differential decay rate as $\bar X_\sigma(\boldsymbol q^{2})$
(see Eqs.~\eqref{eq:diffRate_latt}-\eqref{eq:latt_correlator}). It can be written
as a linear combination of the Euclidean correlator $G_{\mu\nu}$ at different current
separations,
\begin{align}
    \bar X_\sigma(\boldsymbol{q}^2)
    \simeq
    \sum_{\mu,\nu}\sum_{k=0}^{N}
    c_{\mu\nu,k}(\boldsymbol{q};\sigma)\,
    G_{\mu\nu}(k) ,
    \label{eq:barX_poly_reconstruction}
\end{align}
where the coefficients $c_{\mu\nu,k}$ are obtained from the polynomial
approximation to the kernel. In this form, the reconstruction problem can also
be viewed as a variance-reduction problem. The direct use of
Eq.~\eqref{eq:barX_poly_reconstruction} defines a ``naive'' estimator, which is
then supplemented by a correction term,
\begin{align}
    \bar X_\sigma(\boldsymbol{q}^2)
    =
    \bar X_\sigma^{\rm naive}(\boldsymbol{q}^2)
    +
    \delta \bar X_\sigma(\boldsymbol{q}^2) .
    \label{eq:barX_variance_reduction}
\end{align}
The correction $\delta \bar X_\sigma$ can be interpreted as a noisy zero: it
does not change the target observable in the exact-data limit, but it
substantially reduces the statistical error. The HLT and Chebyshev strategies
differ in how this correction is constructed. In the HLT approach, information
on the covariance of the target observable $\bar{X}_\sigma$ is incorporated directly in the
determination of the reconstruction coefficients. 
The Chebyshev approach, instead, controls the noise by refitting the original correlator
in order to impose the mathematical bounds implied by the
polynomial basis.
 In particular, the latter provides a controlled reconstruction of
$\bar X_\sigma(\boldsymbol{q}^2)$ and a systematic way to study the dependence on both the
polynomial order $N$ and the smearing width $\sigma$.

\begin{figure}[h]
    \centering
    \includegraphics[scale=0.45]{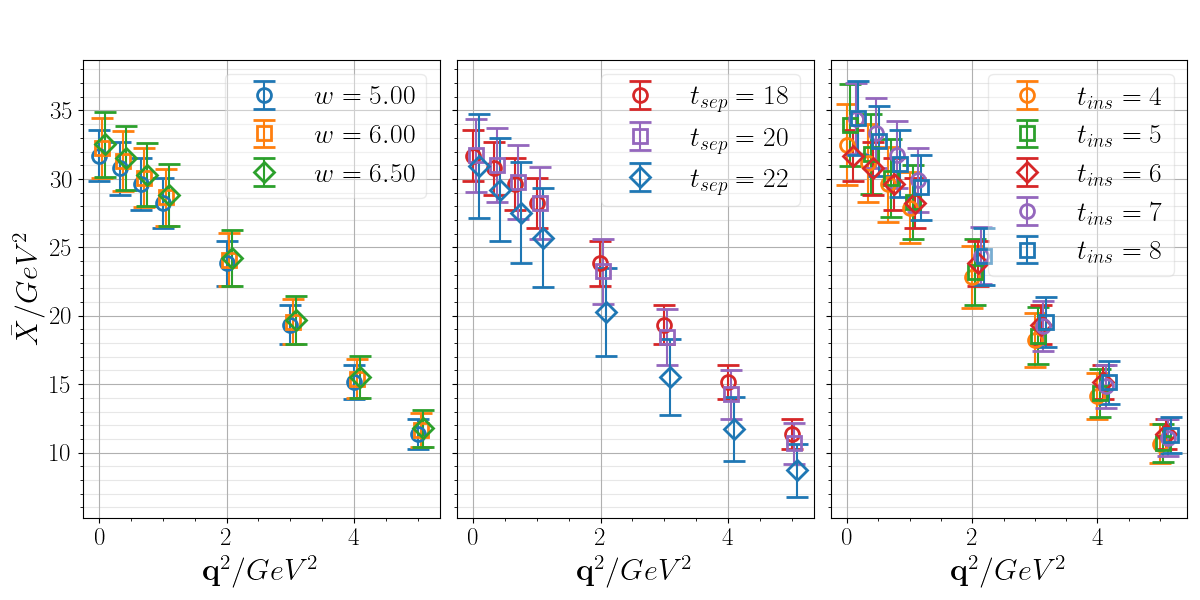}
    \caption{
$\bar X_\sigma(\boldsymbol q^{2})$ across the full kinematic range for different
choices of the four-point-function parameters. The different colours denote the
Jacobi smearing width $w$ applied to the $B_s$ interpolating operators (left),
the source--sink separation $t_{\rm sep}$ in lattice units (middle), and the
position of the fixed current insertion $t_{\rm ins}$ in lattice units (right).
For each value of $\boldsymbol q^{2}$, the results obtained from the different
choices are consistent within statistical uncertainties.
}
    \label{fig:Bs_corr_syst_individual}
\end{figure}

\begin{figure}[h]
    \centering
    \includegraphics[scale=0.45]{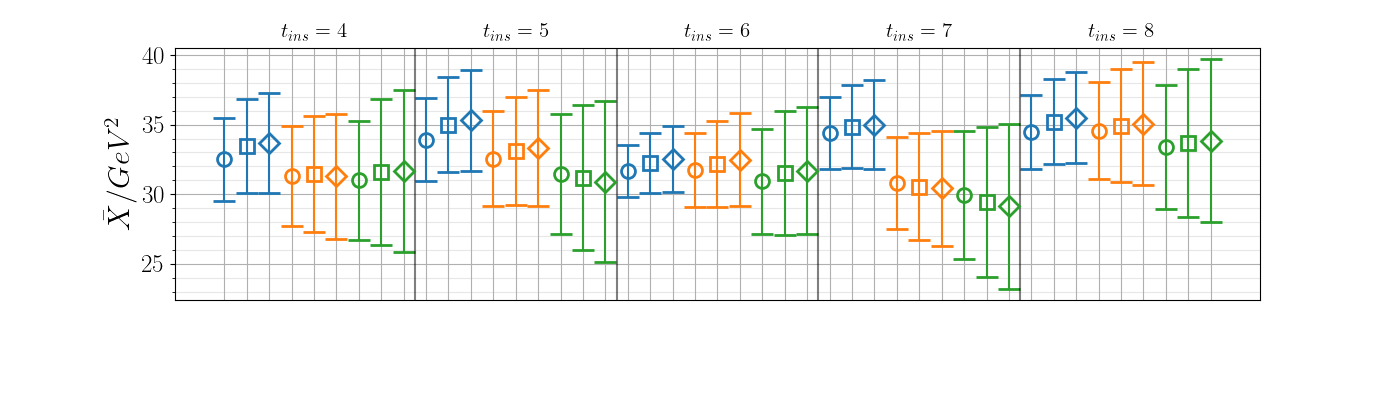}
\caption{
$\bar X_\sigma(\boldsymbol q^{2})$ at $\boldsymbol q^{2}=0.73\,{\rm GeV}^2$ for
different combinations of the four-point-function parameters. The top axis
denotes the position of the fixed current insertion $t_{\rm ins}$. The colours
label the source--sink separations $t_{\rm sep}=18,20,22$, while the circle,
square and diamond markers denote the Jacobi smearing widths
$w=5.0,6.0,6.5$, respectively. All results are compatible within uncertainties.
}
    \label{fig:Bs_corr_syst_combined}
\end{figure}

We first investigate systematic effects associated with the construction of the
four-point correlator. Since the inclusive observable is extracted
from correlation functions with $B_s$ interpolating operators at the source and
sink, possible contamination from excited $B_s$ states must be controlled. To
test this, $\bar X_\sigma(\boldsymbol{q}^{\,2})$ is recomputed across the full
kinematic range while varying the parameters entering the four-point-function
setup. We consider several Jacobi smearing widths $w$ at the $B_s$ source and
sink, different source--sink separations
$t_{\rm sep}=t_{\rm snk}-t_{\rm src}$, and several choices for the position of
the fixed current insertion $t_{\rm ins}$. As shown in
Fig.~\ref{fig:Bs_corr_syst_individual}, the results are stable within
statistical uncertainties under each of these variations separately. The same
conclusion holds when these parameters are varied simultaneously, as illustrated
in Fig.~\ref{fig:Bs_corr_syst_combined}. This indicates that excited-state
contamination from the initial $B_s$ meson is under control at the present level
of precision.

We now turn to the systematic effects intrinsic to the Chebyshev reconstruction.
The physical decay rate involves the sharp kinematic constraint imposed by the
Heaviside function, while in the lattice reconstruction this is replaced by the
smeared function $\theta_\sigma$ contained in $\Theta_\sigma$. The dependence of
$\bar X_\sigma(\boldsymbol q^{2})$ on the smearing width $\sigma$ therefore
provides a direct probe of the reconstruction uncertainty. Smaller values of
$\sigma$ correspond to a sharper approximation to the physical kernel, but they
also require higher-order terms in the Chebyshev expansion. Since the Euclidean
four-point function becomes increasingly noisy at large current separations,
these higher-order contributions cannot be determined with arbitrary precision.
We therefore treat the unresolved higher-order Chebyshev matrix elements
conservatively, using the bounds implied by the Chebyshev basis, and study the
resulting dependence of $\bar X_\sigma$ on $\sigma$ by setting $N=1/\sigma$. 

\begin{figure}[t]
    \centering
    \hbox{
    \includegraphics[scale=0.3]{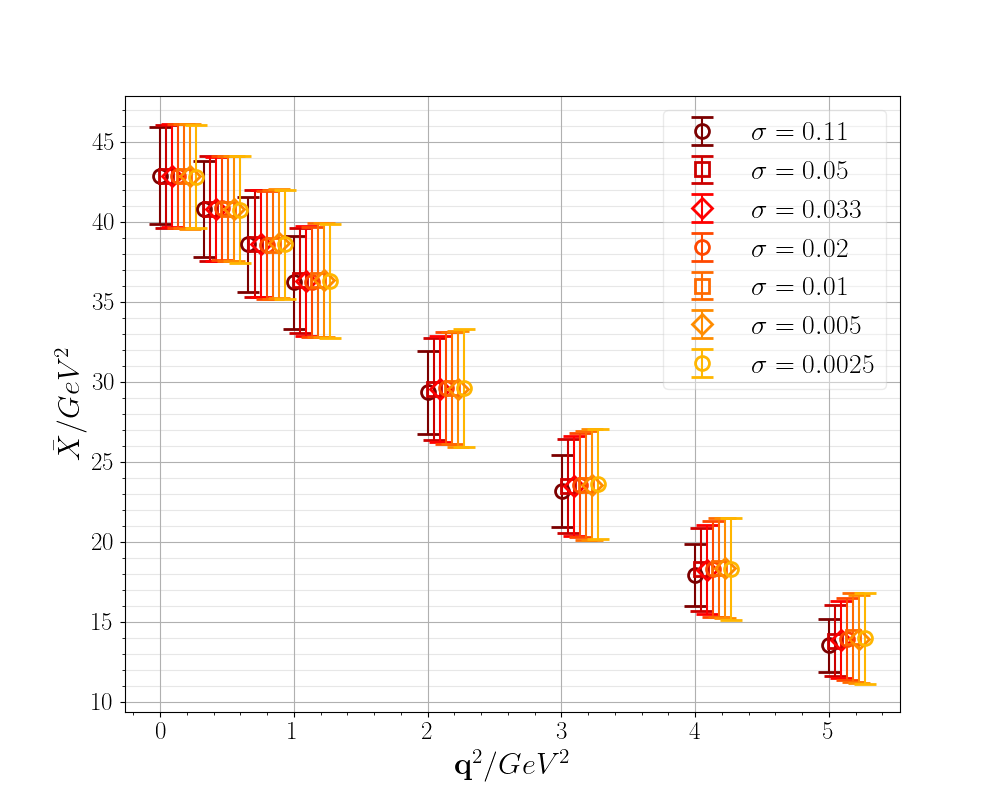}
    \includegraphics[scale=0.3]{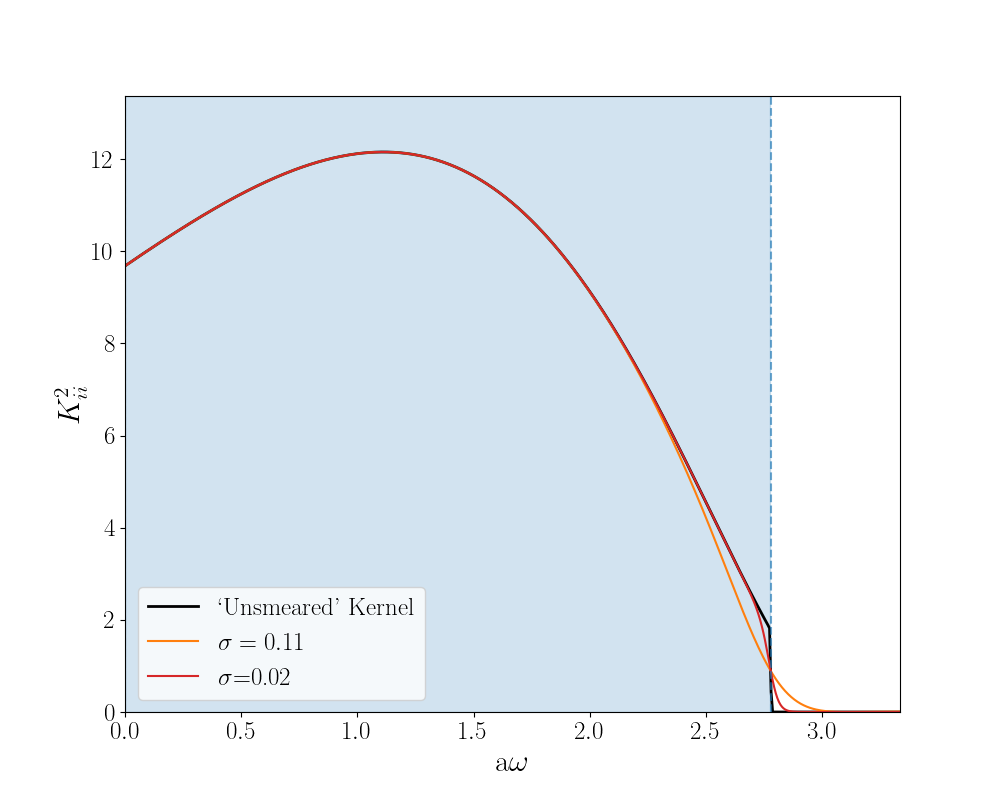}
    }
    \caption{
    Dependence of the reconstructed inclusive observable
    $\bar X_\sigma(\boldsymbol q^{2})$ on the smearing width $\sigma$. The left panel
    shows the total contribution, summed over the different current channels, for
    several values of $\sigma$. The right panel shows the corresponding smeared
    kernels for the dominant $A_i^\dagger A_i$ channel. The mild variation with
    $\sigma$ reflects the fact that the dominant kernel is not sharply peaked.
    }
    \label{fig:Bs_sigma_total}
\end{figure}

As shown in Fig.~\ref{fig:Bs_sigma_total}, the total contribution to
$\bar X_\sigma(\boldsymbol q^{2})$, summed over the different current channels, shows
only a mild dependence on $\sigma$ over the range considered. This behaviour is
related to the form of the dominant kernel entering the decay rate: in the
dominant $A_i^\dagger A_i$ channel, the kernel is not sharply peaked, so the
Chebyshev coefficients vary only mildly as the smearing width is reduced. This
indicates that, for the total rate in the present setup, the reconstruction
uncertainty associated with the $\sigma$ dependence is under control at the
current level of precision.

\begin{figure}[t]
    \centering
    \hbox{
    \includegraphics[scale=0.28]{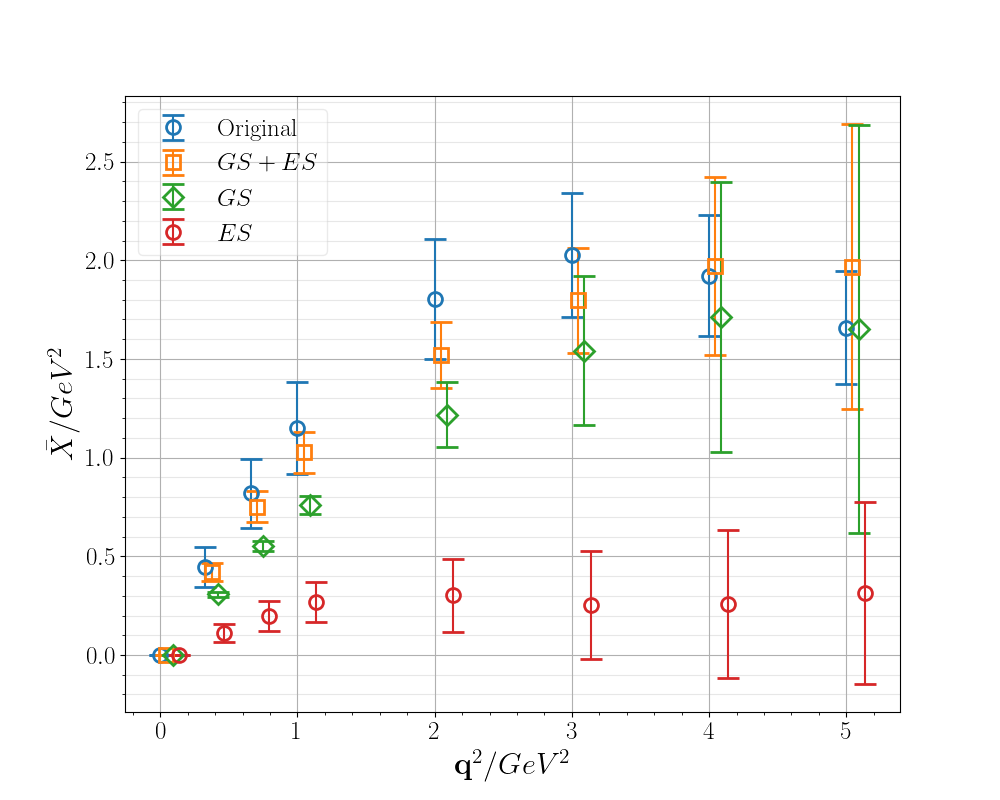}
    \includegraphics[scale=0.18]{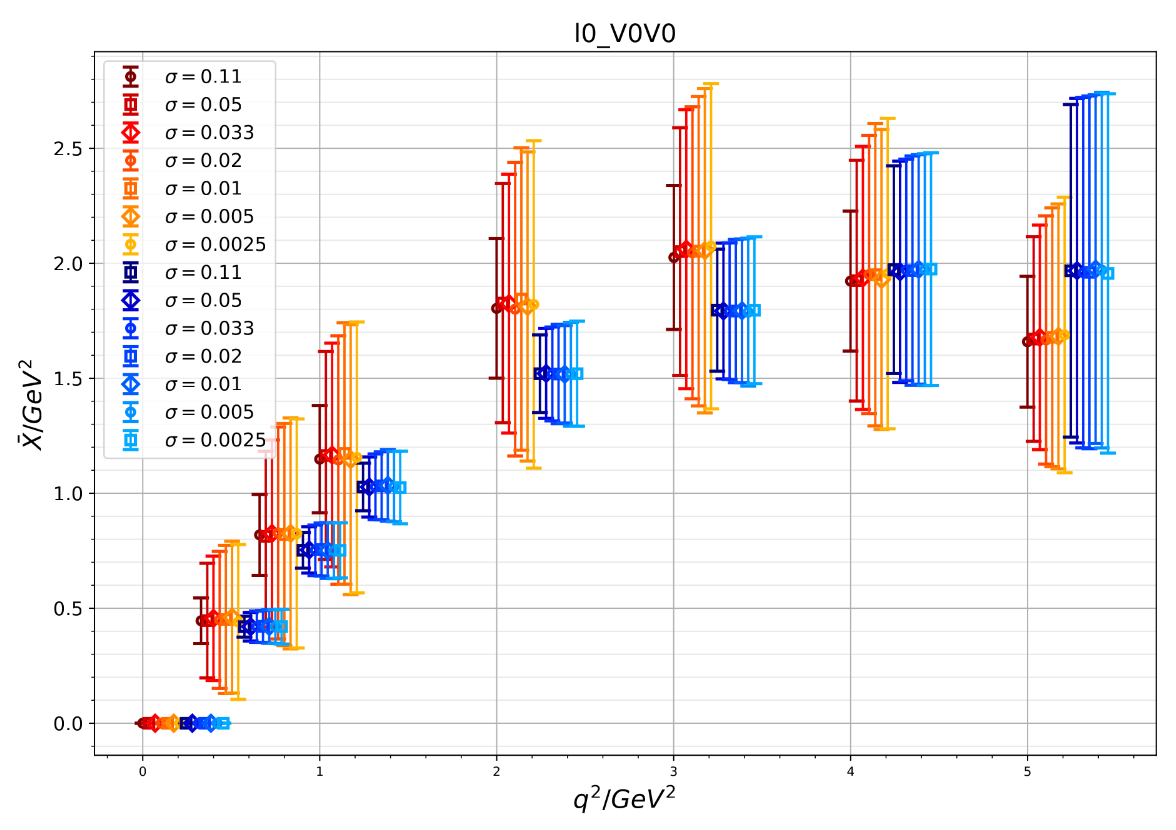}
    }
    \caption{
    Ground-state and excited-state separation in the $V_0^\dagger V_0$ channel.
    The left panel compares the direct reconstruction with the separated
    contributions $\bar X^{\rm GS}$ and $\bar X^{\rm ES}$ and their sum. The
    right panel shows the comparison between the original reconstruction (red) and the
    separated-and-recombined result (blue) in the $\sigma\to0$ limit. The reduced
    dependence on $\sigma$ demonstrates that treating the ground-state
    contribution separately can suppress the reconstruction uncertainty.
    }
    \label{fig:Bs_GS_ES_separation}
\end{figure}

Although the total contribution is rather stable, individual subdominant channels can
display a stronger dependence on $\sigma$.
This occurs in channels where the corresponding kernel is more sharply peaked,
so that the reconstruction becomes more sensitive
to the sharp-kernel limit. The $V_0^\dagger V_0$ channel provides a useful
example of this behaviour. While this channel is subdominant in the present
decay, it is an important stress test of the reconstruction strategy and
motivates a more refined treatment of the low-lying contributions to the
four-point function.
To reduce this source of uncertainty, the four-point function can be separated
into ground-state and excited-state contributions~\cite{Kellermann:2025pzt},
\begin{align}
    G_{\mu\nu}(t)
    =
    G_{\mu\nu}^{\rm GS}(t)
    +
    G_{\mu\nu}^{\rm ES}(t).
\end{align}
The ground-state contribution can be treated using standard exclusive methods,
while the Chebyshev reconstruction is applied only to the excited-state
remainder. The same decomposition then induces a corresponding separation of
the reconstructed observable,
\begin{align}
    \bar X_\sigma(\boldsymbol{q}^{2})
    =
    \bar X_\sigma^{\rm GS}(\boldsymbol{q}^{2})
    +
    \bar X_\sigma^{\rm ES}(\boldsymbol{q}^{2}).
\end{align}
This is advantageous whenever the ground state gives a sizeable contribution,
because the $\sigma$-dependent inclusive reconstruction then affects only the
excited-state part.

This strategy was tested in the $V_0^\dagger V_0$ channel, where the ground
state is identified with the $D_s$ contribution. As shown in
Fig.~\ref{fig:Bs_GS_ES_separation}, the separated-and-recombined result (blue points, right panel) exhibits
a reduced dependence on $\sigma$ compared with the direct inclusive
reconstruction. For most values of $\boldsymbol q^{2}$ this also leads to a
smaller uncertainty in the $\sigma\to0$ limit. The exception is the largest
value of $\boldsymbol q^{2}$, where the extraction of the ground-state
contribution is less stable. This suggests that the separation of exclusive and
excited-state contributions can provide a practical way to suppress
reconstruction systematics, and will be an important ingredient in future
calculations aiming at a full error budget.

This study provides the basis for a full LQCD calculation of inclusive
semileptonic $B_s$ decay rates. The stability under variations of the
four-point-function setup indicates that the initial-state contamination is
under control at the current level of precision, while the study of the
$\sigma$ dependence identifies the channels most sensitive to the inclusive
reconstruction. Together with the separation of ground- and excited-state
contributions, these results define the analysis strategy for future
calculations at multiple lattice spacings and volumes.
\clearpage

%%%%%%%%%%%%%%%%%%%%%%%%%%%%%%%%%%%%%%%%%%%%%%%%%%%%%%%%%%%%%%%%
% b->s l l 
% Experimental and theoretical status at b-factories and LHC
% form factors, hadronic corrections
% lattice and continuum approaches
%%%%%%%%%%%%%%%%%%%%%%%%%%%%%%%%%%%%%%%%%%%%%%%%%%%%%%%%%%%%%%%%%

\section{Experimental status of neutral-current semileptonic decays of b-hadrons}
\label{sec:nc_exp}

Rare $b \to s\ell^+\ell^-$ decays are forbidden at tree level in the SM and proceed only through loop-level electroweak penguin and box diagrams, making them highly sensitive to potential new physics contributions at scales above the electroweak scale. There is a long-standing list of tensions between measurements and SM predictions that have motivated extensive theoretical and experimental effort. The interpretation of the observed discrepancies remains challenging due to the non-trivial role of non-local hadronic contributions, in particular those arising from intermediate charm loops. This section summarises the experimental status of these decays at Belle~II and \lhcb, including searches for modes with missing energy, and the extensive programme of branching-fraction and angular-observable measurements at the LHC.

\subsection{Status of $b \to s \ell \ell$ decays with missing energy at \belle and \belletwo}

{\it Elisa Manoni}\\

The \belletwo detector at the SuperKEKB accelerator, together with its predecessor \belle at KEKB, provides an ideal environment to search for $b \to s \ell \ell$ decays with missing energy. 

As described in Sec.~\ref{sec:experiments}, the production of $B\bar{B}$ pairs at the \FourS threshold provides a highly constrained kinematic environment: the decay products of the two $B$ mesons must be consistent with the known initial-state energy and momentum. Combined with the nearly $4\pi$ angular coverage of the detectors, this enables precision measurements of decays involving missing energy. The latter can be inferred from the difference between the initial energy of the collision and the visible energy recorded in the detector.
These analyses are challenging because the searched-for modes have small (or even zero) branching fractions in the SM and suffer from large backgrounds. In addition, except for the $b \to s \tau \ell$ case, the presence of multiple undetected neutrinos prevents the reconstruction of a narrow signal peak.

The analyses discussed in the following are based on $711~\invfb$ and $363~\invfb$ of data collected by the \belle and \belletwo detectors, respectively.
Different tagging techniques, briefly described in Sec.~\ref{sec:experiments}, are employed to reconstruct the two $\B$ mesons in the event. The signal side is reconstructed either exclusively or semi-exclusively, requiring the presence of a strange meson. In searches involving $\tau$ leptons, only electronic and muonic final states are always considered, while in some analyses $\tau$ decays to charged pions and $\rho$ mesons are also included.
Charged and neutral particles not associated with either the signal or the tag $\B$ ($B_{\rm tag}$) are referred to as "extra".
In exclusive tagging analyses, events containing extra tracks or extraneutral mesons, such as \piz\ and \KS, are usually rejected. On the other hand, extra photons are allowed and their total energy ($E_{extra}$) is one of the most discriminating variables. It is used either as an input to multivariate algorithms or as a fit variable. Mismodeling of the $E_{extra}$ distribution in the simulation is addressed by studying control samples, while any remaining discrepancies are accounted for as systematic uncertainties.
Other discriminating and reasonably well-modelled variables are used for background suppression, including signal-side kinematic quantities, the missing four-momentum, event-shape variables for continuum suppression, and variables designed to reject specific backgrounds, such as $B \to D^{(*)}\ell\nu$ with $D^{(*)}\to K^{(*)}X$.
The dominant systematic uncertainties usually arise from the uncertainty in the background normalization and the limited size of the simulated samples.

A variety of $b \to s \taup \taum$ modes are investigated at \belle and \belletwo. In the SM, branching fractions of the order of $10^{-7}$ are predicted for the ditau modes, while lepton-flavor-violating (LFV) transitions, which can occur through neutrino mixing, are highly suppressed. Several NP scenarios can enhance their branching fractions to levels within experimental reach.

\begin{figure}[!t!]
\centering
\includegraphics[width=0.35\textwidth,angle=0]{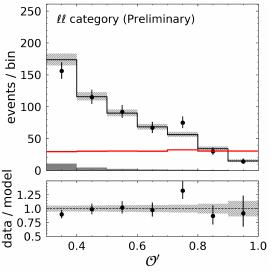}
\includegraphics[width=0.35\textwidth,angle=0]{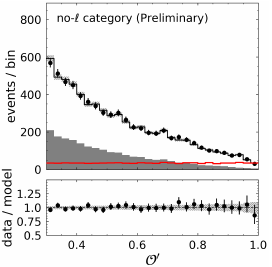}
\includegraphics[width=0.23\textwidth,angle=0]{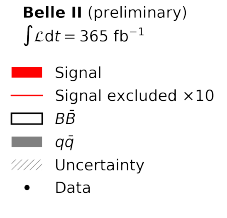}
\hfill
\caption{$B \to \KS \taup\taum$ search with \belle and \belletwo data: distributions of the transformed BDT output with the fit projections. The red filled histogram (not visible) represents signal events at the branching fraction of $1.2 \times 10^{-4}$. The simulated background contributions from $e^+e^- \to q \bar q$ (grey filled histogram) and $e^+e^- \to \BBb$
(open histogram) are stacked and the sum of components is compared with data. The open red histogram (overlaid) illustrates the signal at a level ten times the exclusion limit on the branching fraction. The panels below each distribution show the ratio between the data and model. The uncertainties on the histograms represent the total model uncertainty as determined by the fit. Left (right) panel: both $\tau$'s reconstructed in leptonic (hadronic) modes for \belletwo data. 
}
\label{fig:b2_kstt}
\end{figure}

In the search for $B^0 \to \KS \taup\taum$ using the hadronic tagging technique~\cite{Belle-II:2026bbr}, the $\tau$ leptons are reconstructed in both hadronic and leptonic decay modes. Thanks to the clean $\KS$ reconstruction, this channel is expected to have higher purity than the corresponding $\Kstarz$ mode~\cite{Belle-II:2025lwo}. On the other hand, the presence of a neutral $B$ meson on the tag side results in a lower tagging efficiency than in the $B^+ \to K^+ \taup\taum$ channel~\cite{Belle-II:2026ism}.
The different $\tau$ decay modes are grouped into five orthogonal categories to improve background suppression.
The most discriminating variables are combined in a BDT, trained separately for each of the five categories. The signal region is defined using a transformed BDT output, where the transformation is chosen to produce an almost uniform signal distribution. 
For the categories in which both $\tau$ leptons are reconstructed in leptonic decay modes, the dominant background arises from semileptonic $B$ decays, while hadronic $B\to D^{(*)}X$ decays constitute the main background in the remaining categories. For the backgrounds, the BDT output distribution and the normalization are validated using several control samples, including off-resonance data, the $B_{\rm tag}$ mass distribution, and the $\KS$ mass sidebands. The signal efficiency is validated using events in which the invariant mass of the signal $\KS$ candidate and an additional track in the event is consistent with the $D$-meson mass.
The dominant systematic uncertainties arise from the limited size of the simulated samples and from the knowledge of the branching fractions of the leading background $\epem\to\BBb$ decay modes. Approximately 70 dominant background modes are identified. For the measured modes, the branching fractions used in the simulation are corrected according to the latest measurements~\cite{ParticleDataGroup:2024cfk}, and the corresponding uncertainties are propagated to the systematic uncertainty. For unmeasured modes, a conservative uncertainty of 100\% with respect to the prediction is assigned.
The signal yield is extracted from a three-component (signal, $\BBb$, and $q\bar q$ background) binned maximum-likelihood fit to the transformed BDT output (see Fig.~\ref{fig:b2_kstt} as an example). A simultaneous fit is performed to the \belle and \belletwo data samples, and to the five signal categories, all of which provide comparable sensitivity.
No evidence for signal is observed, and the following upper limit at the 90\% confidence level is obtained: $\mathcal {B}(B^0 \to \KS \taup\taum)<0.84 \times 10^{-3}$. This result represents the first experimental limit on this decay mode.

For the $B^+ \to K^+ \taup\taum$ decay, a combined search on the \belle and \belletwo data samples is performed, with both $\tau$ decays reconstructed in leptonic modes~\cite{Belle-II:2026ism}. A selection on the invariant mass of one of the light leptons and the signal kaons is used to suppress most of the background.
The signal yield is estimated from event-counting in the missing-mass squared, the momentum of the daughter of the $\tau$ leptons, and $E_{extra}$.
No evidence for signal is observed, and the following upper limit at the 90\% confidence level is obtained: $\mathcal {B}(B^+ \to K^+ \taup\taum)<0.87 \times 10^{-3}$, 2.6 times better than the current world best upper limit. 

Searches for the $B^0 \to \Kstarz \taup\taum$~\cite{Belle-II:2025lwo} decay and the LFV decays $B^0 \to \Kstarz \taup\ellm$~\cite{Belle:2025sgv} and $B^0 \to \KS \taup\ellm$~\cite{Belle:2024kur} have been performed using the \belletwo and combined \belle+\belletwo data samples, respectively. No evidence for signal is observed in any of the channels, and upper limits at the 90\% confidence level are set: $\mathcal {B}(B^0 \to \Kstarz \taup\taum)<1.8 \times 10^{-3}$ and at the level of $10^{-5}$ for the LFV modes. For the $B^0 \to \KS \taup\ellm$ channel, these results constitute the only limits currently available.

Modes involving $b \to s \nu \bar \nu$ transitions are searched for with a variety of experimental techniques.
In the SM, the exclusive $B\to K^{(*)}\nu\bar{\nu}$ decays are predicted to have branching fractions in the range $(2$--$10)\times10^{-6}$, with theoretical uncertainties below or around 15\%, depending on the decay mode~\cite{Becirevic:2023aov}. For the fully inclusive decay, the prediction is $\mathcal{B}(B\to X_s\nu\bar{\nu})=(3.48 \pm 0.11)\times10^{-5}$~\cite{Fael:2025xmi}. NP can modify these processes in different ways, depending on the specific final state.

First evidence for the $\Bp \to K^{+}\nu\bar{\nu}$ modes, with $3.5\sigma$ significance with respect to the background-only hypothesis, has been measured by using \belletwo data with inclusive and hadronic tagged methods, the first being more precise, the second cleaner~\cite{Belle-II:2023esi}. The resulting branching fraction is $\mathcal {B}(\Bp \to K^+ \nu\bar{\nu}) = (2.3 \pm 0.5(\mbox{stat})^{+0.5}_{-0.4}(\mbox{syst})) \time 10^{-5}$, which corresponds to a 
$2.7\sigma$ departure from the SM prediction. Future analyses, exploiting additional exclusive decay channels, different tagging techniques, and larger data samples, are expected to confirm this evidence and improve the measurement precision.

A search for the inclusive $B \to X_s \nu\bar{\nu}$ decay on \belletwo data~\cite{Belle-II:2025bho} has been performed using a sum-of-exclusive approach, in which the $X_s$ system is reconstructed in the final states $K n\pi$ ($0 \leq n \leq 4$), with at most two $\pi^0$ mesons, and $3K n\pi$ ($0 \leq n \leq 1$), with at most one $\pi^0$ and one $\KS$ meson, where $K$ denotes either a $K^+$ or a $\KS$ particle. These channels cover approximately 93\% of the total $B \to X_s \nu\bar{\nu}$ branching fraction. The other $\B$ is reconstructed in hadronic final states. Background suppression is achieved using a BDT, and the signal yield is extracted from a two-dimensional binned maximum-likelihood fit to the BDT output and the $X_s$ invariant mass.
No evidence for signal is observed, and an upper limit at the 90\% confidence level is set: $\mathcal {B}(B \to X_s \nu\bar \nu)<3.2 \times 10^{-4}$, which is the most stringent limit reported to date. and is compatible with the hadronic tagged $\Bp \to K^{+}\nu\bar{\nu}$ search.

In summary, $b \to s$ decays with missing energy are excellent NP probes, and the \belle and \belletwo experiments are well suited to study them. The analyses are experimentally challenging, with background suppression and validation representing the main difficulties.
Although the measurements are still statistically limited, systematic uncertainties are becoming increasingly important for all analyses. The $b \to s\nu\bar{\nu}$ channels are unique to \belletwo. The first evidence for the $B^+ \to K^+ \nu\bar{\nu}$ decay has been reported, and future studies will aim to confirm this observation and improve the measurement precision. The first semi-inclusive measurement at the $B$ factories has also been performed, and a similar strategy can be applied to the $B \to X_s\ell\ell$ channel.
Finally, for the $b \to s\tau\tau$ transitions, the first search for the $B \to \KS\tau\tau$ decay was presented.

\subsection{Status of $b \to s \ell \ell$ decays at \lhcb}

{\it Lorenzo Paolucci}\\

The \bsll transition has been investigated extensively with data collected by \lhcb, covering a plethora of observables and decay channels.
Measurements of the properties of \decay{\bquark}{\squark\mup\mun} processes have shown a consistent pattern of discrepancies with their SM predictions.
The observed differential branching fractions of the decays \decay{\Bd}{\Kz\mup\mun}, \decay{\Bu}{\Kp\mup\mun}, \decay{\Bu}{\Kstarp\mup\mun}, \BdToKstmm, and \decay{\Bs}{\phi\mup\mun}~\cite{LHCb-PAPER-2014-006,LHCb-PAPER-2025-041,LHCb-PAPER-2021-014} are in tension with the SM in the \qsq region below the charmonium resonances. The largest disagreement, observed in \decay{\Bs}{\phi\mup\mun} decays, corresponds a local significance above three standard deviations.
A similar discrepancy has also been observed in the differential branching fraction of the decay \decay{\Lb}{\Lz\mup\mun}~\cite{LHCb-PAPER-2015-009}; this tension has been strengthened by the updated measurement of the branching fraction of the \decay{\Lb}{\jpsi\Lz}~\cite{LHCb-PAPER-2025-035} decay, which serves as a normalisation channel to the non-resonant mode.

\begin{figure}[!t!]
\centering
\includegraphics[width=0.52\textwidth,angle=0]{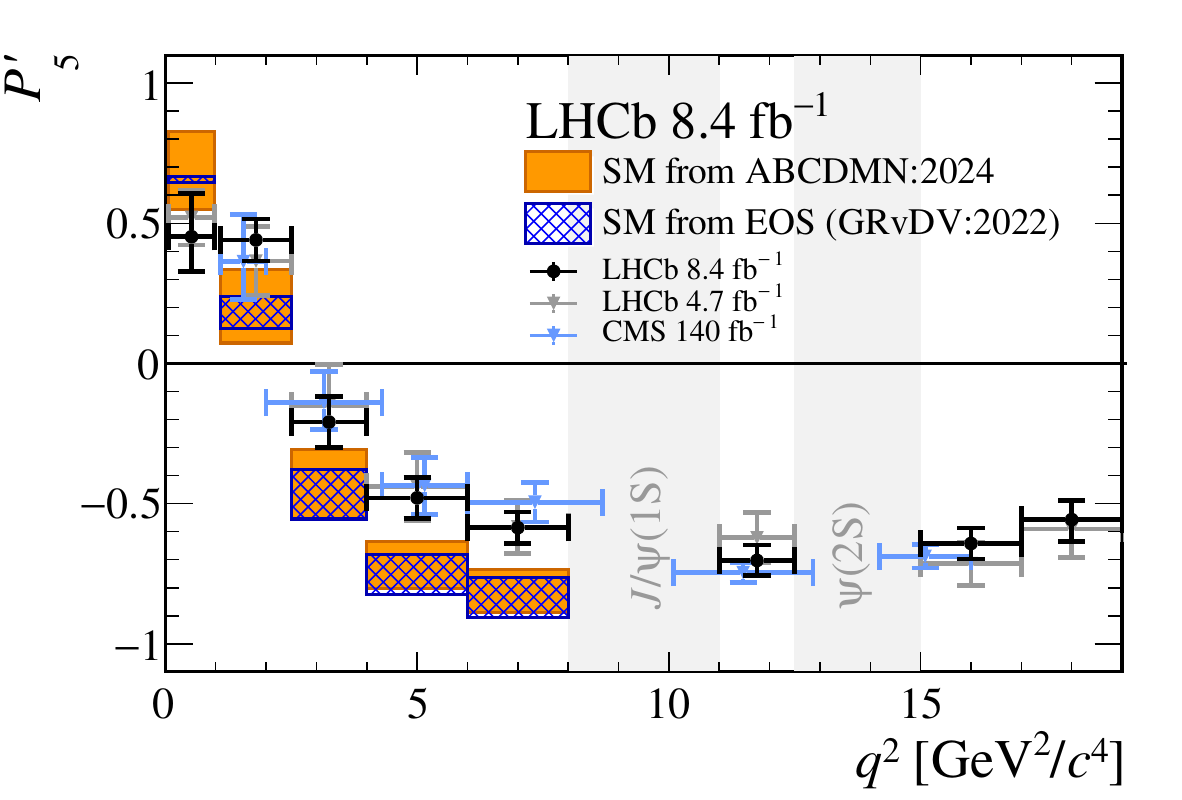}
\includegraphics[width=0.46\textwidth,angle=0]{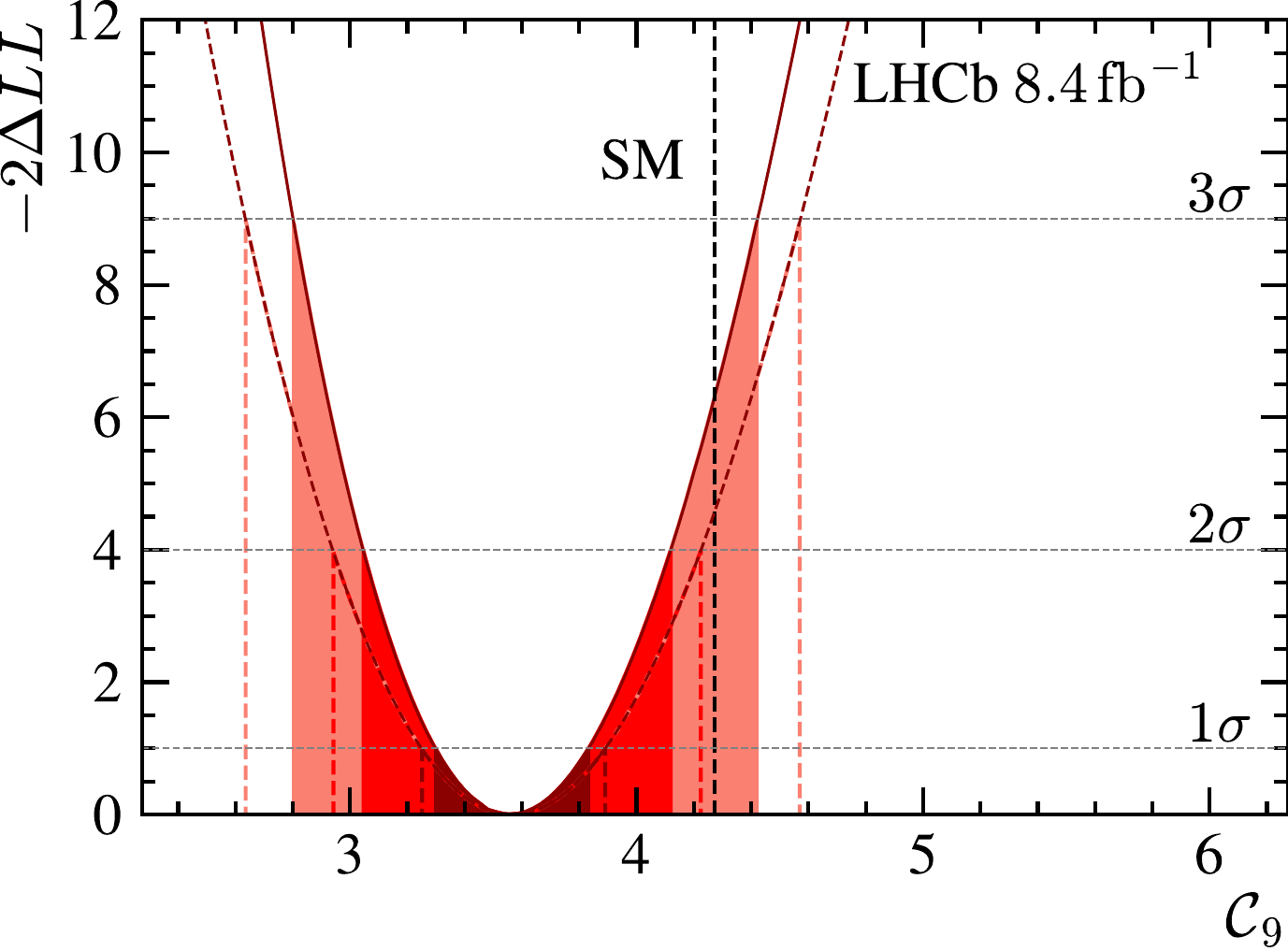}
\hfill
\caption{Left: Measured \CP-averaged observable $P^{\prime}_5$ for the decay \BdToKstmm, in bins of \qsq (black markers), compared to the SM predictions (blue- and orange-filled boxes)~\cite{LHCb-PAPER-2025-041}. Right: Likelihood profile of the Wilson coefficient $\mathcal{C}_9$ from an amplitude fit to \BdToKstmm data. The solid line and shaded areas consider only statistical uncertainties, whilst the dashed curve and dashed vertical lines indicate the same confidence intervals when including systematic effects~\cite{LHCb-PAPER-2024-011}.}
\label{fig:LHCbbsll}
\end{figure}

The latest angular analysis of the \BdToKstmm decay~\cite{LHCb-PAPER-2025-041}, using the full Run 1 and 2 dataset, reported the most precise measurement yet of \CP-averaged angular observables for this channel.
An improved description of the differential decay rate allows for the first time to extract from the angular fit the \CP asymmetries, S-wave observables, as well as to include the effect of massive leptons.
The six-dimensional fit performed in the candidate \Bd mass, $m(\Kp\pim)$, and helicity angles, results in local tensions with the SM predictions at the level of $2-3$ standard deviations.
In particular, it reinforces the long-standing tension in the region $4.0<\qsq<6.0\gevgevcccc$ of the observable $P^{\prime}_5$, as shown in Fig.~\ref{fig:LHCbbsll} (left).
When interpreting the measured observables with Wilson coefficients, using the EOS package~\cite{EOSAuthors:2021xpv}, the best fit corresponds to a shift in $\Real(\mathcal{C}_9)$ from the SM point of $-0.92\pm0.23$.

Given the needed theoretical inputs to interpret the measured branching fractions and angular observables, two amplitude analyses of the \BdToKstmm decay have been performed to help disentangle in data the impact of non-local effects from potential New Physics.
The first analysis~\cite{LHCb-PAPER-2023-032,LHCb-PAPER-2023-033} models the \qsq dependence of the non-local part of the amplitude with a truncated expansion, using theory constraints at $q^2<0$ as well as experimental constraints from previous amplitude analyses of the resonant decays \decay{\Bd}{\jpsi\Kstarz} and \decay{\Bd}{\psitwos\Kstarz}.
The second analysis~\cite{LHCb-PAPER-2024-011} performs a full amplitude fit to the whole \qsq spectrum, measuring the relative amplitudes and phases of one- and two-particle intermediate states, as well as effects due to \decay{\taup\taum}{\mup\mun} rescattering.
These analyses report shifts in $\Real(\mathcal{C}_9)$ of $-0.93^{+0.53}_{-0.57}$ and $-0.71\pm0.33$, respectively.
The likelihood profile of the measured value of $\mathcal{C}_9$ in the full amplitude analysis is shown in Fig.~\ref{fig:LHCbbsll} (right).
The observed shifts are consistent with the binned angular analysis of the same channel. 
However, their uncertainties are increased due to the complexity of the amplitude models used, leading to a reduced tension of about two standard deviations.

While most LHCb measurements focus on muonic final states, which offer the highest experimental sensitivity, decays involving electrons and tau leptons have also been explored.
There are unique challenges in measurements of the \decay{\bquark}{\squark\ep\en} transition: firstly, electrons reconstructed in LHCb suffer from high misidentification rates, leading to significant background pollution from hadronic \bquark-hadron decays. The trigger efficiency in data collected between 2011 and 2018 was significantly lower for electrons than for muons. Furthermore, electrons radiate a significant amount of energy by material interactions, which lowers their reconstruction efficiency, as lower momentum particles are more likely be swept out of the detector's acceptance.
The momentum resolution is improved thanks a dedicated bremsstrahlung recovering procedure.
Branching fractions of the decays \decay{\Bu}{\kaon^{(*)+}\epem}, \decay{\Bd}{\KS\epem}, \decay{\Bd}{\Kstarz\epem}, \decay{\Bs}{\phi\epem}, \decay{\Bu}{\Kp\pip\pim\epem}, \decay{\Lb}{\proton\Km\epem} have been measured relative to their muonic counterparts~\cite{LHCb-PAPER-2019-040,LHCb-PAPER-2021-038,LHCb-PAPER-2022-045,LHCb-PAPER-2022-046,LHCb-PAPER-2024-032,LHCb-PAPER-2024-046,LHCb-PAPER-2024-056}, in bins of \qsq.
The angular observables of the decays \decay{\Bd}{\Kstarz\epem} and \decay{\Bs}{\phi\epem} have also been measured~\cite{LHCb-PAPER-2024-022,LHCb-PAPER-2025-006} and compared to the corresponding measurements of the muonic modes.
All measurements have been found to be consistent with the assumption of lepton flavour universality.

The first search at LHCb for the decays \decay{\Bd}{\Kstarz\taup\taum} and \decay{\Bs}{\phi\taup\taum} has been performed using the full Run 2 dataset~\cite{LHCb-PAPER-2025-048}.
The tau lepton is reconstructed using its muonic decay \decay{\taup}{\mup\neutb\neum}.
Because of the unreconstructed neutrinos, the signal processes are not searched by means of a likelihood fit to the reconstructed \B mass spectrum; instead, the branching fractions are obtained by fitting the score of BDT classifiers trained to discriminate between signal and polluting backgrounds.
World-leading limits have been set for the branching fractions of the \Bd and \Bs decay of $2.8\,(2.5)\times 10^{-4}$ and $4.7\,(4.1)\times 10^{-4}$ at $95\,(90)\%$ confidence level, respectively.

\begin{figure}[!t!]
\centering
\includegraphics[width=0.93\textwidth,angle=0]{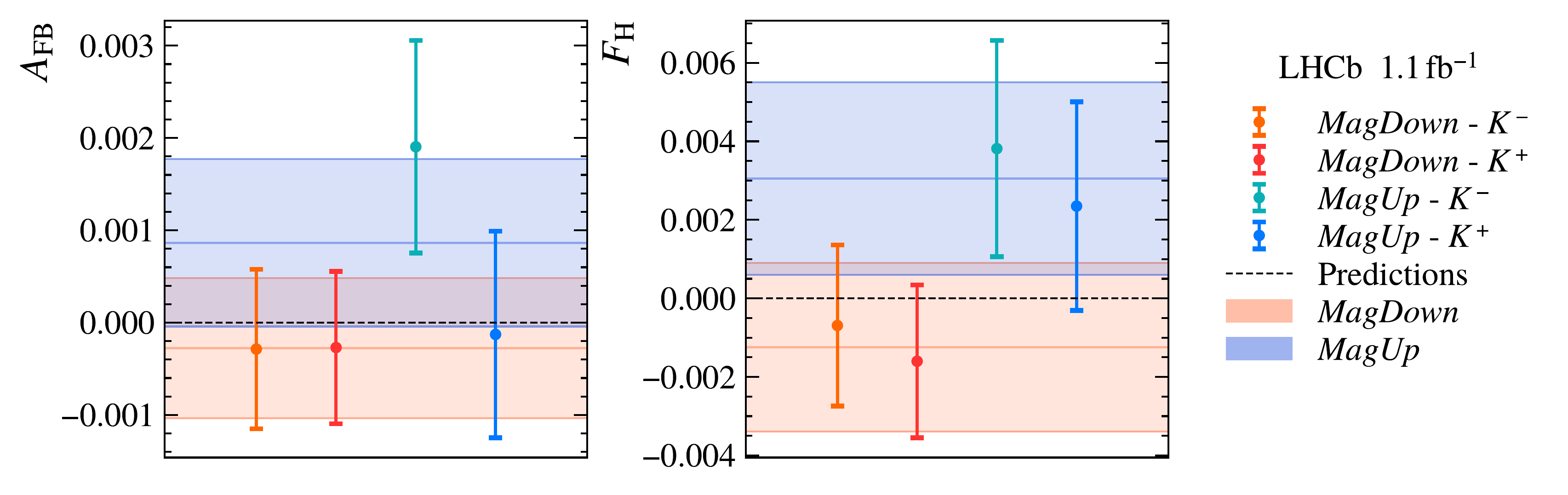}
\hfill
\caption{Measured values of the lepton forward-backward asymmetry $A_{\mathrm{FB}}$ and flatness parameter $F_{\mathrm{H}}$ for \decay{\Bu}{\jpsi \Kp} decays, using LHCb data collected in 2024, obtained by splitting the dataset by kaon charge and magnet polarity (coloured markers).
The horizontal bands indicate the values obtained when only splitting by magnet polarity, and the dashed black line indicates the SM expectation~\cite{LHCb-PAPER-2025-040}.}
\label{fig:LHCbRun3bsll}
\end{figure}

At the current stage, these measurements offer an incomplete picture, due to the limited size of the available datasets.
While some tensions have been present across different decays and observables for many years, it is not yet clear whether these are due to New Physics or non-local QCD effects.
The upgraded LHCb detector, which finished the data-taking campaign in June 2026, will bring significantly larger yields, further reducing uncertainties on key observables and improving our overall understanding.
The robust performance of the upgraded detector has been proven in an angular analysis of the decay \decay{\Bu}{\jpsi \Kp}~\cite{LHCb-PAPER-2025-040}, as shown in Fig.~\ref{fig:LHCbRun3bsll}, paving the way for the future Run 3 analyses of \bsll processes.

\clearpage

\section{Theoretical status of neutral-current semileptonic decays of b-hadrons}
\label{sec:nc_theo}

The interpretation of the experimental results discussed in the previous section relies crucially on theoretical control of the relevant hadronic matrix elements. For the exclusive $B\to K^{(*)}\ell^+\ell^-$ and $B_s\to\phi\ell^+\ell^-$ modes, the local FFs can be computed with good precision using LQCD and LCSR, but the dominant theoretical uncertainty resides in the non-local, long-distance contributions generated by charm-loop effects, whose first-principle treatment remains a major open challenge. LQCD has also begun to address the non-local contributions directly, complementing the local FF calculations with first attempts at a non-perturbative determination of the non-local amplitudes. This section reviews the current theoretical understanding of both local and non-local hadronic contributions to $b\to s\ell^+\ell^-$ transitions, and the prospects for reducing the associated uncertainties from the lattice.

\subsection{Theoretical uncertainties: FFs and hadronic contributions}

{\it Arianna Tinari}\\

Being suppressed in the SM, rare $b \to s \ell^+\ell^-$ decays are sensitive indirect probes of new physics. At low energies, they are described  by an EFT obtained by integrating out heavy degrees of freedom at and above the electroweak scale, 
which is described by the effective Lagrangian given at Eq.~\eqref{eq:L_eff_fcnc}.

The current discrepancy between data on $b \to s \ell^+\ell^-$ observables and SM predictions can be explained by a negative shift in $C_9$ of $\mathcal{O}(20) \%$ with respect to the SM value.
The decay amplitude reads
\begin{equation}
    \begin{aligned}
\mathcal{A}(B \to K^{(*)} \ell^+\ell^-)  &= -\langle K^{(*)}(k) \ell^+ \ell^- | \mathcal{H}_{\text{eff}}|B(k+q) \rangle \\ &=\frac{G_F \alpha V_{tb} V_{ts}^*}{\sqrt{2} \pi} \Big[(C_9 L_V^\mu + C_{10} L_A^\mu) \langle K^{(*)} (k) | \bar{s} \gamma_\mu P_L b | B(k+q)\rangle \\ &- \frac{L_V^\mu}{q^2} \Big(2 i m_b C_7 \langle K^{(*)} (k) | \bar{s} \sigma_{\mu \nu} q^\nu P_L b | B(k+q)\rangle + 16 \pi^2 \mathcal{H}_\mu^{B \to K^{(*)}} \Big) \Big].
\end{aligned}
\end{equation}

The matrix elements explicitly multiplying $C_9, C_{10}$ and $C_7$ are the local FFs, which can be computed with LQCD and LCSR \cite{Gubernari:2023puw, Gubernari:2020eft, Gubernari:2018wyi, Parrott:2022rgu, Bharucha:2015bzk}. The non-local contributions $\mathcal{H}_\mu^{B \to K^{(*)}}$ 
is defined at Eq.~\eqref{eq:nonlocal}, with $M=K^{(*)}$.
The effects of these non-local matrix elements can be effectively taken into account by a shift $\Delta_9^{\lambda}(q^2)$ in $C_9$, that depends on $q^2$ (the dilepton invariant mass) and the polarization of the final state ($\lambda = K, K^*_0, K^*_\parallel, K^*_\perp$):
\begin{equation}
C_9 \to C_9^\lambda(q^2) = C_9^{\text{SM}} + \Delta_9^\lambda(q^2) + C_9^{\text{SD}}.
\end{equation}
Therefore, even though the tension with data can be well described by a deviation from the SM in $C_9$, this shift could come from an inaccurate description of the non-local matrix elements.

\begin{figure}[!t]
    \centering
    \includegraphics[width=0.90\linewidth]{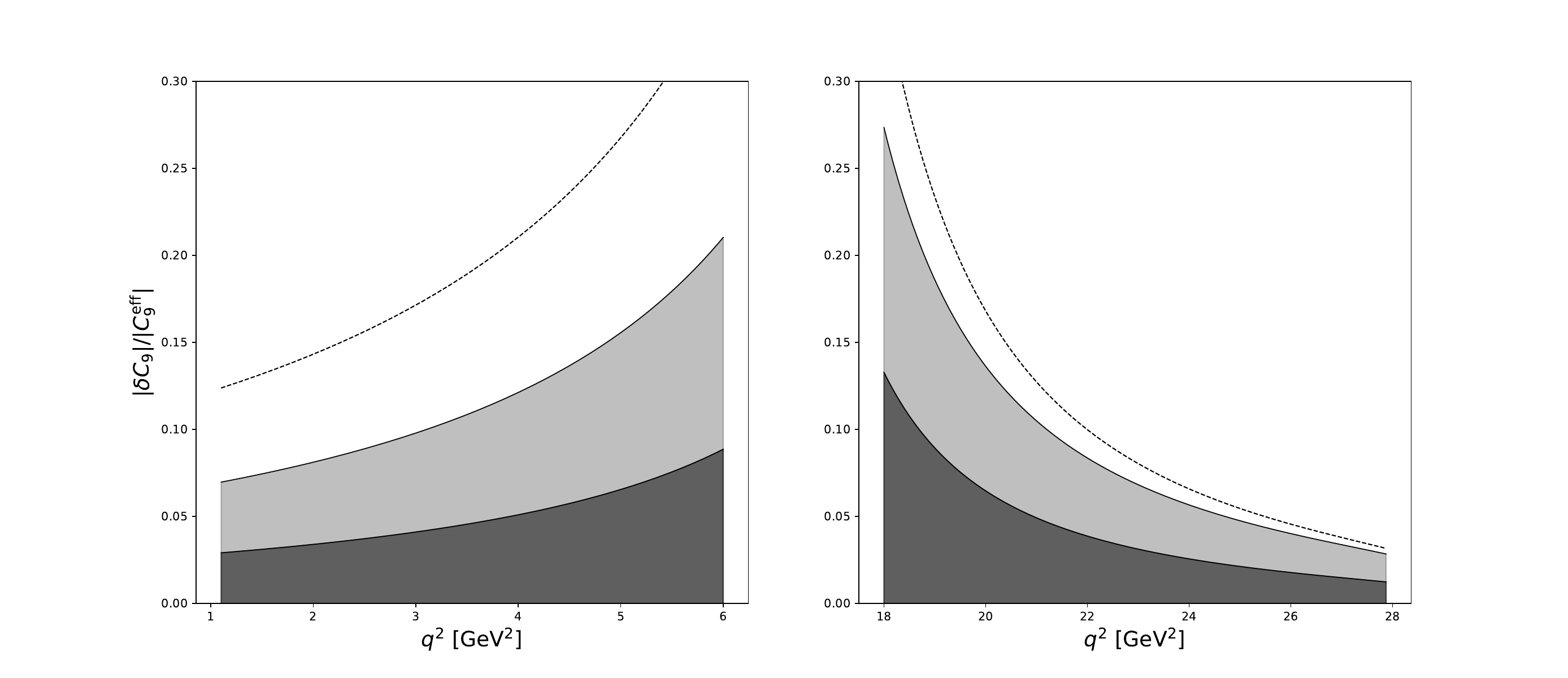}
    \caption{Combined results plotted as $|\delta C_9|/|C_9^{\text{eff}}|$, where $\delta C_9$ is the contribution from charm rescattering triangle diagrams.}
    \label{fig:rescattering}
\end{figure}

The non-local FFs have been computed using light-cone sum rules and dispersive approaches, finding an overall small effect, especially at low $q^2$ \cite{Gubernari:2022hxn, Gubernari:2020eft, Chrzaszcz:2018yza, Bobeth:2017vxj, Khodjamirian:2012rm}. However, it has been suggested that charm rescattering effects associated with triangle topologies are not captured by the dispersive approach \cite{Ciuchini:2022wbq}. To study these contributions, we build a model in terms of hadronic degrees of freedom based on heavy-hadron chiral perturbation theory, inputs from data on $B$ decays, $SU(3)$ light-flavour symmetry, and heavy-quark spin symmetry \cite{Isidori:2024lng, Isidori:2025dkp}. The goal of this model is to estimate the size of these contributions and assess whether they could account for the discrepancy between theory and data. Our estimate is more reliable at high $q^2$; we add appropriate FFs to extrapolate to lower values of $q^2$.
All the diagrams with intermediate states $D D_s$, $D D^*_s$, $D^*_s D$, $D^*_s D^*$ are computed, and the contributions coming from higher excited intermediate states are accounted for by estimating a conservative multiplicity factor. Summing all the contributions in three different scenarios (corresponding to different levels of fine-tuning of the free parameters), we obtain the results shown in Fig.~\ref{fig:rescattering}, plotted as the effective shift in $C_9$ relative to the short-distance value. In the untuned scenario (black region), the rescattering contributions correspond to a correction of order a few \% to $C_9$, especially at high $q^2$, where the estimate is more reliable. In the more tuned scenarios (grey and white regions), the contribution can be larger and exhibits a stronger $q^2$ dependence. Therefore, we conclude that if these rescattering effects are sizeable, they would manifest as a $q^2$ dependence in the value of $C_9$ extracted from data. 

\begin{figure}[!t]
\centering
    \includegraphics[width=0.48\linewidth]{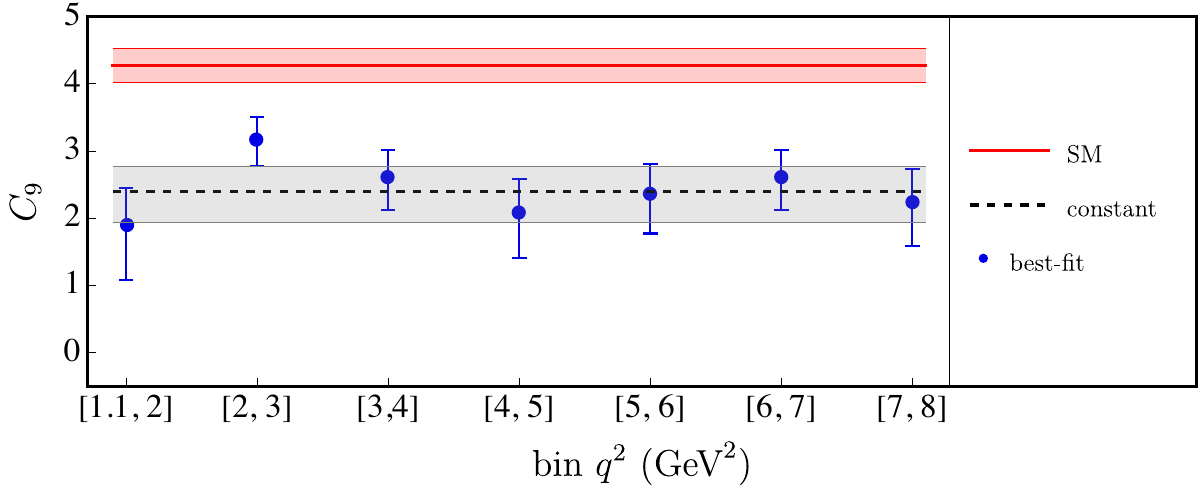}
    \hfill
    \includegraphics[width=0.48\linewidth]{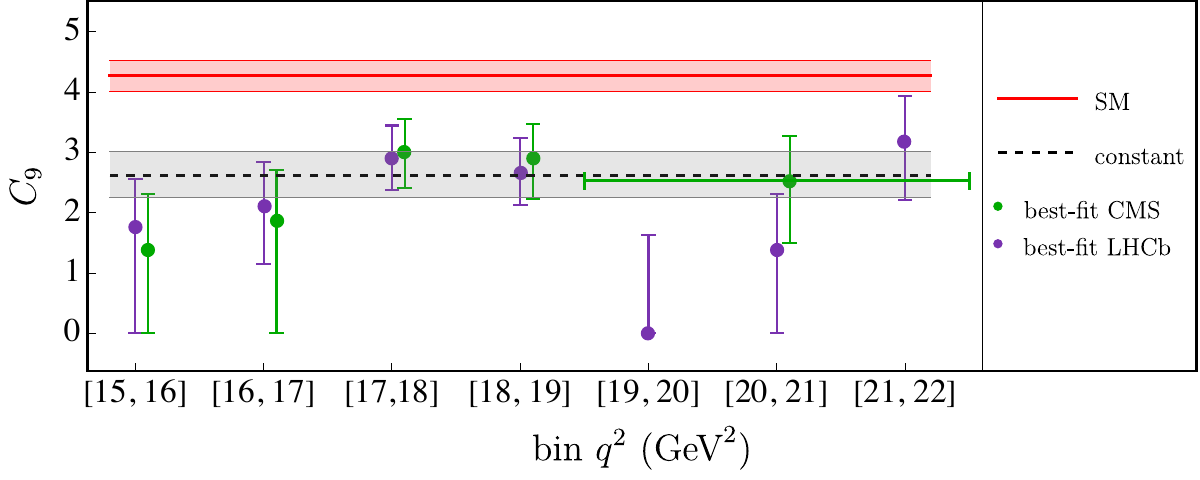}
    \centering
        \centering
        \includegraphics[width=0.48\linewidth]{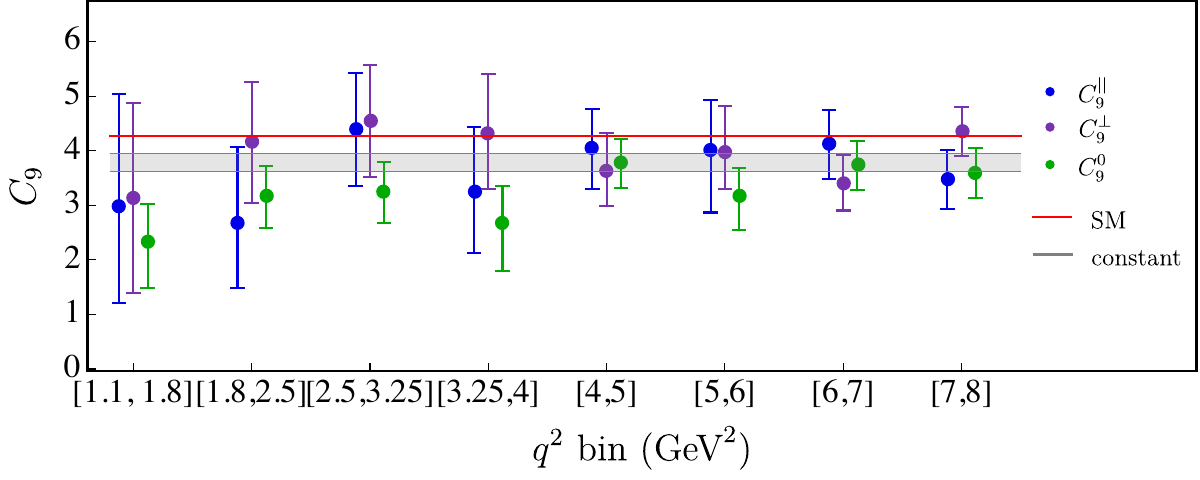}
        \hfill
        \includegraphics[width=0.48\linewidth]{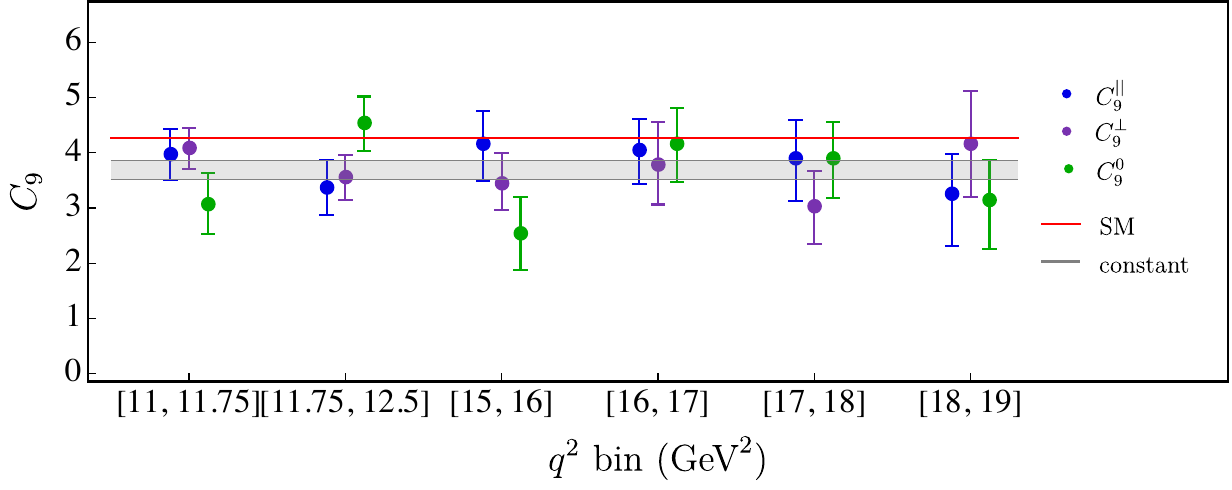}
    \caption{Results from $B \to K^{(*)} \mu^+ \mu^-$ at low and high $q^2$. The gray lines are the results of the fit assuming a constant $C_9$ over the bin and the polarization. For more details, check \cite{Bordone:2024hui, Bordone:2026wov}.}
        \label{fig:fitBKllres}
\end{figure}

To test this, one can fit $C_9$ from data parametrizing all the known $q^2$ dependence \cite{Bordone:2024hui, Bordone:2026wov}. In general,
\begin{equation}
C_9(q^2)= C_9^{\text{SM}} + \delta C_9^{\lambda}(q^2) + Y_\lambda(q^2),
\end{equation}
where $Y_\lambda(q^2)$ includes the perturbative contributions from four-quark operators, and charmonium resonance contributions that we parametrize with a dispersive approach based on a one-time subtracted dispersion relation, where the spectral function is assumed to be dominated by single-particle intermediate states, and where the free parameters are fixed by data \cite{LHCb:2016due, LHCb-PAPER-2024-011}. We extract the residual contribution $\delta C_9^{\lambda}(q^2)$ in different $q^2$ bins and polarizations: the results are shown in Fig.~\ref{fig:fitBKllres} in the $B \to K \mu^+ \mu^- $ and in the $B \to K^* \mu^+ \mu^-$ case, respectively. The relevant experimental data appear in \cite{CMS:2024syx, LHCb-PAPER-2014-006, LHCb-PAPER-2025-041}.
In both cases, the bin-by-bin analyses show values of $C_9$ that are mutually compatible, both at low and high $q^2$, implying that, with the current experimental errors, there is no statistical evidence of unaccounted-for long-distance contributions with $q^2$- and/or helicity-dependence. The value of $C_9$ extracted in the $B\to K \mu^+ \mu^-$ case with the constant assumption is lower, but the comparison is not straightforward given that the overall quality of the fit in the $B\to K$ case is not as good as for $B\to K^*$, especially if $C_{10}$ is fixed to its SM value, as implied by $\mathcal{B}(B_s\to\mu^+\mu^-)$\footnote{Check \cite{Bordone:2024hui, Bordone:2026wov} for the fit quality results. This problem might be related to an underestimate of the uncertainty on the $B\to K$ form factors~\cite{Ciuchini:2026hxy}.}. It is worth mentioning that, within our dispersive model, the angular observable $S_5$ is well described, and $S_7$, which shows clear evidence of non-local phases, is well described by our dispersive treatment as well.  Future data will further test the validity of our dispersive approach. 
To improve our understanding of this puzzle, it is necessary to keep extracting $C_9$ in as many $q^2$ bins as possible to further test the $q^2$ dependence: a $\delta C_9$ different from the SM value and flat in $q^2$ would be strong evidence for new physics presence in the $b \to s \ell^+ \ell^-$ transitions, whereas a significant dependence on $q^2$ and on the polarization would indicate the presence of unaccounted-for long-distance effects.

\noindent
\subsection{$b\to s\ell^{+}\ell^{-}$ decays: status of lattice calculations}

{\it Giuseppe Gagliardi}\\

FCNC $b\to s$ decays arise only at loop level in the SM and therefore provide particularly sensitive probes of possible NP effects. After integrating out the heavy degrees of freedom, they are described by 
the effective Lagrangian given at Eq.~\eqref{eq:L_eff_fcnc}.

In this Report, we briefly discuss two recent LQCD developments by the RM123-Soton collaboration in the study of FCNC decays. The first is the LQCD calculation of the local contributions to the decay $B_s\to\mu^+\mu^-\gamma$ at large dilepton invariant mass~\cite{Frezzotti:2024kqk}. The second is a proposal to compute the  non-local contributions directly on the lattice using spectral-function reconstruction techniques~\cite{Frezzotti:2025hif}.

The decay $B_s\to\mu^+\mu^-\gamma$ is phenomenologically interesting because the additional suppression by $\alpha_{\rm em}$ with respect to $B_s\to\mu^+\mu^-$ is partially compensated by the absence of helicity suppression in the radiative channel. Moreover, unlike $B_s\to\mu^+\mu^-$, the radiative decay is sensitive to additional Wilson coefficients such as $C_7$ and $C_9$.

The non-perturbative structure-dependent information entering the amplitude is encoded in hadronic tensors involving the electromagnetic current and the effective-theory operators. The so-called local contributions, which are those generated by the operators $O_{7,9,10}$, are parametrized in terms of five non-perturbative FFs, $F_{V},F_{A}, F_{TV}, F_{TA}$ and $\bar{F}_{T}$, which have been computed in Ref.~\cite{Frezzotti:2024kqk} on the lattice in the high-$q^2$ region $4.2~{\rm GeV}\lesssim \sqrt{q^2}\lesssim 5.1~{\rm GeV}$. The calculation has been performed using ETMC gauge ensembles at four lattice spacings and for several heavy-strange meson masses below the physical $B_s$ mass. The extrapolation to the physical point was carried out using fit Ans\"atze inspired by heavy-quark effective theory and vector-meson dominance.

\begin{figure}[!t]
\centering
\includegraphics[scale=0.25]{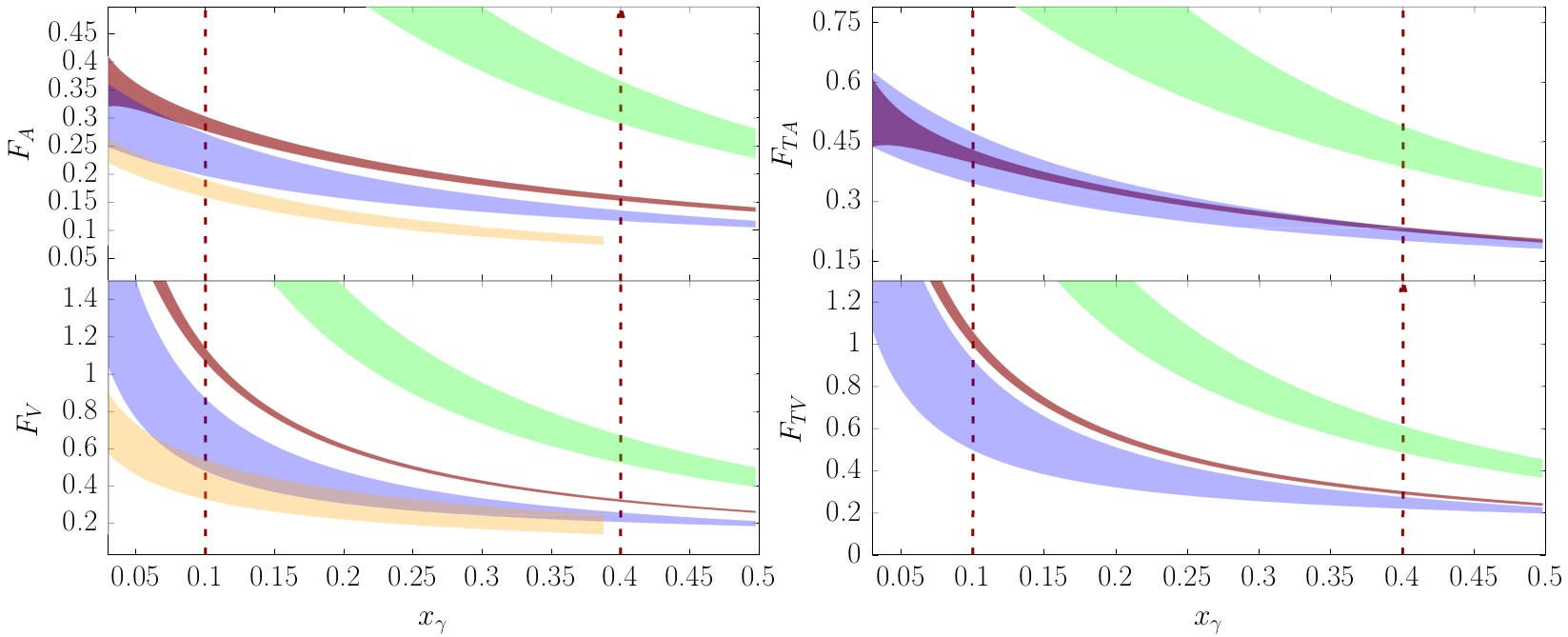}
\caption{Comparison between the lattice determination of the $B_s\to\mu^+\mu^-\gamma$ FFs from Ref.~\cite{Frezzotti:2024kqk} (red)  with existing model-dependent results from Ref.~\cite{Kozachuk:2017mdk} (blue), Ref.~\cite{Guadagnoli:2023zym} (orange) and Ref.~\cite{Janowski:2021yvz} (green).}
\label{fig:ffcomp}
\end{figure}

The lattice results show significant deviations from previous model-dependent estimates, as illustrated in Fig.~\ref{fig:ffcomp}, where $F_{V},F_{A},F_{TV}$ and $F_{TA}$ are plotted as a function of $x_\gamma=2E_\gamma/m_{B_s}$, with $E_{\gamma}$ denoting the photon energy in the $B_s$ rest frame.

Using the lattice determination of the FFs, the structure-dependent contribution to the decay rate was evaluated. For a photon-energy cut corresponding to $x_\gamma\leq 0.166$, the result reads
\begin{align}
\label{eq:res_rate_Bs}
\mathcal{B}_{\rm SD}
=
6.9(9)\times10^{-11},
\end{align}
well below the current LHCb upper bound ($\mathcal{B}^{{\rm LHCb}}_{\rm SD} < 2 \times 10^{-9}$). The result in Eq.~(\ref{eq:res_rate_Bs}) has been obtained using the LQCD determination of the local FFs, while the remaining non-local contributions, and in particular those  generated by $O_{1,2}$, have been estimated using a model parameterization.

More recently, we have started to tackle the problem of computing the non-local contributions to FCNC decays directly on the lattice~\cite{Frezzotti:2025hif}. This is particularly relevant for decays such as $B\to K^{(*)}\ell^{+}\ell^{-}$, where the comparison between theory and experiment currently shows tensions. These discrepancies can be described in terms of a shift of the Wilson coefficient $C_9$ with respect to its SM value. 

At present, however, the theoretical predictions entering this comparison rely on model-dependent estimates of the non-local contributions, rather than on first-principles calculations. Since the contributions from $O_{1,2}$ effectively appear as a $q^2$-dependent correction to $C_9$, it is essential to establish whether the apparent modification of $C_9$ preferred by current phenomenological analyses originates from genuine new-physics effects or instead from an incomplete treatment of long-distance QCD dynamics in the theoretical prediction.

In the case of the $B\to K\ell^{+}\ell^{-}$ decay, the contributions from the $O_{1,2}$ operators are encoded in the amplitudes at Eq.~\eqref{eq:nonlocal}, with $M=K$.

From the LQCD perspective, amplitudes of the type in Eq.~(\ref{eq:nonlocal}), corresponding to the Fourier transform of a time-ordered product of two operators between single-hadron external states, are problematic whenever intermediate states with energies smaller than that of the external states can propagate between the two operators. The amplitude in Eq.~(\ref{eq:nonlocal}) belongs to this class. In this situation, the standard Euclidean-time integration strategy used to reconstruct the amplitude from Euclidean lattice correlators breaks down, reflecting the impossibility of obtaining the physical Minkowskian amplitude through a naive analytic continuation.

In Ref.~\cite{Frezzotti:2025hif}, using the spectral-function reconstruction method of Ref.~\cite{Frezzotti:2023nun}, a strategy to evaluate these contributions directly on the lattice has been developed for FCNC decays involving a single and QCD-stable hadron in the final state, such as $B\to K \ell^{+}\ell^{-}$. The key idea is to promote the Feynman $i\epsilon$ parameter, appearing in the spectral representation of the amplitude in Eq.~(\ref{eq:nonlocal}), to a regulator, i.e. to a non-zero parameter with energy dimension. This defines a smeared version of the physical amplitude, in which the singular behaviour associated with on-shell intermediate states is regulated over an energy scale $\epsilon$. The smeared amplitude can then be reconstructed from Euclidean correlators using spectral-density reconstruction techniques, such as the HLT method of Ref.~\cite{Hansen:2019idp}. The physical amplitude is then finally recovered through a controlled extrapolation to the $\epsilon\to0^+$ limit.

In practice, the extrapolation to the $\epsilon\to0^+$ limit can be numerically challenging, since the smallest accessible values of $\epsilon$ are constrained by the lattice volumes available and by the statistical precision of the lattice correlation functions, although both limitations are systematically improvable. In Ref.~\cite{Frezzotti:2025hif} we carried out a proof-of-principles study of the method, obtaining encouraging results. Further investigations are currently underway in order to assess the feasibility of a complete LQCD determination of these amplitudes with controlled uncertainties.

\clearpage

%%%%%%%%%%%%%%%%%%%%%%%%%%%%%%%%%%%%%%%%%%%%%%%%%%%%%%%%%%%%%%%%
% LFU in b->c & b->s l l 
% Experimemntal status  of LFU seraches
% Sensitivity to new physics
%%%%%%%%%%%%%%%%%%%%%%%%%%%%%%%%%%%%%%%%%%%%%%%%%%%%%%%%%%%%%%%%%

\section{Experimental status of LFU violation}
\label{sec:lfu_exp}

Lepton flavour universality can be tested through the ratios $R(D)$ and $R(D^*)$ in charged-current $b \to c\tau\nu$ transitions, as well as through analogous ratios defined for other $b$-hadrons and for neutral-current $b \to s\ell^+\ell^-$ decays. This section reviews the recent experimental results on these observables from \belletwo and \lhcb. 

After the workshop, both \babar and \belletwo released new results that are not discussed below. In particular \babar released a measurement of $R(D)$ and $R(D^*)$, using the semileptonic tag \cite{BaBar:preliminary_morion2026}. This result reduces the tension between the world average and the SM, to about 2.8$\sigma$. However, its inclusion reduces the $\chi^2$ probability of the world average from 27\% to 0.8\%. The \babar collaboration is currently conducting further checks on the systematics. The \belletwo collaboration, on the other hand, presented a new result from the combined measurement of $R(D^*)$ and $\tau$ polarization $P_\tau(D^*)$ \cite{Belle2:preliminary_ichep26}. The results are consistent with previous measurements and with the SM.

%Review of the current ratio measurements, focusing on experimental techniques used by \babar and \lhcb, challenges and source of systematic uncertainties. Opportunities with "other" $b$-hadron semileptonic decays and spectroscopy.

\subsection{Status at \belletwo}

{\it Tommy Martinov}\\

The \belletwo experiment, operating at the SuperKEKB $e^+e^-$ collider at the
$\Upsilon(4S)$ resonance, is uniquely suited for studying LFU tests in semileptonic beauty meson decays. Its nearly $4\pi$ spatial coverage, combined with the precisely known initial state of the collision, enables measurements involving missing energy and the reconstruction of neutral and low-momentum final-state particles~\cite{10.1093/ptep/ptz106}.

Tests of LFU in semileptonic decays, require an accurate modelling of semileptonic $B$ decays to charm. These decays are modelled as a sum of $B\to D\ell\nu$, $B\to D^{*} \ell\nu$, $B\to D^{**} \ell\nu$ decays and smaller contributions such as $B\to D\pi\pi\ell\nu$. However, the inclusive $B\to X_c \ell\nu$ branching fraction exceeds the sum of the corresponding exclusive branching fractions. To bridge this gap, recent \belletwo measurements include the $B\to D^{(*)} \eta \ell\nu$ channels, commonly referred to as \textit{gap modes}, in the decay cocktail.

\begin{figure}[!t!]
    \centering
    \includegraphics[width=0.75\linewidth]{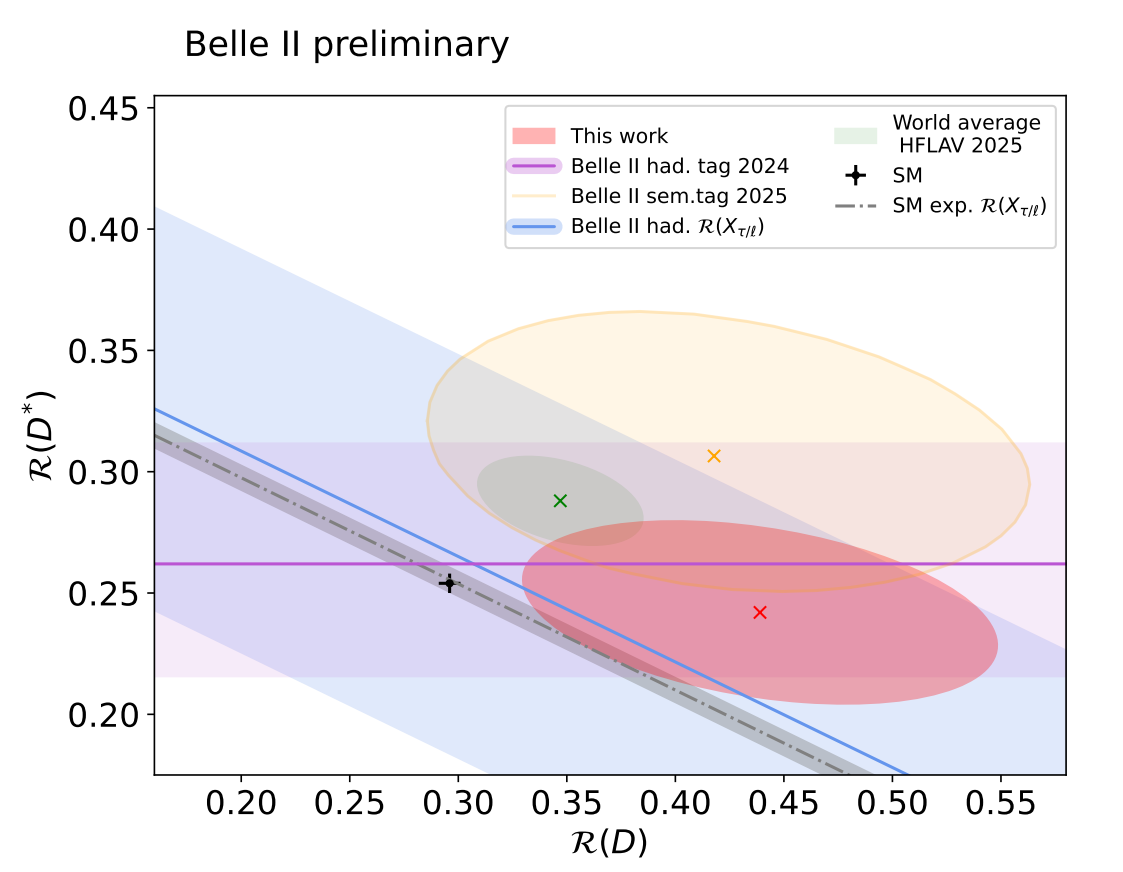}
    \caption{Measured $R(D)$ and $R(D^*)$ with their associated confidence region at 68.3\% confidence level as obtained from the most recent Belle~II measurement (red). These results are compared to the SM prediction (black marker), the world average (green)~\cite{HeavyFlavorAveragingGroupHFLAV:2024ctg}, the Belle~II $R(D^*)$ measurement with hadronic tagging~\cite{PhysRevD.110.072020} which is superseded (purple), the Belle~II $R(D^{(*)})$ measurement with semileptonic tagging~\cite{fmn3-h8fy} (yellow), and the constraint from the Belle~II $R(X_{\tau/\ell})$ measurement~\cite{PhysRevLett.132.211804}. The grey band shows the SM expectation of $R(X_{\tau/\ell})$.}
    \label{fig:rd_rdstar_comp_belle2}
\end{figure}

The most recent LFU result from \belletwo is the simultaneous determination of $R(D)$ and $R(D^{*})$ using hadronic tagging. This analysis supersedes the previous $R(D^{*})$ hadronic-tag measurement~\cite{PhysRevD.110.072020}, doubling the data sample and extending the measurement to $R(D)$. The signal efficiency is maximised by reconstructing all $D^{*}$ decay modes together with 13 $D$ decay channels. Dominant backgrounds are validated using $B\to D^{*} \pi^{0}\ell\nu$ events in a dedicated sideband. 
A fit to this control region indicates that the gap is smaller than expected, a finding confirmed by the main fit, which prefers branching fractions reduced by $40$--$80\%$ relative to the default expectation. The signal yield is extracted from a binned extended maximum-likelihood fit to the two-dimensional distribution of the total energy of neutral clusters not associated with the $B$ tag or the signal hadronic system ($E_{\mathrm{ECL}}$), and the missing mass squared ($M_{\mathrm{miss}}^{2}$). The results are
\begin{align*}
    R(D)   &= 0.439 \pm 0.055\,(\mathrm{stat.}) \pm 0.045\,(\mathrm{syst.}), \\
    R(D^{*}) &= 0.242 \pm 0.019\,(\mathrm{stat.}) \pm 0.016\,(\mathrm{syst.}),
\end{align*}
with a correlation coefficient of $-0.4$ (statistical) and $+0.2$ (systematic). The measured value of $R(D^{*})$ agrees with the SM expectation within $0.5\sigma$, while $R(D)$ lies $2.0\sigma$ above it. The two-dimensional combination is compatible with the world average~\cite{HeavyFlavorAveragingGroupHFLAV:2024ctg} (as of September~2025) at the $1.3\sigma$ level and with the SM prediction at the $1.5\sigma$ level. The leading systematic uncertainty is the finite size of the simulated samples used to model signal and backgrounds. A comparison of the obtained results with past measurements, the SM expectation and the world average is shown in Fig.~\ref{fig:rd_rdstar_comp_belle2}.
\\
\\
A complementary determination of $R(D^{*})$ using semileptonic tagging~\cite{fmn3-h8fy} provides an independent cross-check. Inclusive measurements have also been pursued. Using a hadronic tag and reconstructing the hadronic system $X$ inclusively in $B \to X \ell \nu$ decays, Belle~II has performed the first measurements of both the $e$-to-$\mu$ and $\tau$-to-light-lepton ratios $R(X)$~\cite{PhysRevLett.131.051804, PhysRevLett.132.211804}. In addition, angular asymmetries in $B\to D^{*} \ell\nu$ decays have been measured~\cite{PhysRevLett.131.181801}, providing further tests of LFU beyond the integrated rates.
\\
The current treatment of the gap between the sum of exclusive modes and the inclusive rate is unsatisfactory: the $D^{(*)}\eta$ channels used to fill it are chosen ad~hoc and are unlikely to account for the full missing rate. Belle~II therefore assigns a $100\%$ uncertainty to their branching fractions. Ongoing work is investigating additional decays that could contribute to the gap. Furthermore, multibody $D$ decays are currently simulated using pure phase-space models. Efforts are underway to implement Dalitz-plot models based on BES-III inputs, which should improve the description of such decays in future analyses.

\begin{figure}[!t!]
    \centering
    \includegraphics[width=0.75\linewidth]{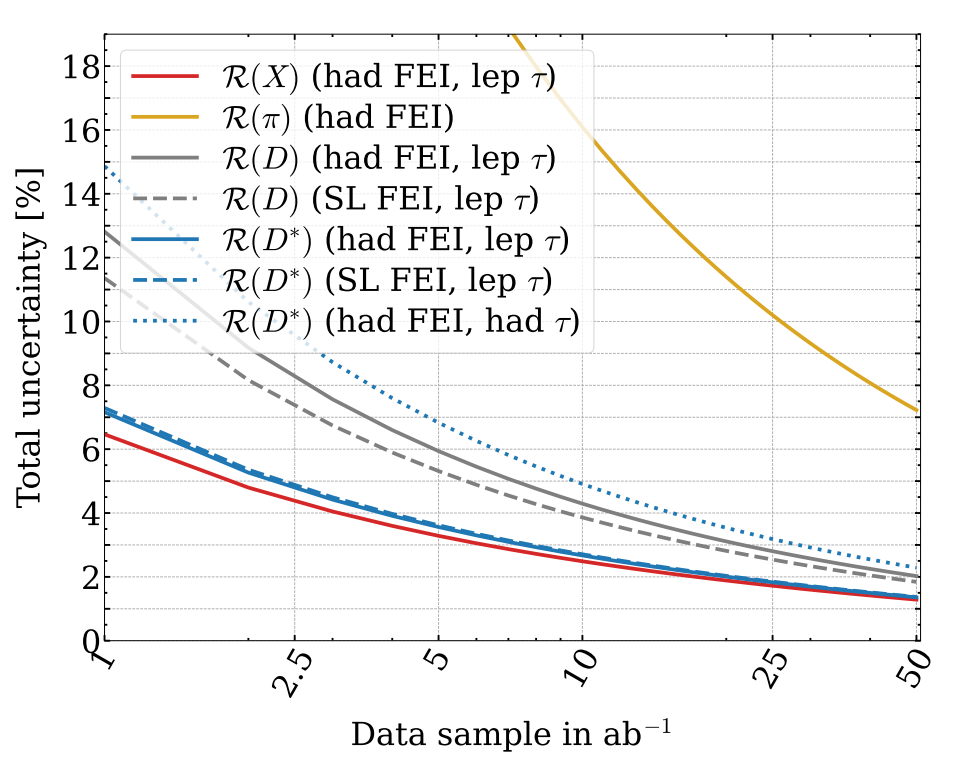}
    \caption{Expected Belle~II sensitivity for various LFU ratio measurements as a function of luminosity.}
    \label{fig:snowmass_belle2}
\end{figure}

\belletwo is continuing to accumulate data, with substantial further integrated luminosity expected from the current running period. Projections based on an integrated luminosity of $1~\mathrm{ab}^{-1}$ indicate that the total uncertainty on $R(D)$ will reach the $13$--$15\%$ level, the uncertainty on $R(D^{*})$ about $8\%$, and the uncertainty on $R(X)$ approximately $16\%$~\cite{SnowMass2022}. The latter can be reduced further by improvements in semileptonic modelling, in particular more precise FFs and better-measured exclusive branching fractions. Projections for larger Belle~II datasets are shown in Fig.~\ref{fig:snowmass_belle2}. Together with complementary measurements at LHCb and the expected theoretical progress, these results will sharpen the experimental picture of lepton-flavour universality in $b\to c$ transitions over the coming years.

\subsection{Status at \lhcb}
\label{LFU-LHCb}

{\it Ricardo Vazquez Gomez} \\

The LHCb experiment provides a unique environment to study these decays thanks to large production rates of $b$ hadrons and excellent vertexing capabilities. 
However, the hadronic collision environment introduces significant experimental challenges, particularly due to missing neutrinos and large backgrounds. 
This contribution summarizes the current status of LFU measurements at LHCb, focusing on recent measurements, and future prospects.
\\
Two complementary strategies are used at LHCb depending on the $\tau$ decay mode. In the case of muonic $\tau$ decays ($\tau \to \mu \nu \nu$), the visible final state is identical to that of the normalization channel ($B \to D^{(*)} \mu \nu$), which allows many systematic effects to cancel. 

For hadronic $\tau$ decays ($\tau \to 3\pi \nu$) only two neutrinos are present, making the kinematics intrinsically simpler, and the excellent vertex resolution of LHCb allows the $\tau$ decay vertex to be resolved. This enables powerful discrimination between signal and background through the separation between the $B$ and $\tau$ vertices. However, the visible final state differs from that of normalization channels, requiring normalization to other hadronic decays and leading to an increased sensitivity to background modelling. In both cases, the analyses rely on kinematic observables, some of which are introduced in Sec.~\ref{sec:experiments}.

The most recent LHCb measurement of $R(D^+)$ and $R(D^{*+})$ using muonic $\tau$ decays is based on 2 fb$^{-1}$ of data collected in 2015--2016, \cite{LHCb:2024jll}. The analysis performs a simultaneous determination of both ratios by fitting three-dimensional distributions involving $q^2$, the missing mass squared, and the muon energy in the $B$ rest frame. 
This analysis benefits from the fact that signal and normalization modes share the same visible final state, which allows for a strong cancellation of systematic uncertainties related to reconstruction efficiencies and detector effects. Nevertheless, significant uncertainties remain. The dominant contributions arise from the modelling of hadronic FFs and from the description of backgrounds, in particular those originating from $D^{**}$ states.

A key ingredient of the analysis is the use of data-driven control samples to constrain these backgrounds. Events with additional tracks are used to isolate contributions from excited charm states, and the resulting constraints are propagated into the signal fit. This approach is crucial to reduce the dependence on simulation and to control systematic effects. The measured values are
\begin{align}
R(D^+) &= 0.249 \pm 0.043 \, (\mathrm{stat.})\pm 0.047 \,(\mathrm{syst.}), \\
R(D^{*+}) &= 0.402 \pm 0.081 \,(\mathrm{stat.}) \pm 0.085 \,(\mathrm{syst.}).
\end{align}

The study of decays involving excited charm states provides an important cross-check of LFU measurements, as these processes can contaminate the $R(D^{(*)})$ measurements through feed-down. LHCb has reported evidence for the decay $B^- \to D^{**0}\tau^- \nu_\tau$ with a significance of $3.5\sigma$, \cite{LHCb:2025fri}.
The analysis uses the full Run 1 and Run 2 dataset and reconstructs the $D^{**}$ states through their decays into $D^* \pi$ combinations. Dedicated multivariate classifiers are employed to suppress backgrounds from fake $D^{**}$ candidates and from multi-body charm decays. The signal extraction is performed using fits to kinematic variables including $q^2$ and mass differences.
\\
An important aspect of this measurement is the sensitivity to the modelling of FFs for excited states. Different theoretical parametrizations can lead to noticeable differences in kinematic distributions, inducing systematic uncertainties of a few percent. The measured ratio is
\begin{equation}
R(D^{**0}) = 0.13 \pm 0.04.
\end{equation}
\noindent The result is consistent with SM expectations and indicates that feed-down from excited states is unlikely to be responsible for the observed LFU tensions. 

The measurement of $R(D^{*-})$ using hadronic $\tau$ decays exploits the decay $\tau^+ \to 3\pi \nu$, which provides a distinctive topology with a displaced vertex. 
In this case, the measurement is performed relative to a hadronic normalization channel, $B \to D^* 3\pi$, whose branching fraction is known from an independent measurement. This approach avoids the need to compare different lepton species directly but introduces additional systematic uncertainties related to external inputs.

The main challenge in this analysis is the presence of large backgrounds from double charm decays, such as $B \to D D_s^+$, where the $D_s^+$ meson decays into three pions. These backgrounds are controlled using dedicated control regions and isolation variables. Extensive validation is performed to ensure that the simulation accurately reproduces the data, and corrections are applied where necessary. The result obtained using 2015--2016 data, \cite{LHCb:2023uiv}, is
\begin{equation}
R(D^{*-}) = 0.260 \pm 0.015\,(\mathrm{stat.}) \pm 0.016\,(\mathrm{syst.}) \pm 0.012\,(\mathrm{ext.}).
\end{equation}

\noindent The ratio $R(J/\psi)$, defined in $B_c$ decays, provides a complementary probe of LFU. A previous measurement using data collected in 2012-2013, yielded
\begin{equation}
R(J/\psi) = 0.71 \pm 0.17\,(\mathrm{stat.}) \pm 0.18\,(\mathrm{syst.}),
\end{equation}
and an updated result using the dataset collected in 2016-2018, \cite{LHCb:2026dev}, gives
\begin{equation}
R(J/\psi) = 0.51 \pm 0.12\,(\mathrm{stat.}) \pm 0.08\,(\mathrm{syst.}),
\end{equation}
showing a significant reduction in the systematic uncertainty, achieved through an improved selection and updated LQCD form factors for the $B_c\to J/\psi$ transition \cite{Harrison:2025yan}. 

One of the main challenges in this channel is the presence of large backgrounds from misidentified muons. These backgrounds are particularly prominent in regions of low transverse momentum, where the modelling becomes difficult. The analysis strategy therefore includes the exclusion of problematic regions and the use of data-driven techniques to determine misidentification rates, improving the robustness of the measurement.

Future measurements of $R(D^0)$ and $R(D^{*0})$ using hadronic $\tau$ decays will face similar challenges to the charged modes, with an additional emphasis on the modelling of double charm backgrounds. Inclusive decays involving $D_s^+ \to 3\pi$ represent a dominant source of background and must be carefully constrained using multiple control samples. Data-driven techniques are essential to validate and correct the simulation.
\\
The study of LFU in the $B_s^0$ sector offers several advantages. The spectroscopy of excited $D_s^+$ states differs significantly from that of non-strange charm mesons, leading to reduced feed-down contamination. In addition, the decay products are often highly collimated due to the boost at LHCb, allowing the use of cone-based reconstruction techniques to reduce combinatorial backgrounds.
\\
Nevertheless, new challenges arise, including potential contributions from poorly understood or unmeasured decay modes. These ``unknown unknowns'' can impact the stability of the fits and require careful investigation through control regions and alternative models.
\\
LFU tests involving electrons, such as $R(D^*)_{\mu/e}$ and $R(D^*)_{\tau/e}$, are becoming increasingly accessible at LHCb. Historically, these measurements were limited by trigger and reconstruction efficiencies. With the Run 3 upgrade, including the removal of the hardware trigger and improvements in electron reconstruction, these channels are now within reach.
\\
A key complication in electron modes is bremsstrahlung radiation, which affects the energy reconstruction and leads to categorization of events depending on whether the emitted photons are recovered. Dedicated strategies are employed to account for these effects and to ensure accurate determination of efficiencies.
\\
The current measurements of $R(D)$, $R(D^*)$, and related observables are consistent with previous results and continue to show tension with SM expectations. Complementary studies, such as those involving excited states and new decay modes, provide important cross-checks and help to clarify the origin of this discrepancy.
\\
Looking ahead, the availability of larger datasets from Run 2 and Run 3, combined with improvements in simulation and reconstruction techniques, is expected to significantly enhance the precision of these measurements. In particular, the extension to new channels, including $B_s^0$ decays and electron modes, will provide a more complete picture of LFU in the heavy flavour sector.

\clearpage

\section{Theoretical status of LFU violation}
\label{sec:lfu_theo}

A precise theoretical treatment of LFU observables must account not only for the hadronic and CKM inputs entering the SM predictions of $R(D^{(*)})$ and $R(K^{(*)})$, but also for higher-order electroweak and QED radiative corrections, which are currently modelled in experimental analyses using the PHOTOS package. This section examines the size and reliability of these radiative corrections, discusses the prospects and remaining challenges for a first-principles, non-perturbative treatment of QED effects in LQCD, and reviews the current status of global fits to charged- and neutral-current LFU data in the search for genuine, lepton-flavour-universality-violating New Physics.

\subsection{Impact of radiative corrections on LFU observables}

{\it Zach Polonsky} \\ 

Although higher-order QED corrections naturally enter processes with charged particles with a $\alpha/\pi\lesssim 1\%$ suppression, several effects can enhance these corrections to the few-percent level. As experimental measurements of semileptonic $B$ decays are pushed below the $10\%$ level, it then becomes important to evaluate how these QED corrections are accounted for in modern experimental analyses.

Currently, these effects are accounted for using the PHOTOS Monte Carlo package~\cite{Barberio:1990ms,Barberio:1993qi,Golonka:2005pn,Golonka:2006tw}. PHOTOS uses a so-called ``afterburner'' algorithm, which takes pre-generated Monte Carlo data and incorporates QED bremsstrahlung effects using leading-logarithmic factorization to alter the final-state kinematics of the Born-level process. Schematically, this factorization takes the form~\cite{Barberio:1990ms,Baier:1980kx}
\begin{equation}\label{eq:qedfactorization}
    d\Gamma = d\Gamma^{(0)}\omega(x, \cos\theta; m_i) dxd\Omega\,,
\end{equation}
where $d\Gamma^{(0)}$ is the Born-level differential rate, $x$ is the energy fraction carried by the radiated photon, $m_i$ is the mass of the charged particle which radiates the photon, and $\cos\theta$ is the angle of the photon with respect to the charged particle. The universal radiation function, $\omega$, entering Eq.~\eqref{eq:qedfactorization} only depends on the photon kinematics and the mass of the final-state charged particles and is normalized so that the fully photon inclusive rate reproduces the Born-level result
\begin{equation}\label{eq:radiationFuncNorm}
    \int dx d\Omega\frac{d\Gamma}{dx d\Omega} = d\Gamma^{(0)}\,.
\end{equation}
The factorization utilized by PHOTOS accounts for emission of soft Bremsstrahlung photons which are generically enhanced by large, real collinear logarithms $\sim\log(m_i/m_B)$. However, there are several other potentially enhanced effects which can enter semileptonic $B$ decays:
\begin{enumerate}
    \item Short-Distance Effects: Renormalization group running in the low-energy EFT from the electroweak scale to the $b$ scale leads to large logarithms $\sim \log(m_b^2/MZ^2)$. Additionally, it has been shown that, while next-to-leading order electroweak matching contributions are generally very small, neglecting these effects can lead to a relatively large theoretical uncertainty associated the choice of electroweak renormalization scheme~\cite{Brod:2010hi,Bobeth:2013tba,Brod:2021qvc,Brod:2022har}. These effects are largely accounted for in semileptonic $B$ decays (c.f.~\cite{Sirlin:1981ie, Huber:2005ig, Brod:2008ss, Gorbahn:2022rgl} and references therein).
    \item Virtual Collinear Logarithms: Similar large logarithms to those arising from real radiation are also present in virtual corrections. While these are not explicitly accounted for in PHOTOS, for a fully inclusive process, the real and virtual collinear logarithms must exactly cancel, and therefore the normalization of the radiation function, Eq.~\eqref{eq:radiationFuncNorm}, \textit{implicitly} takes such logarithms into account.
    \item Coulomb/Threshold Effects: In processes with multiple charged particles in the final state, the emitted charged particles can rescatter via virtual photon exchanges, leading to a $\sim\pi^2/\beta_{ij}$ enhancement, where $\beta_{ij}$ is the relative velocity between charged particles $i$ and $j$. This effect is much more relevant for decays with mutliple heavy particles in the final state, since for $m_i,m_j\to0$, $\beta_{ij}\to 1$.
    \item Initial- and Final-State Inteference: Since PHOTOS only incorporates final-state radiation, additional effects involving the interference between real emission from initial and final states, which are enhanced by logarithms of the photon energy cut, are not included.
    \item Structure-Dependent Effects: Non-perturbative effects arising from the internal structure of hadrons have been shown to give corrections on the order of a few percent in semileptonic kaon decays~\cite{Bijnens:1994me}. These effects are very challenging to quantify, but much progress has been made in recent years in evaluating structure-dependent effects in fully leptonic decays~\cite{Beneke:2017vpq,Beneke:2019slt,Beneke:2021jhp,Beneke:2021pkl,Beneke:2022msp,Cornella:2022ubo,Boer:2023vsg,Cornella:2026lkp}.
\end{enumerate}
Critically, points 2-4 all can potentially introduce additional LFUV effects, and so can in principle lead to percent-level deviations in $R_{D^{(*)}}$ and $R_{K^{(*)}}$ if not properly accounted for.

In order to investigate this in the context of charged-current decays, the authors of Ref.~\cite{deBoer:2018ipi} examined the phenomenological model for semileptonic $B$ decays given in Ref.~\cite{Bernlochner:2010fc} in the soft-photon limit. This limit allows for the explicit resummation of leading soft Bremsstrahlung logarithms, $\sim\alpha^n\log^n(E_{\text{max}}^\gamma)$, as well as the Coulomb contribution, $\sim(\pi\alpha/\beta_{ij})^n$, the latter of which is only relevant for the decay with a charged $D$ meson in the final state. It was found that the Coulomb effect leads to a $\sim 2\%$ shift in the $\bar B^0\to D^+\mu\bar\nu$ mode and a $\sim 4\%$ shift in the $\bar B^0\to D^+\tau\bar\nu$ mode, also changing the sign of the QED correction in the $\tau$ mode. This leads to an overall difference of $1-2\%$ in the QED corrections to $R_{D^+}$ versus $R_{D^0}$.

\begin{figure}[t]
    \centering
    \includegraphics[width=0.45\textwidth]{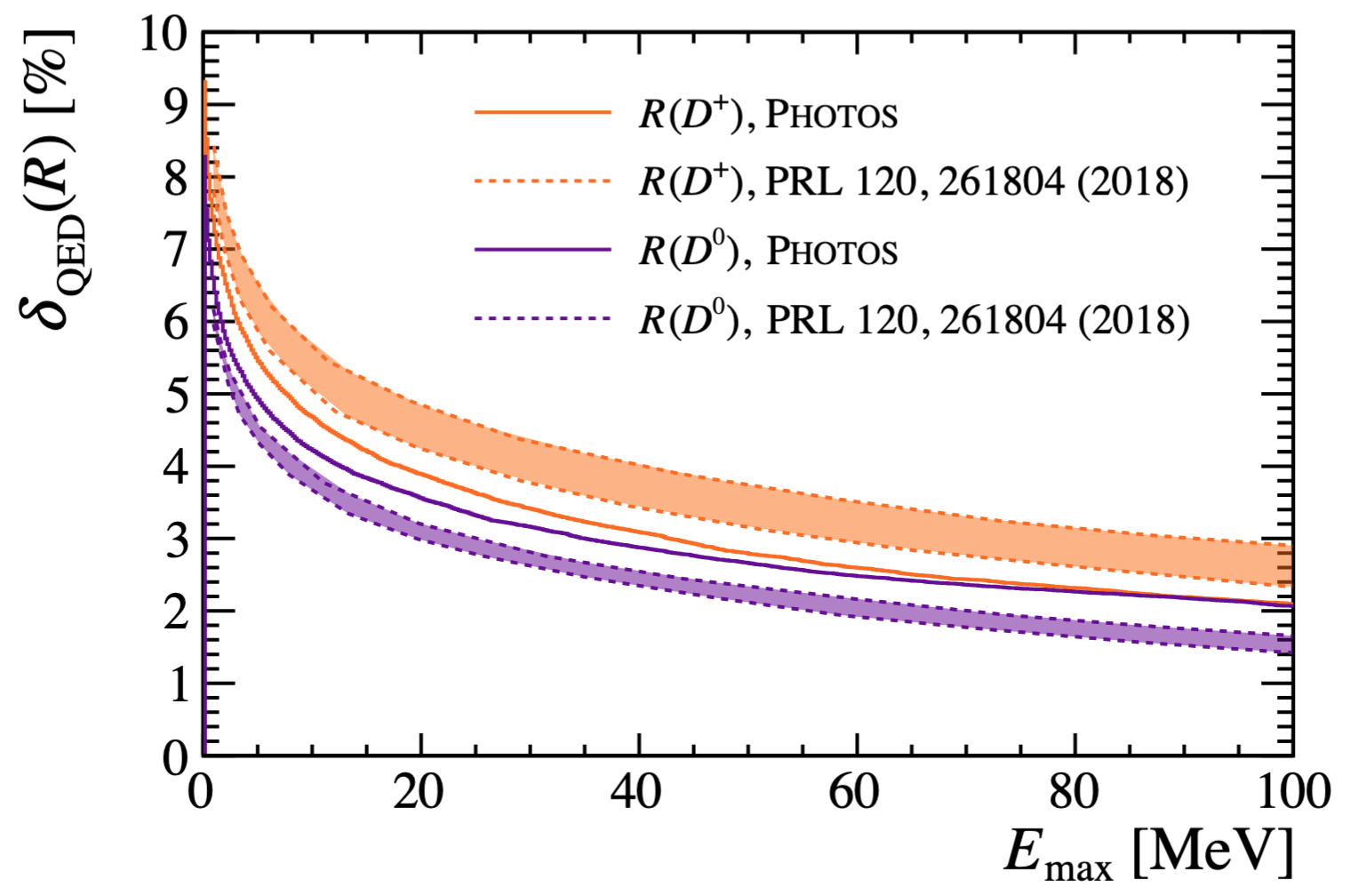}
    \hfill
    \includegraphics[width=0.45\textwidth]{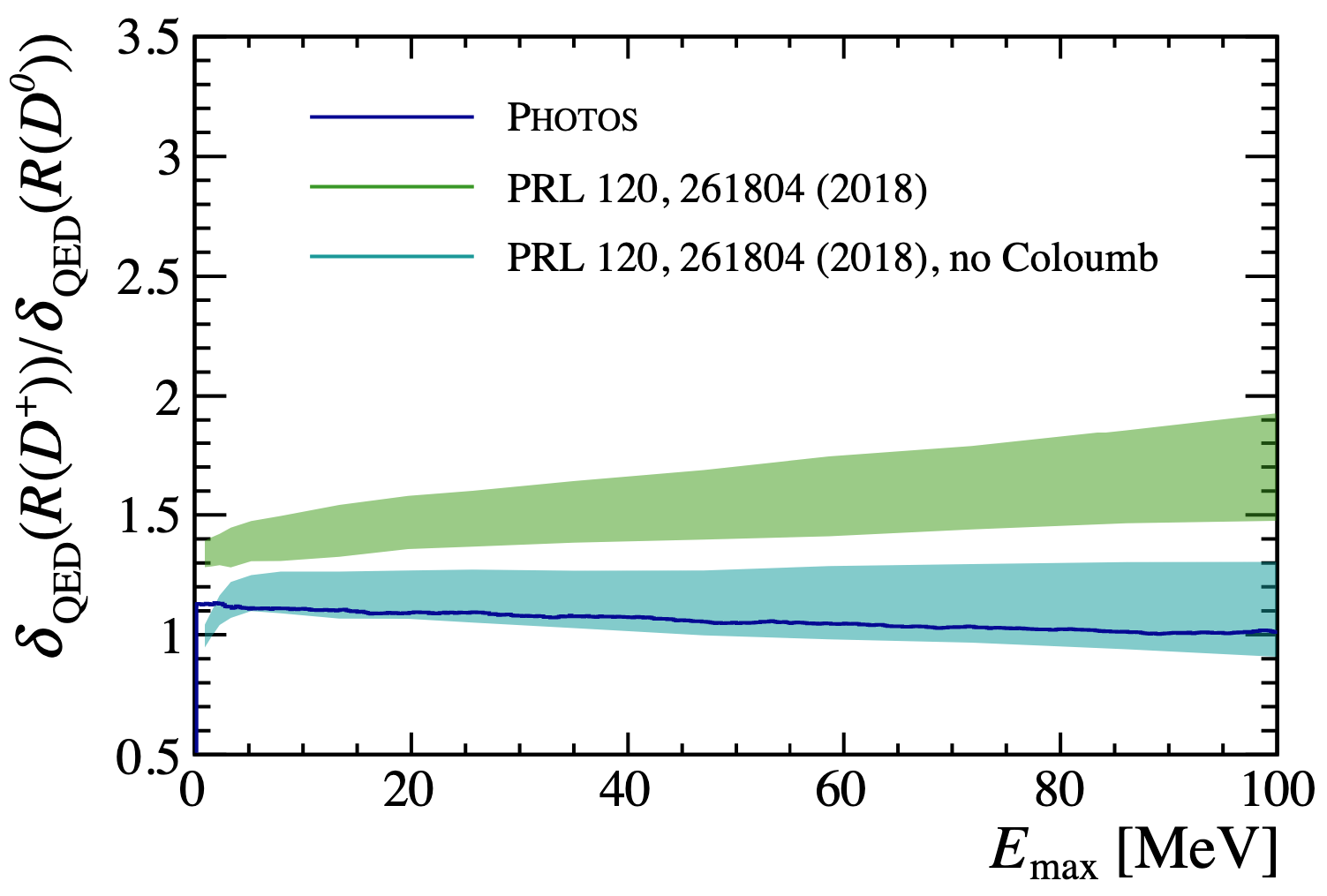}
    \caption{Comparison of $R_{D^{+/0}}$ between model calculation from Ref.~\cite{deBoer:2018ipi} and Monte Carlo simulation using PHOTOS. \textit{Left:} Overall QED corrections to each mode independently. \textit{Right:} Relative correction between the charged and neutral mode. Figures from Ref.~\cite{Cali:2019nwp}.}
    \label{fig:RDPHOTOS}
\end{figure}

The authors of Ref.~\cite{Cali:2019nwp} compared the results of the model used in Ref.~\cite{deBoer:2018ipi} with those from PHOTOS, and found $\sim1\%$ differences in both $R_{D^+}$ and $R_{D^0}$, though the shifts are in opposite directions for the two modes, see Fig.~\ref{fig:RDPHOTOS}. Additionally, Ref.~\cite{deBoer:2018ipi} only considered soft photons, while PHOTOS also incorporates the emission of photons with energies in the range where structure-dependent effects become relevant. In order to estimate the impact of neglecting these contributions, the authors of Ref.~\cite{Cali:2019nwp} also conducted LHCb-like pseudo-experiments, where $E_{\text{max}}$ cuts were placed either on fitted templates or generated data in order to simulate over- or under-estimating QED effects. In the most extreme cases, it was found that this could lead to $\sim7\%$ bias on the measurements of $R_D$.

NLO QED effects were also studied in the context of inclusive $b\to ce\nu$ decays in Ref.~\cite{Bigi:2023cbv}. Here, due to the small mass of the electron, the Coulomb/threshold effect are less enhanced, but still lead to a $\sim 1.6\%$ correction to the inclusive branching fraction and $q^2$ moments. Although this is smaller than the current best measurements from BaBar~\cite{BaBar:2004bij}, it was emphasized that the full NLO QED corrections are larger than the systematic uncertainties attributed to QED effects from PHOTOS.

\begin{figure}[t]
    \centering
    \includegraphics[width=0.45\textwidth]{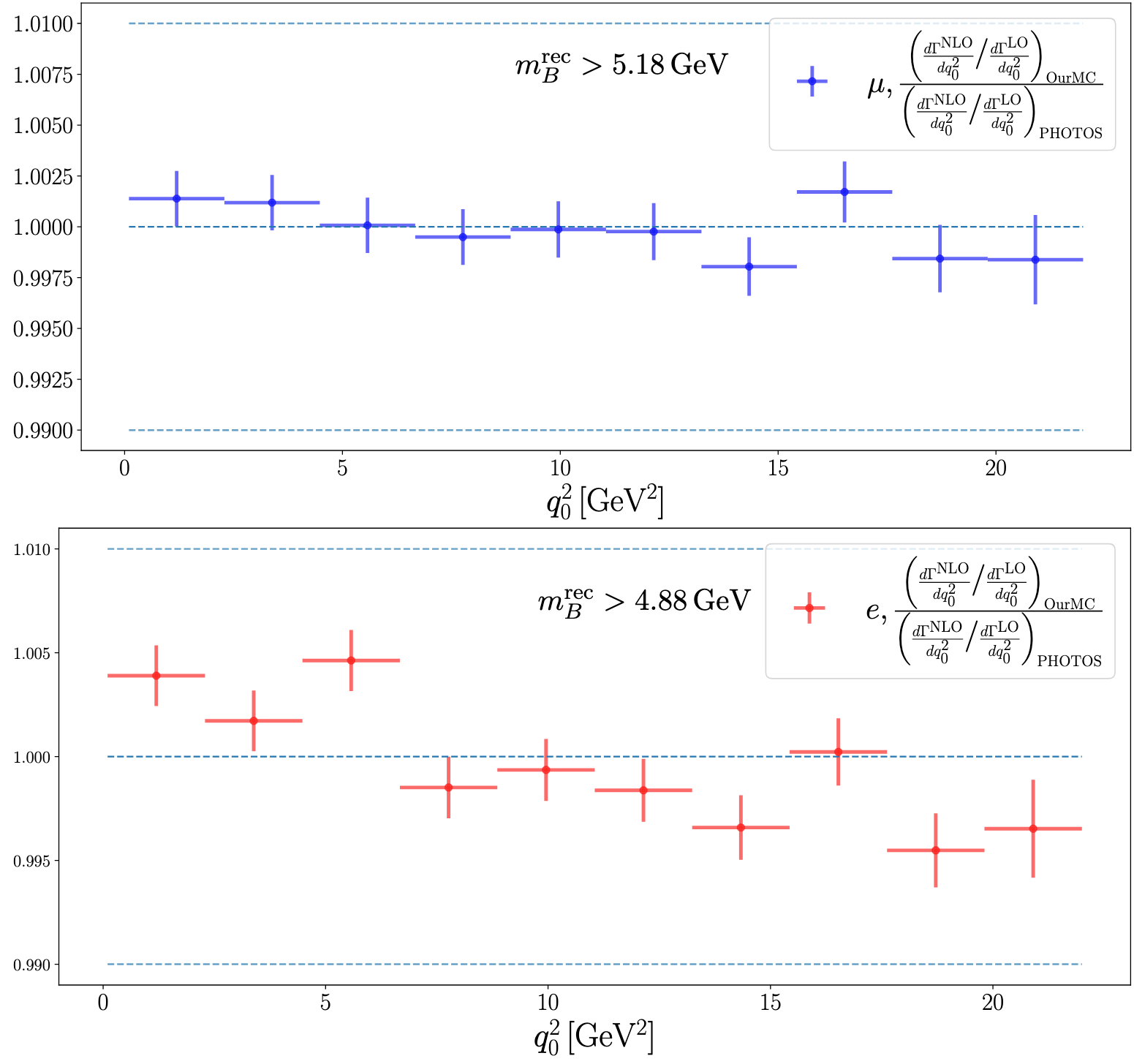}
    \hfill
    \includegraphics[width=0.45\textwidth]{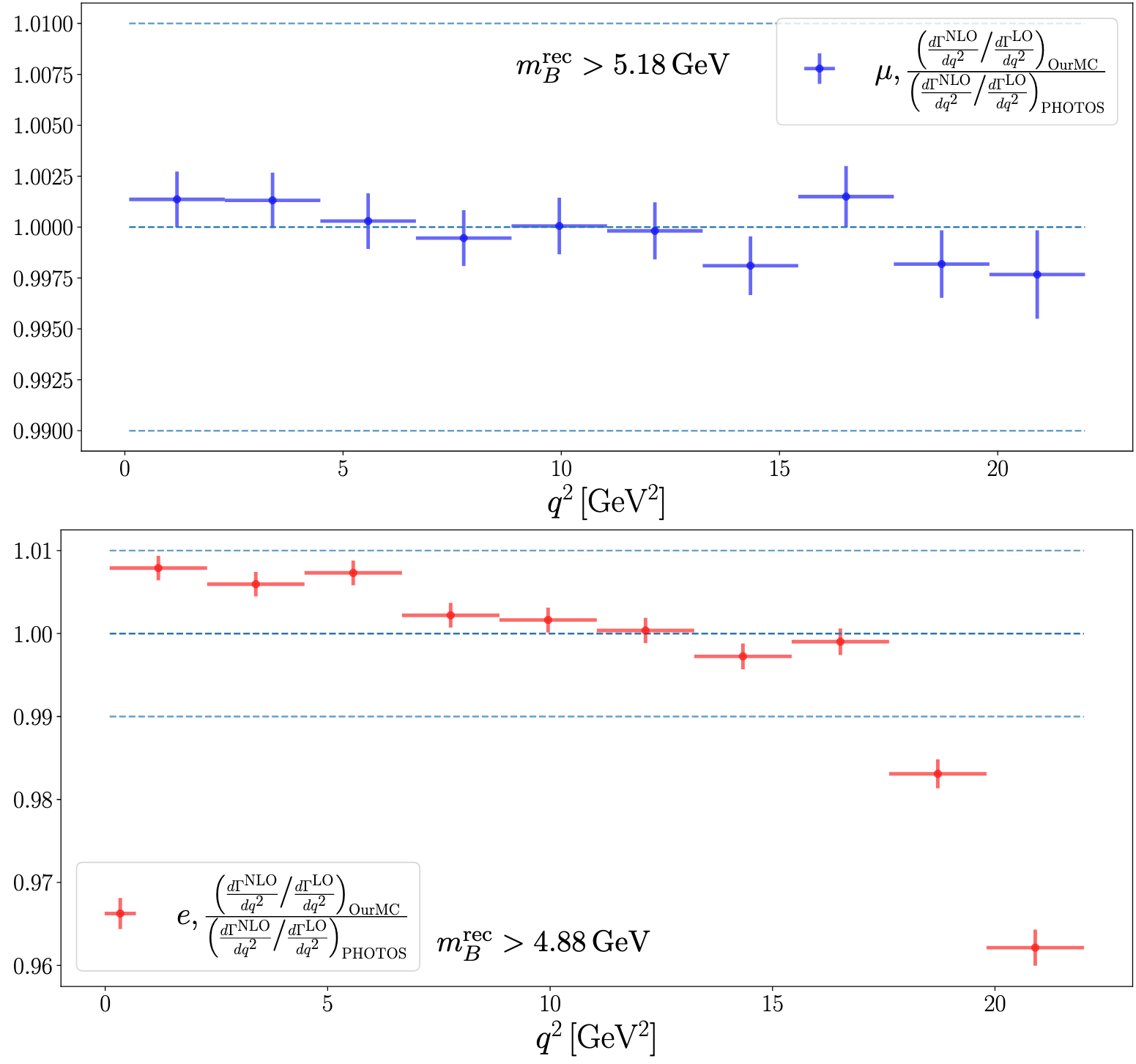}
    \caption{Comparison of FCNC decays between model calculation from Ref.~\cite{Isidori:2020acz} and Monte Carlo simulation using PHOTOS for muon mode (blue) and electron mode (red). \textit{Left:} Comparison in $q_0^2$ bins using the IR safe $B$-factory kinematic parameters. \textit{Right:} Comparison in $q^2$ bins using hadron collider parameters. Note that the disagreement at high $q^2$ in the electron mode arises from the compressed photon spectrum, indicating a breakdown of the NLO calculation in~\cite{Isidori:2020acz} rather than a true disagreement with PHOTOS. Figures from Ref.~\cite{Isidori:2022bzw}.}
    \label{fig:FCNCPHOTOS}
\end{figure}

Similar analyses to those in Refs.~\cite{deBoer:2018ipi} and~\cite{Cali:2019nwp} were performed in Refs.~\cite{Bordone:2016gaq,Isidori:2020acz,Isidori:2022bzw} for the FCNC $B\to K\ell\bar\ell$ decays. Here, a particular emphasis was put on the differences between the IR safe $B$-factory kinematic parameters, and hadron collider parameters. The main point is that, for fully photon inclusive processes, rates in terms of $B$-factory parameters are explicitly independent of collinear logarithms and feature no LFUV, while those in terms of hadron collider parameters are only LFU after integration over $q^2$. In all cases, PHOTOS reproduced the results of the NLO QED corrections to within the percent level (see Fig.~\ref{fig:FCNCPHOTOS}), indicating that radiative corrections to $B\to K\ell\bar\ell$ are well-captured by PHOTOS.

\subsection{Radiative corrections: prospects and challenges on the Lattice}

{\it Davide Giusti}

The determination of CKM matrix elements from heavy-meson semileptonic decays has entered the sub-percent precision era. In this regime, electromagnetic radiative corrections and strong isospin-breaking effects can no longer be neglected. Theoretical control of these corrections is essential for precision extractions of quantities such as $|V_{cb}|$ and $|V_{ub}|$ from processes including $B \to D^{(*)}\ell\nu$, $B \to \pi \ell\nu$, and $D \to K(\pi)\ell\nu$. LQCD calculations of semileptonic FFs have reached accuracies where QED effects represent one of the dominant missing ingredients~\cite{FlavourLatticeAveragingGroupFLAG:2024oxs}.

The inclusion of QED in lattice calculations is substantially more difficult than pure QCD because electromagnetic interactions are long-ranged and generate infrared divergences in decay amplitudes. Physical decay rates become finite only after combining virtual-photon corrections with real-photon emission, according to the Bloch--Nordsieck mechanism~\cite{Bloch:1937pw}.

A major conceptual breakthrough was the formulation of a general framework for treating radiative corrections to hadronic decay amplitudes directly in lattice QCD+QED. In this framework, first introduced in~\cite{Carrasco:2015xwa}, one computes the decay rate of a parent hadron $H_P$ decaying into a daughter particle $H_D$ and a lepton-neutrino pair
\begin{equation}
\Gamma(H_P \to H_D\ell\nu[\gamma]) = \Gamma_0 + \Gamma_1^{\rm pt}(\Delta E_\gamma),
\end{equation}
where the subscripts $0,1$ indicate the number of photons in the final state and soft-photon emission below an experimental energy resolution, $\Delta E_\gamma$, is included.

The decay rate is decomposed into a structure-independent point-like (pt) contribution, calculable analytically in perturbation theory, for which the soft energy cut-off $\Delta E_\gamma$ is small enough not to resolve the internal structure of the hadrons, and a structure-dependent non-perturbative contribution evaluated on the lattice to include all momentum modes due to the exchange of virtual photons. This strategy successfully removes infrared singularities while preserving gauge invariance and allowing finite-volume extrapolations. The method was first developed for leptonic decays~\cite{Carrasco:2015xwa,Lubicz:2016xro,Giusti:2017dwk,Giusti:2018puc,Giusti:2018guw} and later generalized conceptually to semileptonic processes~\cite{Sachrajda:2019uhh}.

A major milestone was the first lattice determination of electromagnetic corrections to light-meson leptonic decay rates, including $\pi_{\mu2}$ and $K_{\mu2}$, by the Rome-Southampton collaboration~\cite{Giusti:2017dwk,DiCarlo:2019thl}. These investigations demonstrated that lattice QCD+QED can achieve phenomenologically relevant precision and provided improved determinations of $|V_{us}|$.
Subsequently, other major lattice groups applied a similar strategy to study leptonic weak decays of mesons~\cite{Boyle:2022lsi,Boccaletti:2024guq,Cotellucci:2026kny} and further approaches have been proposed~\cite{Christ:2023lcc,DiCarlo:2025uyj}.
The extension to semileptonic decays, however, is significantly more challenging because the final-state hadron carries non-trivial momentum and the hadronic matrix elements depend on kinematic variables, such as the squared four-momentum transfer $q^2 = (p_{H_P}-p_{H_D})^2$.

For heavy-meson semileptonic decays, several intertwined challenges arise. Below are the main ones.
\begin{enumerate}
    \item {Infrared Structure and Real-Photon Emission}: In semileptonic decays, the emitted photon can couple not only to external charged particles but also to the hadronic transition current. The resulting structure-dependent radiation introduces non-universal contributions that extend beyond the inner-bremsstrahlung terms, which themselves depend on the semileptonic FFs and their derivatives with respect to $q^2$~\cite{Sachrajda:2019uhh}, and therefore cannot be described solely by soft-photon theorems~\cite{Low:1958sn}.

    In heavy-meson decays this issue is amplified by the richer resonance structure and by the presence of multiple kinematic scales, such as $m_b,\; m_c,\; \Lambda_{\mathrm{QCD}},\; E_\gamma,\; q^2$.
    Unlike leptonic decays, where soft-photon factorization is relatively simple, semileptonic amplitudes require control over generalized Compton amplitudes and non-local current insertions. Thus, the consistent separation between universal infrared effects and hadronic structure-dependent terms remains a crucial ingredient when approaching studies of those processes.

    In this respect, the introduction of the soft energy cut-off $\Delta E_\gamma$ can be avoided by computing amplitudes with a real photon in the final state directly on the lattice. Such pioneering calculations have been performed in~\cite{Kane:2019jtj,deDivitiis:2019uzm,Desiderio:2020oej,Kane:2021zee,Giusti:2023pot,Frezzotti:2024kqk} and are gradually reaching a high level of maturity in the field. The non-perturbative evaluation of $\Gamma_1$ has important practical implications especially for heavy-meson decays. For example, since $m_{B^\star}-m_B \simeq 45 ~\rm MeV$, the hyperfine splitting for heavy mesons provides another small scale, which limits the scope and precision of the perturbative calculations for soft photons.

    An additional complication arises from experimental definitions of inclusive observables. Experimental analyses often apply cuts on photon energy and angular acceptance, and lattice calculations must eventually reproduce these fiducial definitions to enable direct phenomenological comparison.

    \item {Finite-Volume Effects in QED}: Finite-volume effects are especially severe in lattice QED because the photon is massless. Corrections scale only polynomially with the lattice size $L$, typically as $1/L$, rather than exponentially as in pure QCD.

    Different prescriptions for finite-volume QED---such as QED$_{\rm L}$, QED$_{\rm M}$, QED$_{\rm r}$ and $C^\star$ boundary conditions (for a review on the subject we refer the reader to~\cite{Patella:2017fgk})---modify the infrared behaviour and require careful theoretical understanding.

    For semileptonic decays, finite-volume effects become kinematics dependent and may couple nontrivially to momentum transfer and recoil. Universal analytic corrections can remove leading power-law effects, but structure-dependent terms are difficult to estimate reliably, especially for heavy mesons at non-zero recoil.

    Recent developments such as infinite-volume reconstruction methods aim to alleviate these issues by reconstructing long-distance contributions analytically from Euclidean lattice data~\cite{Christ:2023lcc,Christ:2024xzj}. These methods have shown promising results in leptonic and radiative decays and may eventually become central tools for semileptonic applications.

    \item {Renormalization and Matching}: The inclusion of QED modifies operator renormalization substantially. The use of computationally cheaper, lattice fermionic formulations not preserving chiral symmetry at non-zero lattice spacing induces complicated mixing of the weak currents with other operators, making the non-perturbative renormalization of the weak Hamiltonian more challenging in presence of QED~\cite{DiCarlo:2019thl,DiCarlo:2019knp}. On top of that lattice operators must be matched consistently to continuum schemes that include electroweak effects.

    A central conceptual issue is the definition of ``pure QCD'' quantities in the presence of QED. Since electromagnetic interactions shift hadron masses and quark masses, the separation between QCD and QED contributions becomes scheme dependent.

    Hadronic schemes---where selected hadron masses are matched between QCD and QCD+QED theories---have been advocated as physically transparent and practically useful~\cite{DiCarlo:2019thl,Tantalo:2023onv}.

    For heavy quarks, additional complications arise from the interplay between QED corrections and heavy-quark/soft-collinear effective theory, relativistic heavy-quark actions, light-cone sum rules, heavy-hadron ChPT or non-relativistic effective theories such as NRQCD. Precise matching between lattice and continuum descriptions is required for percent-level phenomenology.

    \item {Euclidean-Time Formulation and Non-local Matrix Elements}: Radiative corrections introduce non-local products of weak and electromagnetic currents,
    \begin{equation}
    T\{J^\mu_{\mathrm{em}}(x)\, J^\nu_W(0)\}.
    \end{equation}

    In Euclidean space, extracting the corresponding Minkowski amplitudes is non-trivial because of analytic continuation issues and possible divergent contributions from intermediate states lighter than or close to the mass of the decaying particle. Those states arise due to the electromagnetic rescattering of the final state hadron with the charged lepton, a feature that is not present in leptonic decays.
    
    Semileptonic decays are especially difficult because the hadronic final state carries momentum and may overlap with multiparticle states which, for heavy mesons, become denser in most of phase space. The Maiani--Testa theorem~\cite{Maiani:1990ca} limits the extraction of real-time amplitudes from Euclidean correlators, stating that generic Minkowski scattering amplitudes cannot be reconstructed directly from infinite-volume Euclidean correlators except at thresholds, although inclusive approaches based on smeared spectral functions offer a promising alternative~\cite{Hansen:2019idp,Frezzotti:2023nun}.
    
    Recent work on inclusive semileptonic decay rates has demonstrated that Euclidean four-point functions can be related to physical inclusive observables through controlled smearing procedures~\cite{Gambino:2022dvu,DeSantis:2025qbb}. Although these studies have so far focused primarily on pure QCD aspects, the framework may ultimately provide a natural setting for incorporating electromagnetic corrections.

    \item {Heavy-Quark Discretization Effects}: Heavy-meson semileptonic decays require accurate control of discretization errors associated with charm and bottom quarks. The addition of QED increases the number of relevant scales and can induce cutoff effects proportional to combinations such as $\alpha_{\mathrm{em}} a m_Q$.

    Current heavy-quark formulations were largely optimized for pure QCD applications, and systematic studies of QED-induced lattice artifacts are still limited. Reliable continuum extrapolations therefore require finer lattices and improved heavy-quark actions, leading to significantly increased computational cost~\cite{Boyle:2022uba}. Promising new directions and strategies to alleviate those problems while reaching the physical bottom quark mass in lattice simulations have been recently proposed in~\cite{Giusti:2021rsf}.
\end{enumerate}
    
\noindent Despite the challenges mentioned above--many of which, however, appear to be surmountable thanks to the conceptual and methodological achievements briefly outlined here--the rapid progress in supercomputing over the past decade, which has now fully entered the exascale era, points to the anticipated emergence of first pilot studies on this topic in the imminent future, towards phenomenologically competitive applications.

\subsection{Status of experimental tests and prospects of new physics}

{\it Mauro Valli}\\

Flavour physics is among the most powerful indirect probes of physics beyond the SM. At low energies, the SM 
reduces to $\mathrm{QED} \times \mathrm{QCD}$, a gauge theory that at the renormalizable level is endowed with a large set of \emph{accidental} symmetries: a global $U(1)$ acting on each of the fermion species $N_f$, and the discrete symmetry $CP$ (exact up to possibly tiny violations inducing a non-zero neutron electric dipole moment). Neither is fundamental -- the $U(1)^{N_f}$ flavour symmetry is broken by the Yukawa interactions, while $CP$ violation in the electroweak sector requires three fermion generations and is controlled by the hierarchical CKM matrix \cite{UTfit:2022hsi}. LFU belongs to the same class: it is accidental in the gauge sector of the SM and broken only by the charged-lepton Yukawa couplings, meaning there is no reason a priori for it to be respected by the UV dynamics. 
\begin{figure}[t]
  \centering
  \includegraphics[width=\linewidth]{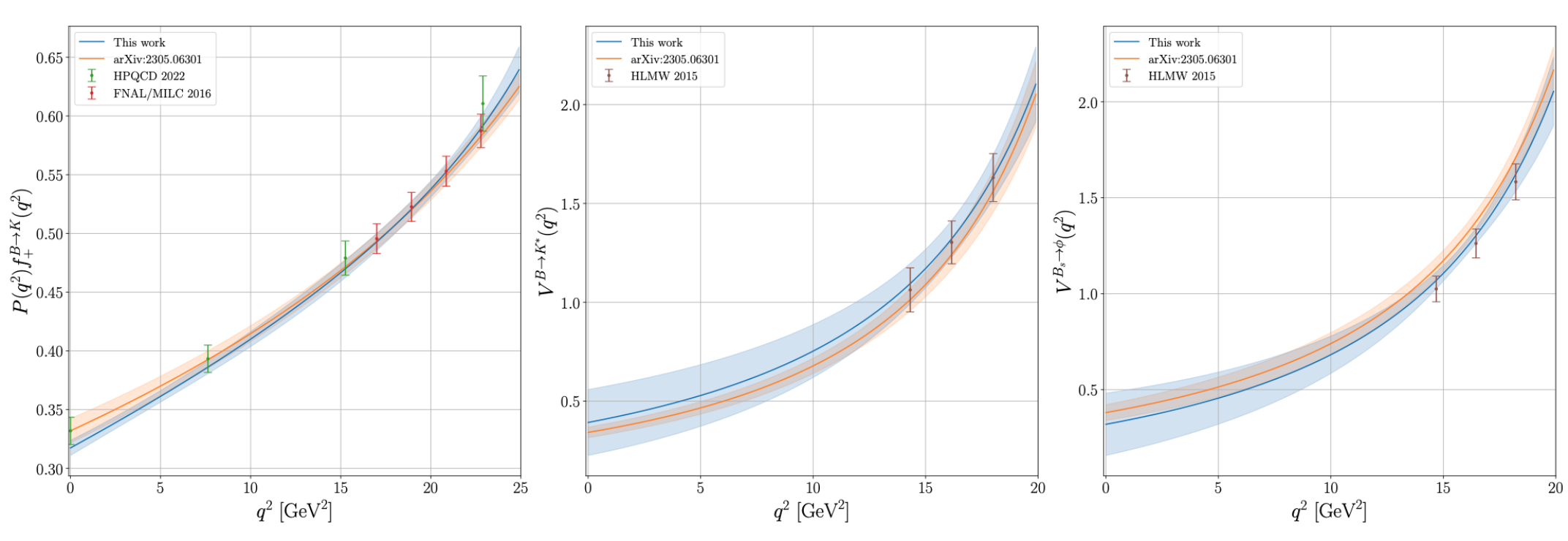}
  \caption{FFs of the vector current for the $B\to K$, $B\to K^*$ and $B_s\to\phi$ transitions, extrapolated to the full kinematic range, see Ref.~\cite{Ciuchini:2026hxy}. The new determination presented here (``This work'', blue) is obtained within the DM approach from LQCD inputs (HPQCD and FNAL/MILC for $B\to K$; HLMW for $B\to K^*$ and $B_s\to\phi$); the orange bands reproduce the analysis of Ref.~\cite{Gubernari:2023puw}.}
  \label{fig:ff}
\end{figure}
This is what makes (approximate) accidental symmetries so valuable: they act as a \emph{lamppost} for NP, where any genuine deviation would be an unambiguous signal of new degrees of freedom. Conversely, the absence of observed deviations translates, in the language of the SMEFT, into stringent lower bounds on the scale of new flavour-violating interactions, reaching $10^{5}$-$10^{7}$~TeV for the most sensitive observables \cite{Silvestrini:2018dos}. In this contribution, I briefly review the status of these tests in semileptonic $B$ decays, discussing in turn the neutral-current transitions $b\to s\ell^+\ell^-$, and the charged-current processes $b\to c\ell\nu$. In particular, the observables connected to these processes provide key predictions in recent model building for a theory of flavour not far above the electroweak scale, e.g., for UV completions based on flavour deconstruction~\cite{Barbieri:2023qpf}.

While waiting for a sensitive measurement also of the tauonic channel, the exclusive transitions $B \to K^{(*)}\ell^+\ell^-$ and $B_s \to \phi\,\ell^+\ell^-$ do not longer feature evidence of LFU violation in light-lepton modes. Nevertheless, they offer interesting observables to test possible hints for NP mainly coupled to third generation as the class of models mentioned above. These exclusive processes are described, in the helicity basis, by amplitudes that combine the short-distance Wilson coefficients $C_7$, $C_9$, and $C_{10}$ -- with SM values at $\sim m_b$ being $C_7^{\text{SM}} \sim -0.3$, $C_9^{\text{SM}} \sim 4.1$, and $C_{10}^{\text{SM}} \sim -4.2$ -- with hadronic quantities. Two qualitatively different hadronic inputs enter the evaluation: the \emph{local} FFs (Fig.~\ref{fig:ff}) and the \emph{non-local} correlators $h_\lambda(q^2)$, which encode virtual charm-quark loops, the so-called charming penguins. The latter represent the main theoretical bottleneck for a clean SM interpretation, since the rescattering of intermediate hadronic states makes their first-principle, analytic treatment extremely challenging, potentially mimicking short-distance NP effects.

\begin{figure}[t]
  \centering
  \includegraphics[width=0.92\linewidth]{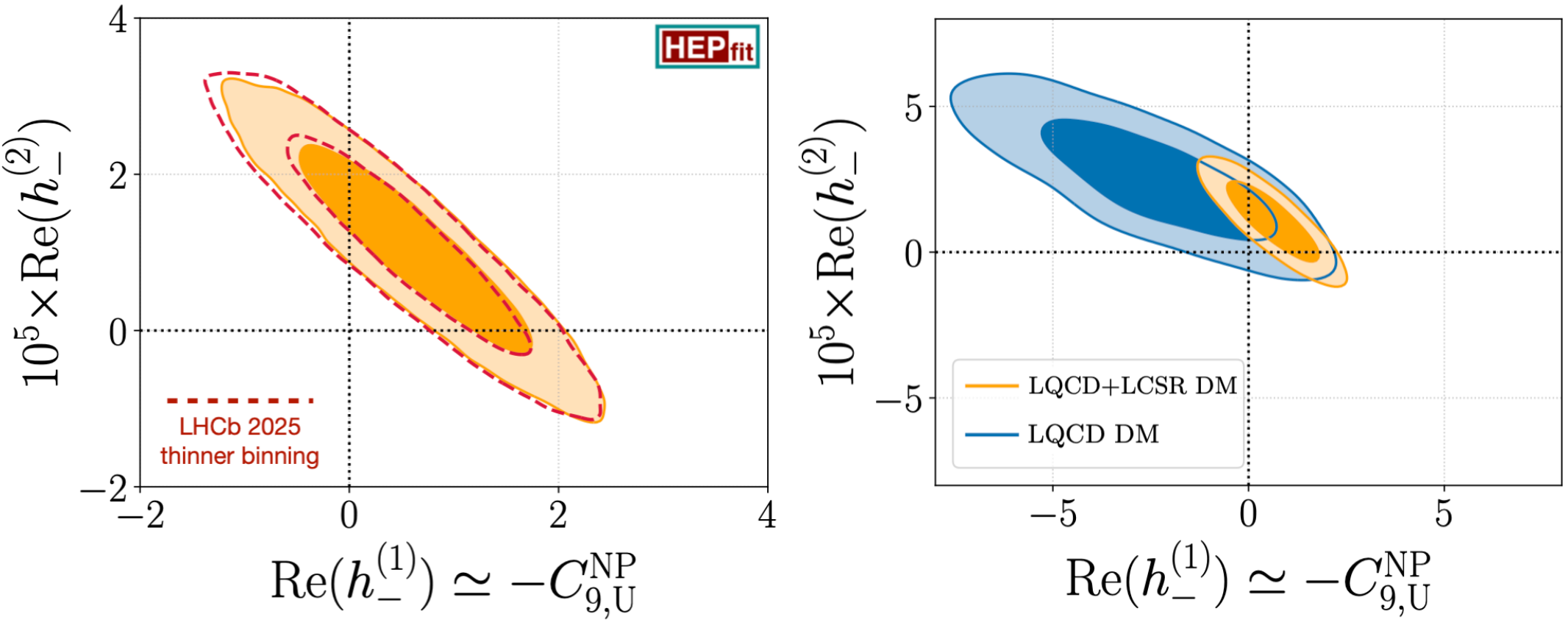}
    \caption{Global fit of $b\to s\ell^+\ell^-$ data in the $\bigl(\mathrm{Re}\,h_-^{(1)},\,10^5 \times \mathrm{Re}\,h_-^{(2)}\bigr)$ plane~\cite{Ciuchini:2026hxy}. \emph{Left}: result of the fit, including the LHCb~2025 measurements with thinner $q^2$ binning (red dashed).
    \emph{Right}: comparison of fits using local FFs from LQCD only
    (blue) and from $\text{LQCD}+\text{LCSR}$ (orange), both within the Dispersive
    Matrix approach. The horizontal axis is degenerate with a lepton-flavour-universal NP shift, $\mathrm{Re}(h_-^{(1)})\simeq-C_{9,U}^{\text{NP}}$.%
    }
  \label{fig:bsll-fit}
\end{figure}

Updating the global analysis of Ref.~\cite{Ciuchini:2022wbq} with the latest LHCb measurements, one finds \emph{no statistically significant evidence for genuine, LFU-violating NP}~\cite{Ciuchini:2026hxy}: the LFU-violating combination $C_{9,\mu-e}$, which is orthogonal to the universal direction $C_{9,U}$, is compatible with zero well within $2\sigma$. What survives is a residual shift $\Delta C_{9,U}$ along the universal direction. The nature of this shift -- whether it stems from large hadronic effects or from a genuinely universal UV physics contribution -- cannot be resolved by the global fit alone, requiring a detailed study of its $q^2$ dependence or independent complementary channels. Notably, assuming the charming-penguin contributions to be negligible, the same data prefer a non-zero shift at more than $5\sigma$ in $C_9$, a pattern naively in favour of a NP LFU scenario~\cite{Bordone:2024hui}. However, independent indications of sizeable hadronic contributions come now also from angular observables like $S_{7,8}$, whose measured values point to non-negligible absorptive parts of the amplitude, and which are interestingly of the correct size required to simultaneously accommodate the tensions in $S_5$ and $A_{\text{FB}}$, see also~\cite{Altmannshofer:2026cwk}.

Sharpening this picture demands a rigorous, improved control also of the local FFs. LQCD provides robust results at high $q^2$, while LCSR cover the low-$q^2$ region; the latter, however, is not a first-principle method and is intrinsically affected by quark-hadron duality assumptions, operator-product-expansion truncation, and systematic uncertainties related to the Borel parameter and continuum threshold. Determinations rooted in analyticity and unitarity -- such as the BGL $z$-expansion and the DM method~\cite{DiCarlo:2021dzg} -- offer a powerful, parameterization-independent framework to combine these inputs in a more controlled and conservative way, as illustrated in Fig.~\ref{fig:ff}. An open question, currently under active investigation, is to what extent the global-fit conclusions are driven by the specific LCSR input. This impact can be quantified by systematically comparing fits based on LQCD-only data against those including the full LQCD${}+{}$LCSR data set (Fig.~\ref{fig:bsll-fit}, right). Eventually, a decisive breakthrough in resolving the hadronic-vs-NP ambiguity is expected from the first non-perturbative computations of the non-local charming-penguin amplitudes~\cite{Frezzotti:2025hif}, which aim to directly evaluate the virtual charm-loop correlator in a more rigorous way via spectral function reconstruction on the lattice.

Crucially, the di-neutrino modes $B\to K^{(*)}\nu\bar\nu$ are theoretically 
cleaner: being free from long-distance charm-loop contamination, meaning that 
hadronic FFs dominate the theoretical uncertainty. Updated SM 
predictions leveraging the DM framework yield $\mathcal{B}(B^+\to K^+\nu\bar\nu) 
= (3.96\pm0.22)\times10^{-6}$ and $\mathcal{B}(B\to K^{*}\nu\bar\nu) \approx 
(9\text{--}10)\times10^{-6}$. The recent Belle~II measurement, 
$\mathcal{B}(B^+\to K^+\nu\bar\nu)=(2.3\pm0.7)\times10^{-5}$~\cite{Belle-II:2023esi}, 
exhibits a $2.7\sigma$ excess over the SM expectation, stimulating some 
theoretical interest. Conversely, the vector mode $B\to K^{*}\nu\bar\nu$ has 
not been observed yet; the existing experimental upper bound remains 
especially relevant for constraining right-handed currents.

\begin{figure}[t]
  \centering
  \includegraphics[width=0.92\linewidth]{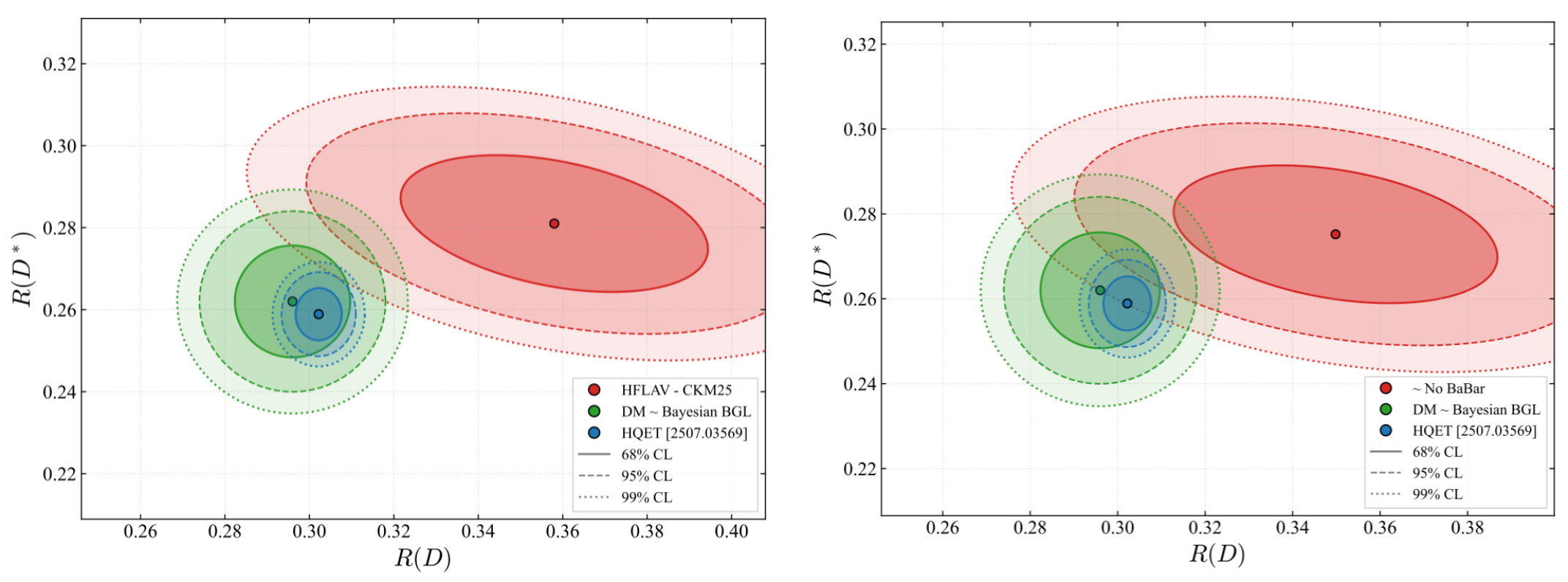}
  \caption{The $R(D)$-$R(D^{*})$ plane. SM predictions obtained with updated
    $b\to c$ FF treatments -- the DM~\cite{Martinelli:2021onb,Martinelli:2023fwm}\,/\,Bayesian BGL~\cite{Bordone:2024weh} approach
    (green) and an HQET-based parametrisation (blue, Ref.~\cite{Bordone:2025jur}) --
    are compared with the experimental average (red). \emph{Left}: HFLAV
    CKM25 average. \emph{Right}: average obtained leaving out the BaBar
    measurement. The updated predictions are mutually consistent and show no
    serious tension with data, robustly against the removal of BaBar.}
  \label{fig:rdrds}
\end{figure}

Finally, the charged-current ratios $R(D)$ and $R(D^{*})$ have historically constituted 
one of the most prominent arenas for LFU tensions. However, recent years have 
witnessed a significant evolution in the treatment of the underlying hadronic 
FFs. Incorporating the DM approach~\cite{Martinelli:2021onb,Martinelli:2023fwm} or  unitary Bayesian implementations of the 
BGL parameterization~\cite{Bordone:2024weh} -- both 
relaxing overly restrictive heavy-quark symmetry assumptions -- as well as updated HQET descriptions~\cite{Bordone:2025jur}, the resulting SM 
predictions have shifted upward toward the experimental data. Consequently, no 
severe tension survives, a conclusion that remains robust even when the BaBar 
measurement is excluded from the world average (Fig.~\ref{fig:rdrds}). As of 
today, the long-standing anomalies in LFU tests of $B$ decays seem to have largely 
faded, extending this statistical consistency of the SM to the $b\to c\ell\nu$ sector.

The precision flavour programme remains exceptionally exciting: only a naive 
interpretation of the current bounds would imply that the NP scale is entirely 
out of reach. Accidental symmetries such as LFU offer a unique, highly 
sensitive setup to stress-test the SM and, acting as a lamppost, to guide the 
search for new degrees of freedom. At present, LFU ``looks good'' with no 
significant tension observed in semileptonic $B$ decays -- a structural 
agreement that itself demands a deeper theoretical explanation, such as flavour 
alignment or screening mechanisms. With the accumulation of more data from 
Belle~II and LHCb, paired with parallel non-perturbative progress on the 
lattice, global analyses will soon be poised to clarify the nature of the 
residual $\Delta C_{9,U}$ shift, definitively disentangling an uncalculated 
hadronic origin from a genuinely universal NP contribution. Should a 
significant mass gap lie above the electroweak scale, the SMEFT will remain 
our most powerful tool, and flavour physics one of the most sensitive 
probes of BSM physics~\cite{deBlas:2025xhe}.

\clearpage

%%%%%%%%%%%%%%%%%%%%%%%%%%%%%%%%%%%%%%%%%
% Future 
%%%%%%%%%%%%%%%%%%%%%%%%%%%%%%%%%%%%%%%%%

\section{Present data taking and future}
\label{sec:future}

%This section cover the status of the present datataking at \belletwo and \lhcb. The future planned upgrades are discussed.
Both \belletwo and \lhcb are currently in active data-taking phases. \belletwo is accumulating luminosity at SuperKEKB towards its target of $50~\text{ab}^{-1}$, while \lhcb is completing its Run3 campaign with the upgraded detector. These larger samples will directly reduce the uncertainties on many key observables. However, to fully exploit them, requires advances in simulation, background modelling, and theoretical inputs. This section summarises the present status of the data-taking, outlines the planned detector upgrades and their expected physics impact.

\subsection{\belletwo}

{\it Vidya Sagar Vobbilisetti}\\

\belletwo operates at the SuperKEKB asymmetric-energy $e^+e^-$ collider, running at
or close to the \FourS resonance, and has been collecting data since 2018. The
nano-beam collision scheme, in which the beams are squeezed to a vertical size of
tens of nanometres at the interaction point, has enabled SuperKEKB to reach a world-record instantaneous luminosity of $5.1\times10^{34}~\mathrm{cm}^{-2}\mathrm{s}^{-1}$\sout{,
achieved on 27 December 2024}\,\footnote{ At the time of writing these proceedings, the instantaneous luminosity increased further, reaching the value of $5.32\times10^{34}~\mathrm{cm}^{-2}\mathrm{s}^{-1}$.} The ultimate goal of the programme is to accumulate an integrated luminosity of multiple tens of \invab, roughly two orders of magnitude beyond the sample available today, which will drive the precision of the semileptonic and leptonic observables reviewed in this Report.

\paragraph{Present data-taking status.}
The experiment collected $427~\invfb$ during its first data-taking period (Run~1,
2019--2022), after which SuperKEKB and \belletwo entered the first long shutdown
(LS1, 2022--2024). During LS1 a number of consolidation activities were carried
out, including the installation of a non-linear collimator to mitigate beam losses,
the replacement of the TOP micro-channel-plate photomultipliers, and the completion
of the two-layer pixel vertex detector (PXD). Data taking resumed with Run~2 in
2024, bringing the total recorded luminosity to about $617~\invfb$.\,\footnote{ At the time of writing these proceedings, the total integrated luminosity reached $914\,fb^{-1}$.} A dominant
operational concern is the beam-induced background~\cite{Belle2:UpgradeCDR}: in particular, high-dose sudden
beam losses (SBL) pose a risk to the innermost detectors. As a mitigation, the
fully installed and functional \sout{second PXD layer} PXD detector is currently kept switched off while
a fast beam-abort system, targeting a response time below $30~\mu$s, is being
developed.
\begin{figure}[!tb]
    \centering
    \includegraphics[width=0.75\textwidth]{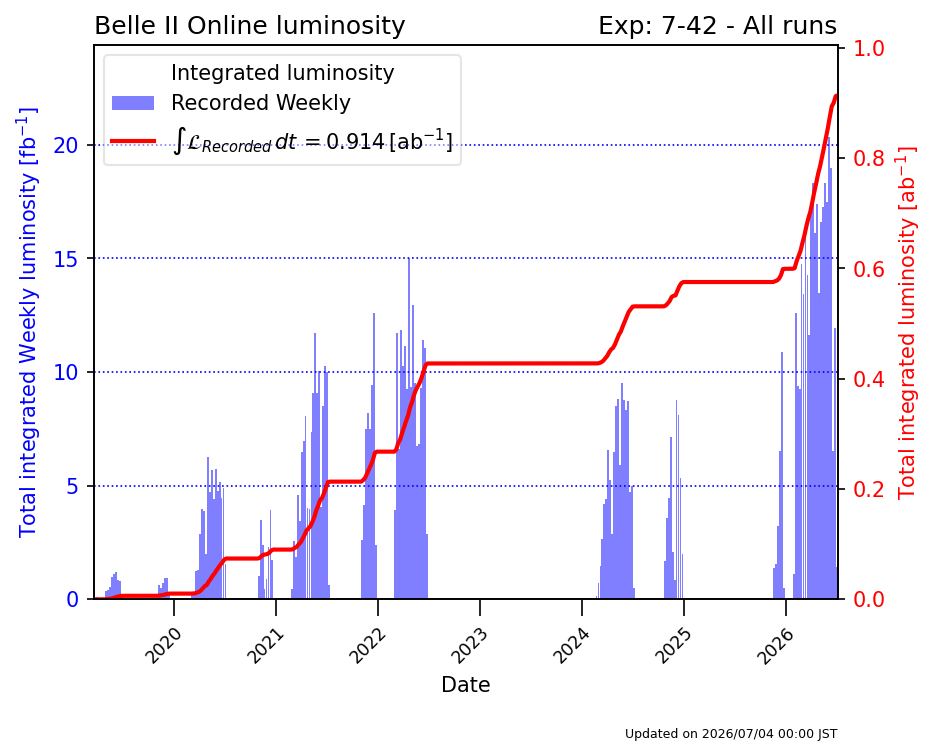}
    \caption{Total integrated luminosity recorded by the \belletwo experiment as a
    function of time, during the Run~1 period and
    the ongoing Run~2.}
    \label{fig:belle2lumi}
\end{figure}

\paragraph{Detector performance and software improvements}
Because semileptonic and leptonic analyses rely heavily on the reconstruction of
low-momentum tracks, of neutral particles, and on the tagging of the accompanying $B$ meson, considerable effort has been devoted to consolidating and improving the detector performance relative to Belle, in the harsher SuperKEKB environment . The
tracking system delivers roughly a factor of two better vertex resolution than Belle, with good data--simulation agreement for both fast tracks and the slow pions
that are crucial for $D^{*}$ reconstruction; graph-neural-network based track finding~\cite{Belle2:graph_based_track_finder}
is being explored to further improve robustness against machine backgrounds. Photon
reconstruction in the electromagnetic calorimeter, aided by dedicated classifiers
that suppress beam-background clusters and hadronic split-offs, reaches a
systematic uncertainty of about $0.3\%$ in the barrel, compared with $2\%$ at Belle.
Charged-particle identification, which combines information from all sub-detectors
into likelihood- and neural-network-based classifiers, provides $\pi$--$K$ and
lepton separation comparable to that of Belle and BaBar despite the more demanding
background conditions~\cite{Belle2:hadronID}; a data--simulation discrepancy in the
CDC $dE/dx$ response, traced to beam-injection backgrounds through its dependence on
the time since the last injection, has been largely mitigated by improved calibration
campaigns.

Of particular importance for measurements with neutrinos in the final state is
hadronic $B$-tagging, performed with the Full Event Interpretation (FEI)
algorithm~\cite{Keck:2018lcd}. Dedicated improvements to the modelling of hadronic
$B$ decays, together with looser pre-selections, have yielded about $40\%$ more
tagged $B$ mesons at fixed purity in the most recent reprocessing campaign,
directly increasing the effective sample size available to tagged analyses. The
performance of hadronic tagging remains ultimately limited by the imperfect
knowledge of hadronic $B$ decays, roughly half of the total hadronic branching
fraction is generated inclusively in simulation rather than from measured exclusive
modes, which motivates an ongoing \belletwo programme to remeasure key
$B\to D^{(*)}h$ channels, many of which are presently known only from measurements
that are decades old and carry large uncertainties. In parallel, inclusive and
semi-inclusive tagging strategies, including graph-neural-network approaches, are
being developed to recover efficiency for channels such as $B\to K^{(*)}\nu\bar\nu$.

\paragraph{Future upgrades of SuperKEKB.}
Reaching the target luminosity requires both an increase in beam currents and a
further squeezing of the beams. In the near term, before the second long shutdown,
these gains are pursued simultaneously but are limited by beam--beam effects and by
the associated rise in machine backgrounds, which is why the CDC and TOP readout
systems are already undergoing upgrades to cope with higher rates. Ultimately,
approaching the design vertical beta function of $\beta_y^{*}\simeq0.3$~mm requires a
modification of the interaction region during a second long shutdown (LS2),
presently foreseen around 2032. The planned interventions include a new QC1
final-focus magnet, moved about $100$~mm closer to the interaction point and using
Nb$_3$Sn in place of NbTi to sustain higher currents, a new compensation solenoid
near the interaction point, and a reduction of the chromatic $X$--$Y$ coupling
between the interaction point and QC1. With these upgrades SuperKEKB is projected to
integrate about $10~\invab$ by the time of LS2 and to reach a target of multiple tens
of \invab by the end of the following decade. The nano-beam scheme also establishes valuable
experience for future high-energy $e^+e^-$ colliders, with clear synergies with
FCC-ee.

\paragraph{Detector upgrades.}
The modification of the interaction region requires a full replacement of the
vertex detector (VXD~$\to$~VTX), which must also cope with the increased backgrounds
and occupancy at the target luminosity~\cite{Belle2:UpgradeCDR}. The upgrade replaces the
present DEPFET pixels with CMOS depleted monolithic active pixel sensors (MAPS): the
OBELIX sensor, derived from the TJ-Monopix2 prototype developed for the HL-LHC ATLAS
upgrade, provides a spatial resolution below $15~\mu$m, a time stamp of $100$~ns,
increased radiation tolerance, and a hit-rate capability of order $100~\mathrm{MHz/cm}^2$.
The baseline is an all-silicon, five-layer geometry with a reduced material budget,
of about $3\%~X_0$ in total. Full-simulation studies at conservative target-luminosity
backgrounds show that the VTX yields roughly $60\%$ higher slow-pion reconstruction
efficiency, together with improved impact-parameter resolution, which benefits both
the signal side, for instance in $R(D^{*})$ measurements, and the tag side of
semileptonic analyses, as well as about $30\%$ better decay-vertex resolution in
$B\to J/\psi\,K_S^0$, albeit with a modestly lower $K_S^0$ reconstruction
efficiency.

Beyond the vertex detector, the central drift chamber shows a larger-than-anticipated
gain loss with accumulated charge and may need to be replaced; complementary
technologies are being explored, including CMOS strip detectors for the supplemental
tracking volume and LGAD detectors providing precise time-of-flight and thus
additional particle identification. Other sub-systems are being consolidated against
higher backgrounds and to extend their lifetime: the electromagnetic calorimeter
electronics are being upgraded for faster shaping (from $500$~ns to $150$~ns) and
better background rejection, the TOP front-end electronics for lower-gain operation
and extended MCP-PMT lifetime, and the barrel KLM resistive plate chambers are moving
to an avalanche (proportional) readout to preserve efficiency at high hit rates. A
longer-term option to polarise one of the beams (``Chiral Belle'') is also under
study, opening additional sensitivity to electroweak and dark-sector physics.

Taken together, the dataset of multiple tens of \invab and the improved detector and tagging
performance will substantially sharpen the \belletwo determinations of $R(D^{(*)})$
and $R(X_{\tau/\ell})$, of the inclusive and exclusive extractions of \Vcb and \Vub,
and of the rare modes with missing energy such as $B\to K^{(*)}\nu\bar\nu$, providing
measurements that are highly complementary to those of \lhcb~\cite{10.1093/ptep/ptz106,SnowMass2022}.

\subsection{\lhcb}

{\it Matteo Palutan}\\

\begin{figure}[!tb]
    \centering
    \includegraphics[width=0.85\textwidth]{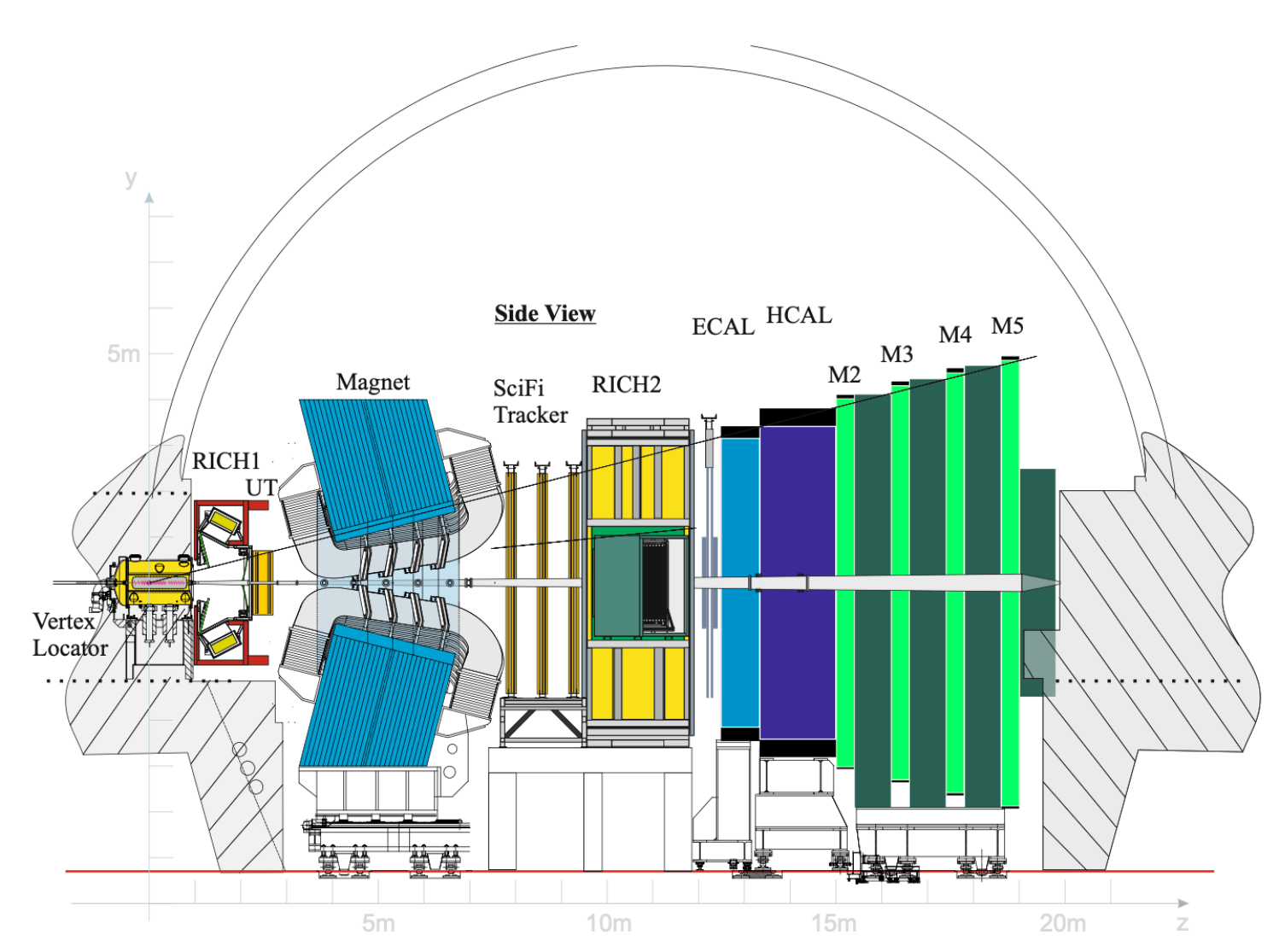}
    \caption{Side view of the LHCb detector as of Run 3 of the LHC.}
    \label{fig:lhcbdet}
\end{figure}

The very successful operation of LHCb during Run 1 and Run 2 of the LHC vindicated the concept and design of a dedicated heavy flavour physics experiment at a hadron collider.
The detector was able to run at an instantaneous luminosity of $4\times 10^{32}$~cm$^{-2}$s$^{-1}$, twice the design value,  and to collect  9~fb$^{-1}$ of data by the end of Run 2.
In order to be able to continue the LHCb physics programme, a first upgrade of the detector (Upgrade~I, see Fig.~\ref{fig:lhcbdet}) was approved in 2012, and brought in operation at the beginning of Run 3, in 2022~\cite{LHCb:U1}.

\begin{figure}[!tb]
    \centering
    \includegraphics[width=0.75\textwidth]{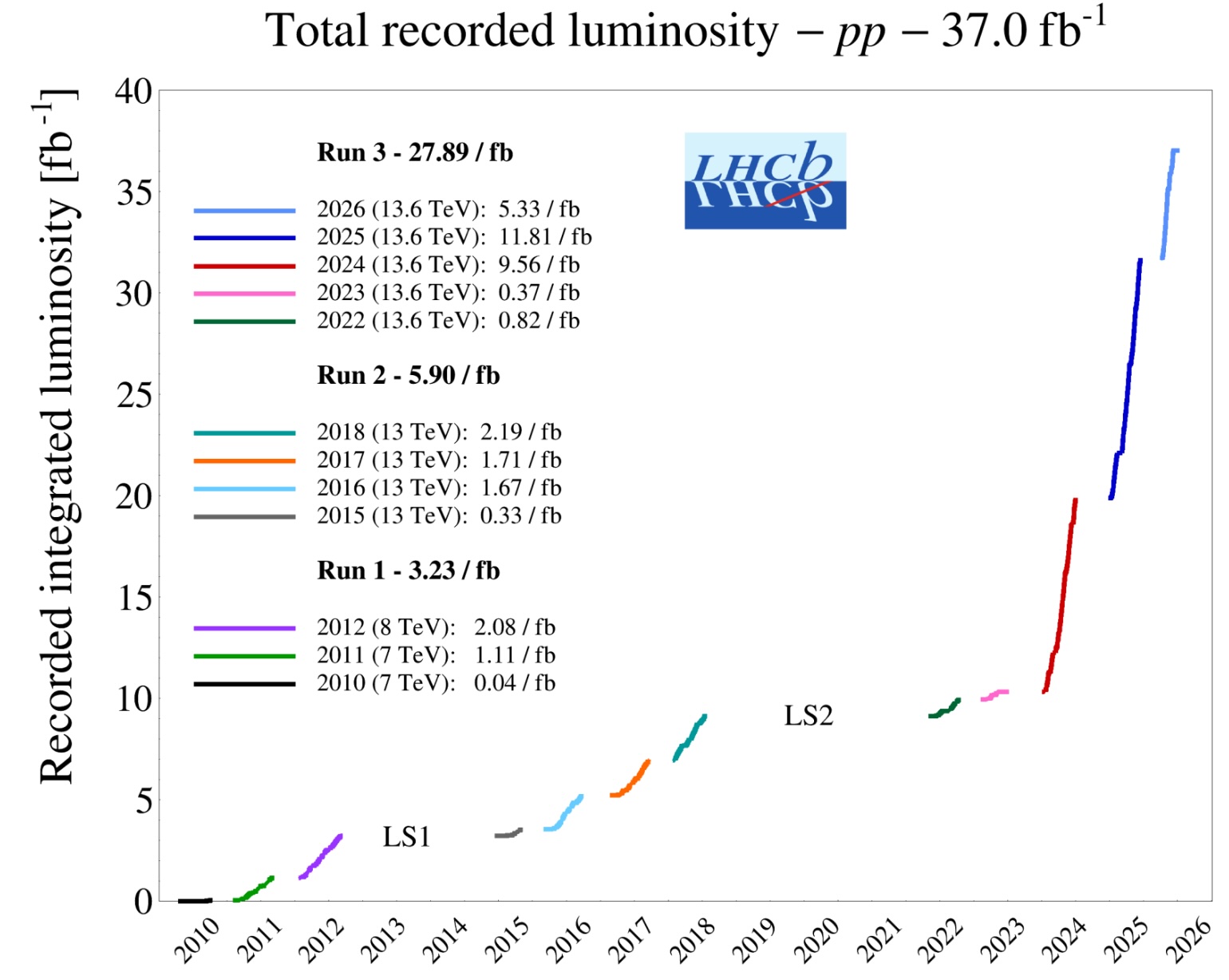}
    \caption{Total integrated luminosity accumulated by the LHCb experiment.}
    \label{fig:lhcblumi}
\end{figure}

The key concept of this upgrade was to remove the  bottleneck of the level 0 hardware trigger. By reading
out the full detector at the LHC bunch crossing rate of 40~MHz and implementing all trigger decisions
in software, it was possible to increase the luminosity without suffering a compensating loss in 
efficiency. By increasing the instantaneous luminosity by a factor of five, to $2\times 
10^{33}$~cm$^{-2}$s$^{-1}$, and improving the trigger efficiency for hadronic  modes by a factor of two, 
the annual yields in most channels increased by an order of magnitude as compared with Run 2. The target 
total integrated luminosity for the Upgrade~I phase was around 35~fb$^{-1}$ at the end of Run 3, which 
was very recently met by the experiment (Fig.~\ref{fig:lhcblumi}). This also paves the road for 
reaching the final goal for the Upgrade~I detector of integrating around 65~fb$^{-1}$ at the end of Run 4.

Further data taking with the Upgrade I detector beyond Run 4 will not be attractive, on account
of the excessive “data-doubling” time, and also due to the fact that many of its components will have 
reached the end of their natural lifespan in terms of radiation exposure.
There is therefore a strong motivation to perform  a second upgrade (Upgrade~II) of the detector~\cite{LHCb-TDR-023}, in order
to fully realise the flavour physics potential of the HL-LHC.
The Upgrade II~is proposed for installation during LHC long shutdown 4, presently scheduled in 2034, 
and it is expected to take data at a maximum instantaneous luminosity of $1.0\times 10^{34}$~cm$^{-2}$s$^{-1}$, 
with the target of accumulating $\sim 300$~fb$^{-1}$ up to end of the HL-LHC era. The flavour physics data sample 
will at this point be significantly larger than that of any other planned experiment, and will lead to improvements 
in the precision of a large number of key observables without being limited by systematic uncertainties.

A subset of the sensitivities expected for key flavour observables in Upgrade~I and Upgrade~II are presented in
Table~\ref{tab:uncerts}, compared to what has been obtained with data accumulated during Run 1 and Run 2.
These includes the magnitudes and phases of CKM elements, the charm CP violation parameters, and various 
observables related to rare decays. For all the extrapolations the basic assumptions are to maintain similar  
performance as for the original LHCb detector, but with the increased luminosity of Upgrade~I and Upgrade~II, 
respectively. This has been achieved already during Run 3 by the Upgrade~I detector.
\begin{table}[!tb]
  \centering
  \resizebox{\textwidth}{!}{
  \begin{tabular}{l c@{\hspace{-3mm}}c c c c }
    \hline
    Observable & \multicolumn{2}{c}{Current LHCb} & \multicolumn{2}{c}{Upgrade~I} & \upgradetwo\ \\
    &  \multicolumn{2}{c}{(up to $9\invfb$)} & ($23\invfb$) & ($50\invfb$) & ($300\invfb$) \\
    \hline
    \multicolumn{6}{l}{\textbf{\underline{CKM and semileptonics}}} \\
    $\;\;\g$ ({\small $B \to DK$, \etc}) & $2.8\degrees$ &    & 1.3\degrees & 0.8\degrees  & 0.3\degrees \\
    %$\;\;\phis$ ({\small $\Bs \to \jpsi \phi$}) & $32\mrad$ & \cite{LHCb-PAPER-2019-013}      & $14\mrad$     & $10\mrad$    & $4\mrad$       \\
    $\;\;\phis$ ({\small $\Bs \to \jpsi \phi$}) & $20\mrad$ &      & $12\mrad$     & $8\mrad$    & $3\mrad$       \\
    $\;\;|\Vub|/|\Vcb|$ ({\small $\Lb \to p\mun\neumb$, \etc}) & $6\%$ &           & 3\%         & 2\%        & 1\%          \\
    $\;\;R(D)-R(\Dstar)$    & 14\%,~6.0\%  &  & ---  &  4.4\%,~3.2\% & 3.3\%,~3.0\% \\
%    $\;\;a_{\rm sl}^d$ ({\small $\Bz\to\Dm\mup\neum$})   & $36 \times 10^{-4}$ & \cite{LHCb-PAPER-2014-053} & $8 \times 10^{-4}$ & $5 \times 10^{-4}$ & $2 \times 10^{-4}$ \\
%    $\;\;a_{\rm sl}^s$ ({\small $\Bs\to\Dsm\mup\neum$})   & $33 \times 10^{-4}$ & \cite{LHCb-PAPER-2016-013} & $10 \times 10^{-4}$ & $7 \times 10^{-4}$ & $3 \times 10^{-4}$ \\
    %\hline
    \multicolumn{6}{l}{\textbf{\underline{Charm}}} \\
    $\;\;\Delta A_{\CP}$  ({\small $\Dz \to \Kp\Km, \pip\pim$})  & $29 \times 10^{-5}$ &  & $13\times 10^{-5}$ & $8\times 10^{-5}$ & $3.3\times 10^{-5}$ \\
    $\;\;A_\Gamma$  ({\small $\Dz \to \Kp\Km, \pip\pim$})  & $11 \times 10^{-5}$ &  & $5\times 10^{-5}$ & $3.2\times 10^{-5}$ & $1.2\times 10^{-5}$ \\
    $\;\;\Delta x$  ({\small $\Dz \to \KS\pip\pim$})  & $18 \times 10^{-5}$ & & $6.3 \times 10^{-5}$ & $4.1 \times 10^{-5}$ & $1.6 \times 10^{-5}$ \\
    %\hline
    \multicolumn{6}{l}{\textbf{\underline{Rare decays}}} \\
    $\;\;\BF{(\Bd\to\mumu)}/\BF{(\Bs\to\mumu)}$\hspace{-6mm} & 69\% &  & 41\% & 27\% & 11\% \\
    $\;\;S_{\mu\mu}$  ({\small $\Bs\to\mumu$})     & ---    &        & ---         & ---        & 0.2 \\
    $\;\;A_{\rm T}^{(2)}$  ({\small $\Bz \to \Kstarz\epem$})  & 0.10 &  & 0.060 & 0.043 & 0.016 \\
%    $\;\;A_{\rm T}^{\rm Im}$  ({\small $\Bz \to \Kstarz\epem$})  & 0.10 & \cite{LHCb-PAPER-2020-020} & 0.060 & 0.043 & 0.016 \\
%    $\;\;{\cal A}_{\phi \gamma}^{\Delta\Gamma} (\Bs\to \phi \gamma)$  & $^{+0.41}_{-0.44}$ & \cite{LHCb-PAPER-2019-015} & 0.124 & 0.083 & 0.033 \\
    $\;\;S_{\phi \gamma}(\Bs\to \phi \gamma)$  & 0.32 &   & 0.093 & 0.062 & 0.025 \\
     $\;\;\alpha_{\gamma}(\Lb\to \Lz \gamma)$  & $^{+0.17}_{-0.29}$ &  & 0.148 & 0.097 & 0.038 \\
%    \multicolumn{6}{l}{\textbf{\underline{Lepton Universality Tests}}} \\
%     $\;\;R_{K}$  ({\small $\Bp \to \Kp\ell^+\ell^-$})         &  0.042 & \cite{LHCb-PAPER-2022-045, LHCb-PAPER-2022-046} & $x$       &  $x$      &  $x$  \\
%    $\;\;R_{\Kstar}$  ({\small $\Bz \to \Kstarz\ell^+\ell^-$})     & 0.072   &  \cite{LHCb-PAPER-2022-045, LHCb-PAPER-2022-046}   & x       &  $x$       &  $x$  \\
%    $\;\;R(\Dstar)$  ({\small $\Bz \to \Dstarm \ell^+ \neul$})    & 0.026  & \cite{LHCb-PAPER-2015-025,LHCb-PAPER-2017-027} & 0.007 & 0.005 & 0.002 \\
    \hline
  \end{tabular}
  }
  \caption{
    Anticipated uncertainties at future upgrades of LHCb for some key flavour observables, modified and updated from Refs.~\cite{LHCb-PII-Physics,LHCb-TDR-023}, as a function of the total integrated luminosity.
    Upgrade~I projections at the end of Run 3 (Run 4) are given for 23~fb$^{-1}$ (50~fb$^{-1}$), which is (may be) surpassed by the present  (expectation), while the Upgrade~II projections are given for 300~fb$^{-1}$. The projections for $R(D)-R(\Dstar)$ are obtained from Ref.\cite{ATLAS:2025lrr}.
  }
  \label{tab:uncerts}
\end{table}

\clearpage

\section{Prospects}
\label{sec:prospects}

{\it Marco Fedele and Marcello Rotondo}\\

The semileptonic decays of $b$-hadrons continue to provide one of the richest
laboratories for testing the flavour sector of the SM and for searching,
indirectly, for physics beyond it. This Report has reviewed the current
experimental and theoretical status of charged-current $b\to c\ell\nu$ and
$b\to u\ell\nu$ transitions, of the neutral-current $b\to s\ell^+\ell^-$ and
$b\to s\nu\bar\nu$ processes, and of the LFU observables that probe both sectors. In
closing, we collect the main messages that emerge from the various
contributions and outline the directions along which further progress is
expected in the coming years.

\paragraph{The precision frontier}
The long-standing tension between inclusive and exclusive determinations of
\Vcb and \Vub remains, at present, one of the most persistent open issues in
flavour physics. On the exclusive side, the availability of three independent
LQCD calculations of the $B\to D^*\ell\nu$ FFs at non-zero recoil, combined
with more flexible BGL-type parametrizations and with dedicated
heavy-quark-expansion treatments of the full set of $\bar B_{(s)}\to
D_{(s)}^{(*)}$ FFs including, for the first time, tensor currents, has
considerably sharpened the theoretical picture. The various LQCD inputs are
found to be mutually compatible, and global fits combining LQCD, LCSR and
experimental differential distributions now return values of \Vcb consistent
with each other at the level of a percent or better. On the inclusive side,
the theoretical description of $B\to X_c\ell\nu$ decays has been extended to
NNLO in several observables, while the perturbative treatment of $B\to
X_u\ell\nu$ decays -- including the fully differential NNLO partonic decay and
the $O(\alpha_s/m_b^2)$ power corrections -- has reached a new level of
maturity. Taken together, these developments have narrowed, without yet fully
resolving, the residual tension between the two determinations of both \Vcb
and \Vub. The next few years should see this issue clarified, thanks to a
combination of larger experimental datasets from Belle~II and LHCb, further
improvements in the LQCD FF calculations at non-zero recoil and in the
heavy-to-light channels, and, notably, the emergence of genuinely
non-perturbative inclusive-rate calculations from the lattice. The latter,
pioneered for inclusive $D_s$ decays and now being extended to $B_s$ decays,
offer for the first time the prospect of an independent, first-principles
cross-check of the HQE itself, rather than a mere consistency test between
different extractions.

\paragraph{The discovery frontier: charged currents}
The status of the ratios $R(D)$ and $R(D^*)$ illustrates how sensitive the
interpretation of these observables remains to the treatment of the
underlying $b\to c$ hadronic FFs. Depending on the theoretical framework
adopted for the SM prediction, the picture that emerges from the current data
differs appreciably: analyses based on the DM approach find the residual
discrepancy reduced to around $2\sigma$, while more traditional
HQET-based treatments still report a tension at the $3\sigma$ level or above,
largely driven by $R(D)$ and by the semitauonic $\tau$-observables such as
$F_{L,\tau}$ and $A_{FB,\tau}$. These different approaches do not yet agree on
how much of the earlier tension should be attributed to underestimated
theoretical uncertainties in the FFs, as opposed to a genuine LFU-violating
effect, and reconciling them is one of the more pressing near-term goals of
the semileptonic precision programme. Belle~II's most recent hadronic-tag
measurement of $R(D)$ and $R(D^*)$, finds $R(D)$
about $2\sigma$ above the SM expectation while $R(D^*)$ remains compatible
with it, with the two-dimensional result agreeing with the current world
average at the $1.3\sigma$ level; this is broadly consistent with both the
more optimistic and the more conservative theoretical pictures, and does not
by itself resolve the disagreement between them. Clarifying this situation
will require, on the theory side, a better understanding of the differences
between LQCD FF determinations at non-zero recoil -- in particular the
so-called slope-of-$F_1$ issue -- and, on the experimental side, first
measurements of observables such as $A_{FB,\tau}$, which the SM predicts to be strongly anti-correlated with $R(D^*)$ and which are not yet available. In parallel, new observables largely insensitive to this FF ambiguity continue to be developed: full angular analyses of $B_{(s)}\to D_{s}^{(*)}\ell\nu$ decays,
model-independent extractions of the angular coefficient functions $I_i(q^2)$,
and dedicated searches for \CP-violating triple-product asymmetries, which
probe combinations of NP couplings not directly accessible through rate
measurements and are subject to comparatively small hadronic systematics. The
LHCb Run~2 and Run~3 datasets, together with Belle~II's continuing
accumulation of luminosity, will allow several of these analyses to reach a
level of precision competitive with, and complementary to, the traditional
$R(D^{(*)})$ measurements.

\paragraph{The discovery frontier: neutral currents}
The situation in $b\to s\ell^+\ell^-$ transitions shares a similar character:
here too the dominant obstacle is theoretical rather than statistical, though
the nature of the ambiguity differs. Rather than depending on the choice of
FF parametrization, as for $R(D^{(*)})$, the interpretation of $b\to
s\ell^+\ell^-$ data hinges on the long-distance, non-local contributions
generated by charm-loop rescattering, which remain the dominant source of
uncertainty in the observed tensions in angular observables and branching
fractions, most notably in $B\to K^*\mu^+\mu^-$ and $B_s\to\phi\mu^+\mu^-$.
Global fits to the available data find no significant evidence for a
genuinely LFU-violating contribution, while a residual, lepton-flavour-universal
shift in $C_9$ persists whose origin -- new universal physics or
underestimated hadronic effects -- cannot yet be disentangled on the basis of
the fit alone. Recent studies of charm-rescattering effects through triangle
topologies, and of the $q^2$- and polarization-dependence of the extracted
shift in $C_9$, both point towards a non-negligible role for long-distance QCD
dynamics, a conclusion reinforced by the emerging picture from angular
observables such as $S_{7,8}$. Settling this question requires, on the one
hand, tighter control of the local FFs entering the SM predictions, through
further LQCD input at high $q^2$ and a better assessment of duality
violations in LCSR at low $q^2$, and on the other hand a genuinely
first-principles treatment of the non-local amplitudes. First steps in this
direction, based on lattice spectral-function reconstruction techniques
applied to $b\to s\ell^+\ell^-$ amplitudes, have already been demonstrated in
proof-of-principle studies and constitute one of the most promising
theoretical developments discussed in this Report. In parallel, the
theoretically cleaner di-neutrino modes $B\to K^{(*)}\nu\bar\nu$ offer an
increasingly important complementary handle: the mild excess recently
reported by Belle~II in $B^+\to K^+\nu\bar\nu$, together with the still
unobserved vector mode, will be closely watched as larger datasets become
available.

\paragraph{Methodological advances}
Beyond the specific observables discussed above, this Report has highlighted
a number of methodological developments that are expected to shape the field
over the next several years. On the lattice side, these include: the
systematic inclusion of tensor FFs and their unitarity bounds in the
heavy-quark expansion; the first fully non-perturbative calculations of
inclusive semileptonic decay rates, both for charm and, prospectively, for
beauty mesons; and continued progress towards a lattice treatment of
electromagnetic radiative corrections and isospin-breaking effects in
semileptonic decays, building on the successes already achieved for purely
leptonic modes. On the experimental and phenomenological side, they include:
full multidimensional angular analyses exploiting the growing LHCb and
Belle~II datasets; model-independent extractions of FFs and angular
coefficients that minimize reliance on specific parametrizations; a
quantitative assessment of the size and reliability of the QED radiative
corrections currently modelled via PHOTOS, and of the residual theoretical
uncertainty this introduces into LFU ratios at the percent level; and
increasingly sophisticated statistical and machine-learning tools, from
neural-network parametrizations of inclusive shape functions to global fits
combining CKM, HQE and NP degrees of freedom simultaneously.

\paragraph{Experimental outlook}
Both Belle~II and LHCb are, at the time of writing, in the midst of extensive
data-taking campaigns that will substantially increase the available
statistics over the coming decade. Belle~II is progressing towards its design
integrated luminosity, which will improve the precision of $R(D^{(*)})$,
$R(X_{\tau/\ell})$, and of the inclusive and exclusive determinations of \Vcb
and \Vub, while also enabling first measurements of increasingly rare decay
modes with missing energy. LHCb, having recently surpassed the target
integrated luminosity for the Upgrade~I phase, is expected to accumulate
around 65\,fb$^{-1}$ by the end of Run~4, with the Upgrade~II detector,
targeting some 300\,fb$^{-1}$ over the HL-LHC era, poised to push measurements
of CKM parameters, LFU ratios, and rare-decay observables to a precision
ultimately limited by theory rather than by statistics. This confluence of
larger datasets and improving theoretical control -- across LQCD, HQE,
LCSR, and QED radiative corrections -- gives good reason to expect that many
of the open questions surveyed in this Report, from the exclusive-inclusive
\Vcb puzzle to the long-distance dynamics of $b\to s\ell^+\ell^-$ decays, will
be substantially clarified within the next several years.

\paragraph{Closing remarks}
Semileptonic $b$-hadron decays sit at the intersection of the precision and
discovery programmes of flavour physics: the same processes that determine
fundamental SM parameters also provide some of the most sensitive tests of
its accidental symmetries. The progress summarized in this Report --
achieved through close collaboration between the experimental and theoretical
communities, and between the continuum and lattice approaches to hadronic
dynamics -- illustrates both how much ground has been covered and how much
remains to be done. As larger datasets and sharper theoretical tools become
available, semileptonic $b$-hadron decays will remain, for years to come, one
of the most powerful windows onto the flavour structure of the Standard Model
and a sensitive probe of what may lie beyond it.

%%%%%%%%%%%%%%%%%%%
% THIS IS THE END
%%%%%%%%%%%%%%%%%%%

%\appendix
%\section{This is an appendix}
%Please always give a title also for appendices.

\acknowledgments

We thanks all the participants to the workshop,
Adlene  Hicheur,
Alejandro       Vaquero,
Alessandro      Barone,
Alfredo Stanzione,
Anna    Lupato,
Anna-Maria      Heyn,
Anton   Poluektov,
Antonio Smecca,
Arianna Tinari,
Bernat  Capdevila,
Biljana Mitreska,
Davide  Giusti,
Elisa   Manoni,
Elisa   Minucci,
Federico        Manganella,
Federico        Mescia,
Flavio  Archilli,
Francesco       Sanfilippo,
Fulvia  De Fazio,
Gael    Finauri,
Giovanni        Gaudino,
Giuseppe        Finocchiaro,
Giuseppe        Gagliardi,
Guido   Martinelli,
J. Tobias       Tsang,
Joan    Ruiz-Vidal,
Lorenzo Paolucci,
Luca    Silvestrini,
Lucine  Tabatt,
Ludovico        Vittorio,
Lukas   Allwicher,
Léa     Dreyfus,
Marcello        Rotondo,
Marco   Fedele,
Maria Laura     Piscopo,
Martin  Jung,
Matteo  Cabibbo,
Matteo  Fael,
Matteo  Palutan,
Mauro   Valli,
Michel  De Cian,
Nico    Gubernari,
Paolo   Gambino,
Patrizia        De Simone,
Prodromos       Polychroniadis,
Ricardo Vazquez Gomez,
Roberto Di Palma,
Silvano Simula,
Tommy   Martinov,
Vidya Sagar     Vobbilisetti,
Vittorio        Lubicz,
Zachary Polonsky,
Zhi     Hu,
for actively contributing to the interesting discussions during the sessions.

The workshop associated to this report, https://agenda.infn.it/event/49469/, has been part of the Theoretical Phenomenology Visiting Institute @ LNF, running from 16 to 27 February 2026, and supported by CSN4 to promote the participation of early-career scientists. The local organizers are also partially supported by the PRIN 2022 project
"Semileptonic B decays at the junction of experiment and theory", funded by the Italian Ministry of University and Research (MUR) and the European Union (EU). PRIN 2022, CUP I53D23001320006.\\ 
We thank the LNF Research Division for support.
This work has  also been carried out within the INFN projects APINE, ENP, LQCD123, SPIF, TASP.

%\paragraph{Note added.} Notes added after the paper has been written.

% Bibliography

%% [A] Recommended: using JHEP.bst file

\bibliographystyle{JHEP}
\bibliography{biblio.bib}

\end{document}

%% file: symbols-def.tex
\usepackage{ifthen} 
\newboolean{uprightparticles}
\setboolean{uprightparticles}{false} %Set true for upright particle symbols
\usepackage{xspace} 
\usepackage{upgreek}

\def\lhcb   {\mbox{LHCb}\xspace}

\def\cms    {\mbox{CMS}\xspace}

\def\babar  {\mbox{BaBar}\xspace}
\def\belle  {\mbox{Belle}\xspace}
\def\belletwo {\mbox{Belle~II}\xspace}

\def\upgradetwo {\mbox{Upgrade~II}\xspace}

\def\MagUp {\mbox{\em Mag\kern -0.05em Up}\xspace}

\ifthenelse{\boolean{uprightparticles}}%
{
 
 \def\Pgamma      {\ensuremath{\upgamma}\xspace}

 \def\Pmu         {\ensuremath{\upmu}\xspace}                 
 \def\Pnu         {\ensuremath{\upnu}\xspace}                 
                  
 \def\Ppi         {\ensuremath{\uppi}\xspace}

 \def\Ptau        {\ensuremath{\uptau}\xspace}

 \def\Ppsi        {\ensuremath{\uppsi}\xspace}

 \def\PDelta      {\ensuremath{\Delta}\xspace}                 
 \def\PXi         {\ensuremath{\Xi}\xspace}                 
 \def\PLambda     {\ensuremath{\Lambda}\xspace}                 
 \def\PSigma      {\ensuremath{\Sigma}\xspace}                 
 \def\POmega      {\ensuremath{\Omega}\xspace}                 
 \def\PUpsilon    {\ensuremath{\Upsilon}\xspace}
 \let\oldPi\Pi
 \def\PPi         {\ensuremath{\oldPi}\xspace}

 \def\PB      {\ensuremath{\mathrm{B}}\xspace}                 
 \def\PD      {\ensuremath{\mathrm{D}}\xspace}                 
 \def\PJ      {\ensuremath{\mathrm{J}}\xspace}                 
 \def\PK      {\ensuremath{\mathrm{K}}\xspace}                 
 \def\Pb      {\ensuremath{\mathrm{b}}\xspace}                 
 \def\Pc      {\ensuremath{\mathrm{c}}\xspace}                 
                  
 \def\Pe      {\ensuremath{\mathrm{e}}\xspace}                 
 \def\Pp      {\ensuremath{\mathrm{p}}\xspace}                 
 \def\Pq      {\ensuremath{\mathrm{q}}\xspace}                 
 \def\Ps      {\ensuremath{\mathrm{s}}\xspace}                 
                  
 \def\Pu      {\ensuremath{\mathrm{u}}\xspace}                 
 \def\thebaroffset{0.0em}
}
{
 
 \def\Pgamma      {\ensuremath{\gamma}\xspace}

 \def\Pmu         {\ensuremath{\mu}\xspace}                 
 \def\Pnu         {\ensuremath{\nu}\xspace}                 
                  
 \def\Ppi         {\ensuremath{\pi}\xspace}

 \def\Ptau        {\ensuremath{\tau}\xspace}

 \def\Ppsi        {\ensuremath{\psi}\xspace}                 
                  
 \mathchardef\PDelta="7101
 \mathchardef\PXi="7104
 \mathchardef\PLambda="7103
 \mathchardef\PSigma="7106
 \mathchardef\POmega="710A
 \mathchardef\PUpsilon="7107
 \mathchardef\PPi="7105
 \def\PB      {\ensuremath{B}\xspace}                 
 \def\PD      {\ensuremath{D}\xspace}                 
 \def\PJ      {\ensuremath{J}\xspace}                 
 \def\PK      {\ensuremath{K}\xspace}                 
 \def\Pb      {\ensuremath{b}\xspace}                 
 \def\Pc      {\ensuremath{c}\xspace}                 
                  
 \def\Pe      {\ensuremath{e}\xspace}                 
 \def\Pp      {\ensuremath{p}\xspace}                 
 \def\Pq      {\ensuremath{q}\xspace}                 
 \def\Ps      {\ensuremath{s}\xspace}                 
                  
 \def\Pu      {\ensuremath{u}\xspace}                 
 \def\thebaroffset{0.18em}
}
\newcommand{\offsetoverline}[2][\thebaroffset]{\kern #1\overline{\kern -#1 #2}}%

\makeatletter
\ifcase \@ptsize \relax% 10pt
  \newcommand{\miniscule}{\@setfontsize\miniscule{4}{5}}% \tiny: 5/6
\or% 11pt
  \newcommand{\miniscule}{\@setfontsize\miniscule{5}{6}}% \tiny: 6/7
\or% 12pt
  \newcommand{\miniscule}{\@setfontsize\miniscule{5}{6}}% \tiny: 6/7
\fi
\makeatother

\DeclareRobustCommand{\optbar}[1]{\shortstack{{\miniscule (\rule[.5ex]{1.25em}{.18mm})}
  \\ [-.7ex] $#1$}}

\def\en         {{\ensuremath{\Pe^-}}\xspace}   % electron negative (\em is taken)
\def\ep         {{\ensuremath{\Pe^+}}\xspace}

\def\epem       {{\ensuremath{\Pe^+\Pe^-}}\xspace}

\def\mup        {{\ensuremath{\Pmu^+}}\xspace}
\def\mun        {{\ensuremath{\Pmu^-}}\xspace} % muon negative (\mum is taken)

\def\mumu       {{\ensuremath{\Pmu^+\Pmu^-}}\xspace}

\def\taup       {{\ensuremath{\Ptau^+}}\xspace}
\def\taum       {{\ensuremath{\Ptau^-}}\xspace}

\def\ellm       {{\ensuremath{\ell^-}}\xspace}

\def\neu        {{\ensuremath{\Pnu}}\xspace}
\def\neub       {{\ensuremath{\overline{\Pnu}}}\xspace}
\def\neum       {{\ensuremath{\neu_\mu}}\xspace}
\def\neumb      {{\ensuremath{\neub_\mu}}\xspace}

\def\neutb      {{\ensuremath{\neub_\tau}}\xspace}

\def\neulb      {{\ensuremath{\neub_\ell}}\xspace}
\def\g      {{\ensuremath{\Pgamma}}\xspace}

\def\quark     {{\ensuremath{\Pq}}\xspace}
\def\quarkbar  {{\ensuremath{\overline \quark}}\xspace}

\def\uquark    {{\ensuremath{\Pu}}\xspace}

\def\squark    {{\ensuremath{\Ps}}\xspace}

\def\cquark    {{\ensuremath{\Pc}}\xspace}

\def\bquark    {{\ensuremath{\Pb}}\xspace}
\def\bquarkbar {{\ensuremath{\overline \bquark}}\xspace}

\def\pion   {{\ensuremath{\Ppi}}\xspace}
\def\piz    {{\ensuremath{\pion^0}}\xspace}
\def\pip    {{\ensuremath{\pion^+}}\xspace}
\def\pim    {{\ensuremath{\pion^-}}\xspace}
\def\pipm   {{\ensuremath{\pion^\pm}}\xspace}

\def\kaon    {{\ensuremath{\PK}}\xspace}
\def\KorKbar {\kern \thebaroffset\optbar{\kern -\thebaroffset \PK}{}\xspace}
\def\Kz      {{\ensuremath{\kaon^0}}\xspace}

\def\Kp      {{\ensuremath{\kaon^+}}\xspace}
\def\Km      {{\ensuremath{\kaon^-}}\xspace}
\def\Kpm     {{\ensuremath{\kaon^\pm}}\xspace}

\def\KS      {{\ensuremath{\kaon^0_{\mathrm{S}}}}\xspace}

\def\Kstarz  {{\ensuremath{\kaon^{*0}}}\xspace}

\def\Kstarp  {{\ensuremath{\kaon^{*+}}}\xspace}

\def\D       {{\ensuremath{\PD}}\xspace}

\def\DorDbar {\kern \thebaroffset\optbar{\kern -\thebaroffset \PD}\xspace}
\def\Dz      {{\ensuremath{\D^0}}\xspace}

\def\Dp      {{\ensuremath{\D^+}}\xspace}
\def\Dm      {{\ensuremath{\D^-}}\xspace}

\def\DpDm    {\ensuremath{\Dp {\kern -0.16em \Dm}}\xspace}
\def\Dstar   {{\ensuremath{\D^*}}\xspace}

\def\Dstarp  {{\ensuremath{\D^{*+}}}\xspace}

\def\Ds      {{\ensuremath{\D^+_\squark}}\xspace}

\def\Dspm    {{\ensuremath{\D^{\pm}_\squark}}\xspace}

\def\Dss     {{\ensuremath{\D^{*+}_\squark}}\xspace}

\def\B       {{\ensuremath{\PB}}\xspace}
\def\Bbar    {{\ensuremath{\offsetoverline{\PB}}}\xspace}
\def\BBb       {{\ensuremath{B{\Bbar}}}\xspace}

\def\BorBbar {\kern \thebaroffset\optbar{\kern -\thebaroffset \PB}\xspace}
\def\Bz      {{\ensuremath{\B^0}}\xspace}
\def\Bzb     {{\ensuremath{\Bbar{}^0}}\xspace}
\def\Bd      {{\ensuremath{\B^0}}\xspace}

\def\BdorBdbar {\kern \thebaroffset\optbar{\kern -\thebaroffset \Bd}\xspace}
\def\Bu      {{\ensuremath{\B^+}}\xspace}

\def\Bp      {{\ensuremath{\Bu}}\xspace}

\def\Bs      {{\ensuremath{\B^0_\squark}}\xspace}

\def\BsorBsbar {\kern \thebaroffset\optbar{\kern -\thebaroffset \Bs}\xspace}
\def\Bc      {{\ensuremath{\B_\cquark^+}}\xspace}

\def\jpsi     {{\ensuremath{{\PJ\mskip -3mu/\mskip -2mu\Ppsi}}}\xspace}
\def\psitwos  {{\ensuremath{\Ppsi{(2S)}}}\xspace}

\def\Y#1S{\ensuremath{\PUpsilon{(#1S)}}\xspace}

\def\FourS {{\Y4S}\xspace}

\def\proton      {{\ensuremath{\Pp}}\xspace}

\def\Lz          {{\ensuremath{\PLambda}}\xspace}

\def\LorLbar     {\kern \thebaroffset\optbar{\kern -\thebaroffset \PLambda}\xspace}

\def\Lb           {{\ensuremath{\Lz^0_\bquark}}\xspace}

\def\BF         {{\ensuremath{\mathcal{B}}}\xspace}

\newcommand{\decay}[2]{\ensuremath{\mathinner{#1\!\to #2}}\xspace}

\def\to                 {\ensuremath{\rightarrow}\xspace}

\newcommand{\as}{{\ensuremath{\alpha_s}}\xspace}

\newcommand{\lqcd}{{\ensuremath{\Lambda_{\mathrm{QCD}}}}\xspace}
\def\qsq       {{\ensuremath{q^2}}\xspace}

\def\CP                {{\ensuremath{C\!P}}\xspace}

\def\Vub  {{\ensuremath{V_{\uquark\bquark}^{\phantom{\ast}}}}\xspace}
\def\Vcb  {{\ensuremath{V_{\cquark\bquark}^{\phantom{\ast}}}}\xspace}

\newcommand{\phis}{{\ensuremath{\phi_{\squark}}}\xspace}

\def\BdToKstmm    {\decay{\Bd}{\Kstarz\mup\mun}}

\def\bsll     {\decay{\bquark}{\squark \ell^+ \ell^-}}

\def\AT#1     {\ensuremath{A_{\mathrm{T}}^{#1}}\xspace}           % 2

\def\C#1      {\ensuremath{\mathcal{C}_{#1}}\xspace}                       % 9
\def\Cp#1     {\ensuremath{\mathcal{C}_{#1}^{'}}\xspace}                    % 7
\def\Ceff#1   {\ensuremath{\mathcal{C}_{#1}^{\mathrm{(eff)}}}\xspace}        % 9  
\def\Cpeff#1  {\ensuremath{\mathcal{C}_{#1}^{'\mathrm{(eff)}}}\xspace}       % 7
\def\Ope#1    {\ensuremath{\mathcal{O}_{#1}}\xspace}                       % 2
\def\Opep#1   {\ensuremath{\mathcal{O}_{#1}^{'}}\xspace}                    % 7

\newcommand{\aunit}[1]{\ensuremath{\text{\,#1}}}       
\newcommand{\tev}{\aunit{Te\kern -0.1em V}\xspace}
\newcommand{\gev}{\aunit{Ge\kern -0.1em V}\xspace}
\newcommand{\mev}{\aunit{Me\kern -0.1em V}\xspace}
\newcommand{\kev}{\aunit{ke\kern -0.1em V}\xspace}
\newcommand{\ev}{\aunit{e\kern -0.1em V}\xspace}
 
\newcommand{\mevc}{\ensuremath{\aunit{Me\kern -0.1em V\!/}c}\xspace}
\newcommand{\gevc}{\ensuremath{\aunit{Ge\kern -0.1em V\!/}c}\xspace}
\newcommand{\mevcc}{\ensuremath{\aunit{Me\kern -0.1em V\!/}c^2}\xspace}
\newcommand{\gevcc}{\ensuremath{\aunit{Ge\kern -0.1em V\!/}c^2}\xspace}
\newcommand{\gevgevcccc}{\ensuremath{\gev^2\!/c^4}\xspace} % for q^2

\def\cm   {\aunit{cm}\xspace}

\def\fm   {\aunit{fm}\xspace}

\def\fb   {\ensuremath{\aunit{fb}}\xspace}
\def\invfb   {\ensuremath{\fb^{-1}}\xspace}
\def\ab   {\ensuremath{\aunit{ab}}\xspace}
\def\invab   {\ensuremath{\ab^{-1}}\xspace}

\def\gsim{{~\raise.15em\hbox{$>$}\kern-.85em
          \lower.35em\hbox{$\sim$}~}\xspace}
\def\lsim{{~\raise.15em\hbox{$<$}\kern-.85em
          \lower.35em\hbox{$\sim$}~}\xspace}

\newcommand{\Real}{\ensuremath{\mathcal{R}e}\xspace}

\def\sPlot{\mbox{\em sPlot}\xspace}

\def\sqs   {\ensuremath{\protect\sqrt{s}}\xspace}

\def\pt         {\ensuremath{p_{\mathrm{T}}}\xspace}

\def\degrees{\ensuremath{^{\circ}}\xspace}

\def\mrad{\aunit{mrad}\xspace}

\def\tell1  {TELL1\xspace}
\def\ukl1   {UKL1\xspace}

\newcommand{\eg}{\mbox{\itshape e.g.}\xspace}
\newcommand{\ie}{\mbox{\itshape i.e.}\xspace}

\newcommand{\etc}{\mbox{\itshape etc.}\xspace}

\newcommand{\lhcborcid}[1]{\href{https://orcid.org/#1}{\hspace*{0.1em}\raisebox{-0.45ex}{\includegraphics[width=1em]{LHCb/Figures/orcidIcon.pdf}}}}